\documentclass[amsmath,amssymb,11pt]{article}
\pdfoutput=1 % if your are submitting a pdflatex (i.e. if you have
\usepackage{jheppub2} % for details on the use of the package, please
\usepackage[T1]{fontenc} % if needed

\usepackage{extarrows}
\usepackage{epsfig}
\usepackage{braket}
\usepackage{ulem}
\usepackage{graphicx}
\usepackage{float}
\usepackage{dsfont}
\usepackage{subcaption}
\usepackage{amsmath}
\usepackage{amsmath,amsopn}
\usepackage{xcolor }
\usepackage{bbold}
\usepackage{tikz}
\usepackage{mathrsfs}
\usepackage{amsmath}
\usepackage{amssymb}
\usepackage{color}
\usepackage[utf8]{inputenc}
\usepackage{graphicx}
\usepackage{bm}
\usepackage{mathrsfs} %Skri symbols
\usepackage{braket}
\usepackage{dsfont}
\usepackage{tikz}
\usepackage{tikz}
\usetikzlibrary{arrows}
\usetikzlibrary{decorations.markings, decorations.pathmorphing}

\usetikzlibrary{arrows}

\newcommand{\rmd}{\mathrm{d}}

\DeclareMathOperator{\arctanh}{arctanh}

\definecolor{pink1}{rgb}{0.858, 0.188, 0.478}

\newcommand{\beq}{\begin{equation}}

\newcommand{\eeq}{\end{equation}}

\title{\boldmath Static sphere observers and geodesics in Schwarzschild-de Sitter spacetime}

\author[a,b,c]{Mir Mehedi Faruk,}
\author[b,c]{Edward Morvan,}
\author[b,c]{Jan Pieter van der Schaar}

\affiliation[a]{Department of Physics, McGill University, Montreal, QC, H3A 2T8, Canada}
\affiliation[b]{Institute of Physics, University of Amsterdam, Science Park 904, PO Box 94485, 1090 GL Amsterdam, the Netherlands}
\affiliation[c]{Delta Institute for Theoretical Physics, Science Park 904, PO Box 94485, 1090 GL Amsterdam, the Netherlands}

\emailAdd{ 
mir.faruk@mail.mcgill.ca, edward.morvan@gmail.com,j.p.vanderschaar@uva.nl}

\abstract{We analyze null- and spacelike radial geodesics in Schwarzschild-de Sitter spacetime connecting two conjugate static sphere observers, i.e. free-falling observers at a fixed radius in between the two horizons. We explicitly determine the changes in the causal structure with respect to these natural observers as a result of the inward bending of the black hole singularity, as well as the outward bending of asymptotic infinity. Notably, the inward and outward bending changes as a function of the black hole mass, first increasing towards a maximum and then decreasing to vanish in the extreme Nariai limit. For a generic mass of the black hole this implies the existence of finite size (temporal) windows for the presence of symmetric radial geodesics between the static sphere observers probing the interior region of the black hole, as well as the exterior de Sitter region. We determine the size of the interior (black hole) and exterior (de Sitter) temporal windows in $4$, $5$ and $6$ spacetime dimensions, finding that they are equal in $D=5$, and compute the proper lengths of the symmetric radial geodesics. We comment on the implications for information exchange and the potential role of the symmetric radial geodesics in a geodesic approximation of static sphere correlators in Schwarzschild-de Sitter spacetime.}
\begin{document} 
\maketitle
\flushbottom

\newpage
\section{Introduction}

Recent studies of quantum gravity in de Sitter \cite{Banks:2000fe,Witten:2001kn,Spradlin:2001pw,Goheer:2002vf,Klemm:2004mb,Anninos:2012qw,Susskind:2021dfc,Susskind:2021omt,Morvan:2022ybp,Galante:2023uyf,Svesko:2022txo,Balasubramanian:2002zh} suggest that, inspired by recent progress in our understanding of quantum gravity in Anti-de Sitter space \cite{Maldacena:2013xja,Penington:2019npb,Almheiri:2019psf,
Balasubramanian:1999zv}, non-perturbative contributions in the (Euclidean) path integral could play an important role \cite{Susskind:2021dfc, Anninos:2022ujl, Anninos:2023epi, Mirbabayi:2023vgl, Cotler:2024xzz}. %complete the list with our own papers%
The simplest, most elementary, solutions of relevance in this context are black holes in de Sitter space, whose contribution to the Euclidean path integral can be (partially) understood using the formalism of constrained instantons \cite{Cotler:2020lxj,Morvan:2022ybp}. %complete the reference list% 
Very concretely, it would be interesting to understand how such non-perturbative contributions affect specific correlators in de Sitter space. Notably, if one assumes the existence of a discrete Hilbert space of states in de Sitter, as suggested by the finite entropy of a de Sitter static patch causal region, one would expect on general grounds that the semi-classical late-time behaviour of large mass correlators between conjugate static-patch observers is modified from exponential decay to a growing `ramp' type behavior that ultimately lands on a constant `plateau' average \cite{Saad:2018ramp, Mirbabayi:2023vgl}. In the absence of a holographic description and inspired by the important role of non-perturbative effects in Anti-de Sitter quantum gravity, our ultimate goal is to understand the contributions of de Sitter black holes to these late-time, large-mass, conjugate correlators.

As a first step in this direction, in this paper we present results on the detailed causal structure of the Schwarzschild-de Sitter solution, from the perspective of a pair of natural conjugate observers. We call these special free-falling observers `static-sphere' observers, as they are located at a unique radius in between the black hole and cosmological horizons. We are particularly interested in the associated (timelike) geodesics, as they correspond to the natural (radial) locations for two conjugate large mass (scalar) field operators \cite{Bousso:1996au,Aalsma:2020aib,Morvan:2022ybp,Svesko:2022txo}. By analyzing the causal structure from the perspective of these natural observers we compute the finite temporal windows for the appearance of spacelike geodesics going through the black hole interior, as well as the de Sitter exterior. These windows are the result of the inward bending of the black hole singularity and the outward bending of asymptotic de Sitter infinity, resulting in a smaller interior, and at the same time a larger exterior causal diamond as compared to empty de Sitter space. These windows depend on the black hole's mass and specific dimension under consideration, and we compute and compare their interior and exterior size in spacetime dimensions $4$, $5$ and $6$, for which the behavior changes qualitatively.   
% Added sentence below to point the reader immediately to the relevant figures % 
These qualitative differences between $D=4,5$ and $6$ are most clearly and conveniently expressed in terms of the Penrose diagrams, see figures \ref{fig:SdS}, \ref{fig:LeoComp5} and \ref{fig:TimeComp466}.
% Good place for comment 2. Yes, I have added a comment below. %
Notably, in $5$ spacetime dimensions the interior inward and exterior outward bending, and consequently their temporal windows, are the same, which can be traced back to the fact that in $D=5$ the black hole small $r$ limit of the geometry scales in the same way as the large $r$ de Sitter limit. And in every dimension the windows show an initial increase towards a maximum, but then decrease to vanish again in the extreme Nariai limit. This detailed shape of the causal Penrose diagram is also of interest for information recovery and exchange protocols involving gravitational shockwaves \cite{Cornalba:2006xk, Freivogel:2019lej, Freivogel:2019whb,Geng:2020kxh,Geng:2021wcq}, which in de Sitter requires disposing of positive energy \cite{Aalsma:2021kle, Morvan:2023tele} and therefore relies, under the assumption of spherical symmetry, on the Schwarzschild-de Sitter geometry as an initial condition.

The opening of temporal windows implies the presence of spacelike geodesics going through the interior black hole, as well as the exterior de Sitter, region connecting the two static sphere observers. We compute the lengths of these spacelike geodesics and consider their limits as we approach the boundaries of the temporal windows. In general, assuming a holographic description exists, understanding the appearance and structure of (spacelike) geodesics can be associated to features of holographic entanglement. And as is well known \cite{Fidkowski:2003nf}, using a geodesic approximation for (large mass) correlators the limits towards the boundaries of the temporal windows imply a transition in the correlator, due to the expected appearance of complex geodesics. As such the specific behavior of geodesics between these static sphere observers probe the black hole singularity, as well as the structure of (exterior) de Sitter asymptotic infinity. We study these limits and show that they behave differently for the interior and exterior spacelike geodesics. 

The organisation of this paper is as follows. We start with a short reminder of computing geodesics in spherically symmetric spacetimes, using (complex) static coordinates, ending with the Anti-de Sitter black hole and empty de Sitter space as useful examples. We then introduce the Schwarzschild-de Sitter geometry, explain the special role of 'static sphere' observers, and then study interior and exterior null geodesics to determine the temporal window for the existence of spacelike geodesics in $4$, $5$ and $6$ spacetime dimensions. Having done that we then look at symmetric spacelike geodesics connecting the static sphere observers, determining their length and analyzing the limits to the boundaries of the temporal windows. We end with some conclusions and suggestions for further research.

\section{Geodesics in static spherically symmetric spacetimes}	

In this introductory section we will set up our notation and conventions for studying radial geodesics. We will remind the reader how to probe the details of the causal structure, in particular the (inward) bending of the black hole singularity, using null geodesics and the corresponding existence of symmetric spacelike radial geodesics probing the interior of the black hole. After setting up the general structure of the geodesic analysis, we end with a reminder of two well-known basic examples of interest, namely Schwarzschild-Anti-de Sitter and pure de Sitter spacetime, before we move on and study the particular spacetime of interest, Schwarzschild-de Sitter.

\subsection{Preliminaries}

To start, let us briefly review the analysis of geodesics in a general curved spherically symmetric background. Geodesics can be understood as stationary points for the action of a point particle
\beq
I = \int\rmd \lambda\, g_{\mu\nu}\rmd\dot x^\mu\rmd\dot x^\nu ~,
\eeq
here the $\dot{x}^\mu$ represents the derivative with respect to the affine parameter $\lambda$. This naturally results in the well known geodesic equation
\beq
\ddot x^\mu + \Gamma_{\nu\sigma}^\mu \dot x^\nu\dot x^\sigma = 0\,,\label{geodesic eq}
\eeq
where $\Gamma_{\nu\sigma}^\mu$ is the usual Levi-Civita connection symbol (torsion-free). As $\lambda$ is an affine parameter, defining $ \epsilon\equiv g_{\mu\nu}\dot x^\mu\dot x^\nu$, we can classify the different types of geodesics as
\beq
 \epsilon=
\begin{cases}
~~0 \qquad \text{null-like}\\
+1 \qquad \text{space-like} \\
-1 \qquad \text{time-like}
\end{cases}
\eeq
Let us now focus on spherically symmetric line elements of the following form
\begin{equation}
    ds^2=-f(r)dt^2+\frac{1}{f(r)}dr^2+ r^2 \, d\Omega^2_{D-2}\,,\label{general black}
\end{equation}
where $d\Omega_{D-2}$ is the volume element of an unit sphere in $D-2$ dimensions. For radial geodesics we then obtain the following equation
% from the  geodesic equation \eqref{geodesic eq},
\begin{equation}
 -f(r)\dot t^2 + f(r)^{-1}\dot r^2 =\epsilon\,.
\label{eq:geogen}
\end{equation}
These geometries feature a timelike Killing vector $\partial_t$, one can therefore infer the existence of a conserved quantity $\mathcal{E}$ along a geodesic given by
\beq
\mathcal{E}=\dot{t}\,f(r)\,.
\label{eq:Et}
\eeq 
Using this in equation (\ref{eq:geogen}), one then obtains
\begin{equation}
 \dot r^2 -\mathcal{E}^2=\epsilon f(r)\,,
\label{kuddus}
\end{equation}

% %%%%%%%%%%%%%%%%%%%%%%%%%%%%%%%%%%%%%%%%%%%

%\subsection{Null and spacelike geodesics}

Let us now concentrate on null and spacelike geodesics. Studying null geodesics ($\epsilon=0$) allows one to map out the causal structure of the spacetime under consideration, which can be transformed into a Penrose diagram. For radial null geodesics the two relevant equations (\ref{eq:Et}) and (\ref{kuddus}) turn into
\begin{eqnarray}
%&&\dot r^2 = E^2 ~\nonumber\\
%\implies &&
\dot{r}=\pm \mathcal{E}\quad\Longleftrightarrow \quad \frac{\dot{r}}{\dot{t}} = \frac{dr}{dt} = \pm f(r) \,,
\end{eqnarray}
where the sign determines the direction of the radial null geodesic (incoming or outgoing). 
% Clearly, this implies
This equation can be solved for $t(r)$, for an outgoing radial geodesic ($r'>r_{\cal O}$), giving
\beq
t(r')-\tau
=\int_{\tau}^{t(r')}\rmd t=\int_{r_{\mathcal{O}}}^{r'}\frac{\dot{t}}{\dot{r}}\rmd r=\int_{r_{\mathcal{O}}}^{r'}\frac{\rmd r}{f(r)}\,,
\label{eq:nullgeo}
\eeq
where $\tau$ labels the observer time at which the null trajectory intersects $r'=r_{\mathcal{O}}$ (see figure \ref{fig:SAdspenrose}). Inserting the value of $r'$ in the above integral fixes the endpoint of the null radial geodesic under consideration, so for an incoming radial geodesic ($r'<r_{\cal O}$) one needs to add a minus sign on the right. For the Schwarzschild-de Sitter geometry, we will be interested in radial null geodesics passing through the black hole horizon and reaching the singularity at $r'=0$, as well as radial null geodesics crossing the cosmological horizon and reaching asymptotic future infinity $\mathcal{I}^+$ where $r'\rightarrow\infty$.

% %%%%%%%%%%%%%%%%%%%%%%%%%%%%%%%%%%%%%%%%%%%

%\subsection{Space-like geodesics}
Space-like separated points ($\epsilon=+1$) satisfy the
% are of particular interest, as their lengths $\mathcal{D}$ can be used to obtain the correlator of a field $\phi$ with mass $m$
%\begin{equation}
 %   \braket{\phi \phi}=\sum_i^N e^{-m\,\mathcal{D}_i}\,.
%\end{equation}
%equation \ref{kuddus} implies
following (radial) geodesic relation, 
\begin{equation}
 \dot r^2 -\mathcal{E}^2= f(r)\quad\Longleftrightarrow\quad\frac{\dot{r}}{\dot{t}}=\frac{dr}{dt}= \pm \frac{f(r)\sqrt{f(r)+\mathcal{E}^2}}{\mathcal{E}}\,,
\label{eq:spacelike}
\end{equation}
where once again we made use of equation (\ref{eq:Et}). For outgoing radial spacelike geodesics (selecting the positive sign and assuming $\mathcal{E}$ is positive) this then results in the following (coordinate) time difference 
\beq
t(r')-\tau
% =\int_{\tau}^{t(r)}dt
=\int_{r_{\mathcal{O}}}^{r'}\frac{\dot{t}}{\dot{r}} \, dr=\int_{r_{\mathcal{O}}}^{r'}\frac{\mathcal{E}\,dr}{f(r)\sqrt{f(r)+\mathcal{E}^2}}\,.
\label{eq:spgeot}
\eeq
As before $\tau$ refers to the initial starting time of the space-like geodesic
at the position of the observer $r=r_{\mathcal{O}}$. Again, for incoming space-like radial geodesics one should add a minus sign. Also note that for space-like geodesics a sign flip of the conserved quantity $\mathcal{E}$ changes the sign of the (coordinate) time difference. Clearly, the initial slope $\frac{dr}{dt}$ at $r=r_\mathcal{O}$ of a radial spacelike geodesic is given by the expression on the right-hand side of equation \eqref{eq:spacelike} evaluated at $r=r_\mathcal{O}$, which will depend on (the sign of) $\mathcal{E}$, asymptoting to the slope of a null geodesic in the limit $\mathcal{E}\rightarrow \pm \infty$. 

As we will soon discover, for symmetric radial space-like geodesics the time $\tau$ is bounded, for real values of the parameter $\mathcal{E}$, where the boundaries are given by special null geodesics ($\mathcal{E}\rightarrow \pm \infty$). Given a starting time $\tau$ and the conserved quantity $\mathcal{E}$, the length of the associated symmetric, radial, space-like geodesic is given by
\beq
 \mathcal{D}=2\int_{r_{\mathcal{O}}}^{r_t}\frac{dr}{\sqrt{f(r)+\mathcal{E}^2}}\,,
\label{eq:spgeoL}
\eeq
where we introduced the turning point radius $r_t$, which is obtained by solving
\beq
f(r)+\mathcal{E}^2=0\,.
\eeq
As will become clear soon, the turning points correspond to the minimal (interior) or maximal (exterior) radii a symmetric spacelike geodesic can probe as a function of the conserved quantity $\mathcal{E}$. For vanishing $\mathcal{E}$, the turning points actually coincide with the horizon radii, implying that the radial geodesics never cross the interior or exterior region. In principle, combining equations \eqref{eq:spgeot} and \eqref{eq:spgeoL}, we can compute the length of these symmetric radial geodesics for allowed $\tau$ and (related) conserved quantity $\mathcal{E}$.

%however in practice the function $t(E)$ is very difficult to invert to obtain $E(t)$ in full generality.
%\newpage

\subsection{Radial geodesics in Schwarzschild-Anti-de Sitter spacetime}

In this section we want remind the reader of the techniques and results described in the paper by Fidkowski et. al. \cite{Fidkowski:2003nf}, as they will play a key role in our study of Schwarzschild-de Sitter. 
In \cite{Fidkowski:2003nf} it was shown explicitly that the causal structure of Schwarzschild Anti-de Sitter (SAdS) spacetimes in dimensions bigger than $3$ cannot be drawn as a perfect square with respect to the two timelike boundaries, i.e. the black hole singularity at $r=0$ is bending inwards. This then meant that the boundary CFT correlators probe the singularity from a particular (boundary) time forward. To briefly review these results, we start with the five dimensional AdS blackening factor %given by \cite{},
\begin{eqnarray}
f(r)=1+\frac{r^2}{\ell^2}-\frac{r_b^2}{r^2}\left( 1+\frac{r_b^2}{\ell^2}\right)
\end{eqnarray}
where, $\ell$ is the AdS radius and $r_b$ is the horizon radius. To further simplify, let us consider the infinite mass limit of the black hole, i.e. $\frac{r_b}{\ell}\rightarrow\infty$, for which the blackening factor reduces to
\begin{eqnarray}
    f(r)=\frac{r^2}{\ell^2}-\frac{\ell^2}{r^2}. \label{blackeningfactorads}
\end{eqnarray}
Putting the relevant observer at the AdS boundary $r_\mathcal{O}=\infty$ equation \eqref{eq:nullgeo} dictates that
\begin{eqnarray}
    t_{SAdS}=\tau+
    \frac{\ell}{2}\left[(1-i)\frac{\pi}{2}+\arctanh\left(\frac{r}{\ell}\right)-\arctan\left(\frac{r}{\ell}\right)\right]\,,
\end{eqnarray}
From this expression we learn that a null geodesic starting from the AdS boundary reaches the black hole singularity $r=0$ at a coordinate time that now contains an imaginary contribution  
\begin{eqnarray}
 t_{SAdS}(r=0)=\tau+\frac{\pi}{4}(1-i)\,\ell\,.
\label{eq:tfSAdS}
\end{eqnarray}
The origin of the imaginary contribution can be understood by noting that by introducing a complex time coordinate one can cover all the four different patches of this global SAdS geometry using the same static coordinates, with the imaginary part identifying the appropriate region. As one crosses the event horizon, the imaginary part shifts by $-i \beta/4$, where $\beta$ is the inverse temperature of the geometry, and the (real) time and space coordinates swap roles. Going around all four patches (counter-clockwise, starting in the left causal patch), each time crossing an event horizon, the imaginary part changes from $0$ to $- i \beta$, consistent with the periodicity of the imaginary time coordinate (i.e. Euclidean time). The inverse temperature $\beta=\pi \ell$ in the infinite mass limit of the SAdS geometry. This structure applies more generally of course: one encounters the same type of behavior for the (complex) time coordinate in all (global) static metrics that feature event horizons.  

This behavior of the complex time coordinate implies that a null geodesic starting from the boundary at $\tau = 0$ reaches the singularity at $r = 0$ off-center in the Kruskal or Penrose diagram, as the real part of $t_{SAdS}(r=0)$ is nonzero. This actually happens for SAdS geometries in $D>3$ \cite{Fidkowski:2003nf,
Festuccia:2005pi}. In fact, in general $D$ dimensions \cite{Fidkowski:2003nf}, a null radial geodesic reaches the the singularity exactly at the center of the diagram only when emitted at a boundary AdS time $\tau = -{\pi\ell }((D-1){\text{tan} (\frac{\pi}{D-1})})^{-1}$, denoted as $-\mathcal{T}$ in figure \ref{fig:penrose-a}. 
So apart from the $3$-dimensional BTZ black hole, the black hole singularity in the SAdS geometry in all higher dimensions is bend inwards, with respect to asymptotic AdS observers (drawn as straight AdS boundary lines), see figures \ref{fig:penrose-b} and \ref{fig:penrose-a}.\\\\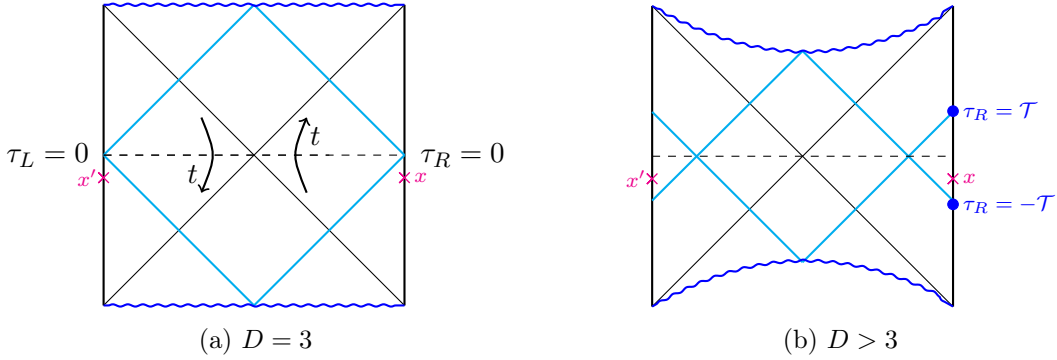
\begin{figure}[h]
\centering
\hfill
\begin{subfigure}[t]{0.5\textwidth}
\centering
\begin{tikzpicture}[scale=1]
\draw[thick] (0,0) -- (0,4);
\draw[thick] (4,0) -- (4,4);
\draw (0,0) -- (4,4);
\draw (0,4) -- (4,0);

\draw[magenta, line width=0.5pt] (4,1.7) -- ++(-2pt,-2pt) -- ++(4pt,4pt);
\draw[magenta, line width=0.5pt] (4,1.7) -- ++(-2pt,2pt) -- ++(4pt,-4pt);
\node[magenta, anchor=west] at (4,1.7) {\scriptsize $x$};

\draw[magenta, line width=0.5pt] (0,1.7) -- ++(-2pt,-2pt) -- ++(4pt,4pt);
\draw[magenta, line width=0.5pt] (0,1.7) -- ++(-2pt,2pt) -- ++(4pt,-4pt);
\node[magenta, anchor=west] at (-0.48,1.7) {\scriptsize $x'$};
% \draw[decoration={markings, mark= at position 0.5 with {\arrow{latex}}}, postaction={decorate}, thick, color=red] (0,2) -- (2,4) node[midway, above right]{};
% \draw[decoration={markings, mark= at position 0.5 with {\arrow{latex}}}, postaction={decorate}, thick, color=red] (4,2) -- (2,4) node[midway, above left]{};

% \draw[decoration={markings, mark= at position 0.5 with {\arrowreversed{latex}}}, postaction={decorate}, thick, color=red] (0,2) -- (2,0) node[midway, above right]{};
% \draw[decoration={markings, mark= at position 0.5 with {\arrowreversed{latex}}}, postaction={decorate}, thick, color=red] (4,2) -- (2,0) node[midway, above right]{};

\draw[thick, cyan] (0,2) -- (2,4);
\draw[thick, cyan] (4,2) -- (2,4);
\draw[thick, cyan] (0,2) -- (2,0);
\draw[thick, cyan] (4,2) -- (2,0);

\draw[dashed] (0,2) -- (4,2) node[pos=1.02,right]{$\tau_R=0$};
\draw[dashed] (0,2) -- (3,2) node[pos=-0.02,left]{$\tau_L=0$};

\draw[->,thick] (2.7,1.5) .. controls (2.5,2) .. (2.7,2.5) node[pos=0.8,right]{$t$};
\draw[<-,thick] (1.3,1.5) .. controls (1.5,2) .. (1.3,2.5) node[pos=0.2,left]{$t$};

% Zigzag line bending inward (inside the square)
\tikzset{decoration={snake,amplitude=.15mm,segment length=2mm,
                       post length=.6mm,pre length=.6mm}};
\draw[blue,thick,decorate] (0,0) -- (4,0);
\draw[blue,thick,decorate] (0,4) -- (4,4);
\end{tikzpicture}
\caption{$D=3$}
\label{fig:penrose-b}
\end{subfigure}\hfill
\begin{subfigure}[t]{0.5\textwidth}
\centering
\begin{tikzpicture}[scale=1]
\draw[thick] (0,0) -- (0,4);
\draw[thick] (4,0) -- (4,4);
\draw (0,0) -- (4,4);
\draw (0,4) -- (4,0);

% \draw[decoration={markings, mark= at position 0.5 with {\arrow{latex}}}, postaction={decorate}, thick, color=green] (2,0.59) -- (0,2.59) node[midway, above right]{};
% \draw[decoration={markings, mark= at position 0.5 with {\arrow{latex}}}, postaction={decorate}, thick, color=green] (2,0.59) -- (4,2.59) node[midway, above right]{};

% \draw[decoration={markings, mark= at position 0.5 with {\arrowreversed{latex}}}, postaction={decorate}, thick, color=green] (2,3.4) -- (4,1.41) node[midway, above right]{};

% \draw[decoration={markings, mark= at position 0.5 with {\arrowreversed{latex}}}, postaction={decorate}, thick, color=green] (2,3.4) -- (0,1.41) node[midway, above right]{};

\draw[thick, cyan] (2,3.4) -- (0,1.41);
\draw[thick, cyan] (2,3.4) -- (4,1.41);
\draw[thick, cyan] (2,0.59) -- (4,2.59);
\draw[thick, cyan] (2,0.59) -- (0,2.59);

\draw[magenta, line width=0.5pt] (4,1.7) -- ++(-2pt,-2pt) -- ++(4pt,4pt);
\draw[magenta, line width=0.5pt] (4,1.7) -- ++(-2pt,2pt) -- ++(4pt,-4pt);
\node[magenta, anchor=west] at (4,1.7) {\scriptsize $x$};

\draw[magenta, line width=0.5pt] (0,1.7) -- ++(-2pt,-2pt) -- ++(4pt,4pt);
\draw[magenta, line width=0.5pt] (0,1.7) -- ++(-2pt,2pt) -- ++(4pt,-4pt);
\node[magenta, anchor=west] at (-0.48,1.7) {\scriptsize $x'$};

% \draw[decoration={markings, mark= at position 0.5 with {\arrow{latex}}}, postaction={decorate}, thick, color=red] (0,2) -- (1.45,3.45) node[midway, above right]{};

% \draw[decoration={markings, mark= at position 0.5 with {\arrow{latex}}}, postaction={decorate}, thick, color=red] (4,2) -- (2.55,3.45) node[midway, above left]{};

% \draw[decoration={markings, mark= at position 0.5 with {\arrowreversed{latex}}}, postaction={decorate}, thick, color=red] (0,2) -- (1.45,0.55) node[midway, above right]{};
% \draw[decoration={markings, mark= at position 0.5 with {\arrowreversed{latex}}}, postaction={decorate}, thick, color=red] (4,2) -- (2.55,0.55) node[midway, above right]{};

% \draw[thick, cyan] (0,2) -- (1.45,0.55);

 \draw[dashed] (0,2) -- (4,2);
% \draw[dashed] (0,2) -- (4,2) node[pos=1.02,right]{$t_R=0$};
% \draw[dashed] (0,2) -- (3,2) node[pos=-0.02,left]{$t_L=0$};

\filldraw[blue] (4,2.6) circle (2pt) node[anchor=west]{ \scriptsize
$\tau_R=\mathcal{T}$};

\filldraw[blue] (4,1.36) circle (2pt) node[anchor=west]{ \scriptsize
$\tau_R=-\mathcal{T}$};

% Label for t_S = -t_C
%\node at (4.3, 2.7) {$t_C$};
%\node at (4.37, 1.2) {$-t_C$};
%\node at (-0.4, 2.6) {$t_C$};
%\node at (-0.5, 1.4) {$-t_C$};
% Zigzag line bending inward (inside the square)
\draw[thick, color=blue, decorate, decoration={snake,amplitude=.15mm,segment length=2mm,
                       post length=.6mm,pre length=.6mm}] (0,4) to[out=-30, in=-150] (4,4);
\draw[thick, color=blue, decorate, decoration={snake,amplitude=.15mm,segment length=2mm,
                       post length=.6mm,pre length=.6mm}] (0,0) to[out=30, in=150] (4,0);
\end{tikzpicture}
\caption{$D>3$}
\label{fig:penrose-a}
\end{subfigure}
\caption{Penrose diagrams of the SAdS geometry. The black $45^{\circ}$ lines stand for black hole horizon, whereas the blue ones stand for null rays. The left panel shows the $D=3$ case, whereas for any dimensions $D>3$ the situation is described by the right panel. The singularity appears to bend with respect to (straight) vertical lines standing for the asymptotic AdS boundary observers.}
\label{fig:SAdspenrose}
\end{figure}

After studying the null geodesics and the associated causal structure, we can now also look for radial space-like geodesics connecting the symmetric points $x$ and $x'$ on the opposite boundaries (see figure \ref{fig:penrose-b}). For a specific boundary time, one then finds a unique symmetric radial spacelike geodesic, labeled by $\mathcal{E}$. Using \eqref{eq:spgeot} and assuming real $\mathcal{E}$, the  turning points of these symmetric spacelike radial geodesics are given by
\begin{equation}
    r_t=\sqrt{\sqrt{1+\mathcal{E}^4/4}-\mathcal{E}^2/2}\,.
\end{equation}
Not only do the turning points have a vanishing derivative, but for real $\mathcal{E}$ they always lie exactly in the middle of the Penrose diagram, i.e. they correspond to a coordinate time whose real part vanishes, ensuring that the complete spacelike geodesic is a continuous differentiable function, connecting the two opposite static patches of the SAdS geometry. Plugging in the blackening factor \eqref{blackeningfactorads}, one finds that the integral \eqref{eq:spgeot} identifies the following initial boundary time of the symmetric radial spacelike geodesics
\cite{Fidkowski:2003nf},
\begin{equation}
    \tau=\text{Re} \left[ \frac{1}{4}\ln{\left(\frac{1-\mathcal{E}+\mathcal{E}^2/2}{\sqrt{1+\mathcal{E}^4/4}}\right)}-\frac{i}{4}\ln{\left(\frac{1+i\mathcal{E}-\mathcal{E}^2/2}{\sqrt{1+\mathcal{E}^4/4}}\right)} \right] \,.
    \label{eq:TfSAds}
\end{equation}
From this relation between the boundary time and $\mathcal{E}$, one then concludes that (for real $\mathcal{E}$) symmetric radial geodesics only exist within a finite-size temporal window given by $-\mathcal{T}<\tau<\mathcal{T}$. For $\mathcal{E}=0$ the endpoints of the symmetric spacelike radial geodesic is $\tau_R=\tau_L=0$, however, when $\mathcal{E}$ approaches $\pm\infty$ the initial boundary time equals $\tau\rightarrow \pm \mathcal{T}$ and the turning points get arbitrarily close to the singularity. In this limit the radial geodesic becomes null and bounces off the black hole singularity at a finite boundary time and its (appropriately renormalized) length diminishes to zero. 

For boundary times outside of this window the only geodesics connecting symmetric points on the boundary of AdS are labeled by complex values of $\mathcal{E}$, necessarily implying that their length has real and imaginary parts as well \cite{Fidkowski:2003nf,Fischetti:2014uxa}. Although we considered an infinite mass limit to simplify the equations, the same conclusions can be derived more generally \cite{Fidkowski:2003nf}, but it is not as straightforward to derive analytic results using \eqref{eq:tfSAdS} and \eqref{eq:TfSAds}. Specifically, for the Schwarzschild-de Sitter black hole we will not need to rely on special limits to derive analytical results.

\subsection{Radial geodesics in de Sitter spacetime}

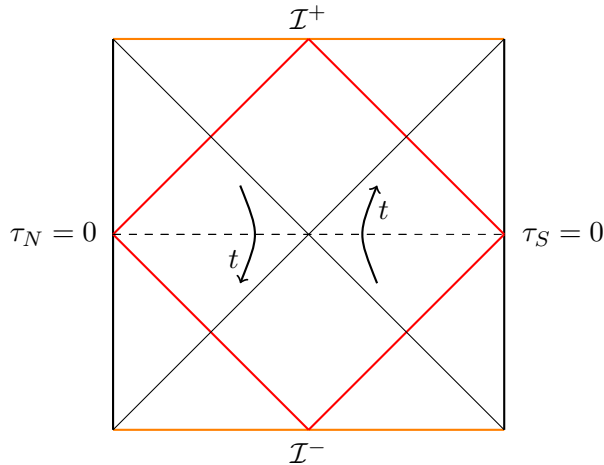
\begin{figure}[t]
\centering
\begin{tikzpicture}[scale=1.3]
\draw[thick,orange] (0,0) -- (4,0);
\draw[thick] (0,0) -- (0,4);
\draw[thick,orange] (0,4) -- (4,4);
\draw[thick] (4,0) -- (4,4);
\node[above] at (2,4) {$\mathcal{I}^+$};
\node[below] at (2,0) {$\mathcal{I}^-$};
\draw[thick,red] (2,4) -- (0,2);
\draw[thick,red] (2,4) -- (4,2);
\draw[thick,red] (2,0) -- (0,2);
\draw[thick,red] (2,0) -- (4,2);

%\draw[decoration={markings, mark= at position 0.5 with {\arrow{latex}}}, postaction={decorate}, thick, color=red] (4,2) -- (2,4) node[midway, above left]{};

%\draw[decoration={markings, mark= at position 0.5 with {\arrowreversed{latex}}}, postaction={decorate}, thick, color=red] (0,2) -- (2,0) node[midway, above right]{};
%\draw[decoration={markings, mark= at position 0.5 with {\arrowreversed{latex}}}, postaction={decorate}, thick, color=red] (4,2) -- (2,0) node[midway, above right]{};

\draw[dashed] (0,2) -- (4,2) node[pos=1.02,right]{$\tau_S=0$};
\draw[dashed] (0,2) -- (3,2) node[pos=-0.02,left]{$\tau_N=0$};

\draw[->,thick] (2.7,1.5) .. controls (2.5,2) .. (2.7,2.5) node[pos=0.8,right]{$t$};
\draw[<-,thick] (1.3,1.5) .. controls (1.5,2) .. (1.3,2.5) node[pos=0.2,left]{$t$};

\draw (0,0) -- (4,4);
\draw (0,4) -- (4,0);

\end{tikzpicture}
\caption{Penrose diagram of the de Sitter spacetime geometry. Null rays sent from the north or south pole at $\tau=0$ meet at future infinity $\mathcal{I}^+$. Similarly, null rays emanating from the center of past infinity $\mathcal{I}^-$ reach the north or south pole at $\tau=0$.}
\label{fig:SymmetricPenrose}
\end{figure}

Another basic example of interest, one that also serves as a limiting case of the full Schwarzschild-de Sitter geometry that we will soon address, is pure de Sitter spacetime. The de Sitter metric, written in terms of static patch coordinates, has following blackening factor in eq. \eqref{general black}
\begin{eqnarray}
  &&  f(r)=1-\frac{r^2}{\ell^2},
  \label{eq:dsblackening}
\end{eqnarray}
with the coordinate singularity at $r=\ell$ corresponding to the cosmological horizon. In this metric the only free-falling observer at fixed radius, determined by  
\begin{equation}
f'(r)\big|_{r=r_{\mathcal{O}}}=0 \, . 
\label{eq:posobs}
\end{equation} 
This of course gives the expected result $r_\mathcal{O}=0$, i.e. the observer situated at the center of the static patch that is surrounded by the cosmological horizon. In terms of a global slicing, one can identify this position as one of the poles on the global $3$-sphere (for $4$-dimensional de Sitter).  %\cite{Morvan:2022ybp}. 
The global geometry features another, causally disconnected, static patch observer, for which time runs in the opposite direction, and which can then be identified with the other pole of the spatial $3$-sphere slicing the (global) de Sitter geometry. Let us now introduce radial null geodesics originating from, or arriving at, the two static patch centers, depicted as the red curves in the de Sitter Penrose diagram \ref{fig:SymmetricPenrose}. 
For de Sitter space equation \eqref{eq:nullgeo} turns into
\begin{equation}
t_{dS}(r)=\tau + \int_0^{r}\frac{ \ell^2\,\rmd r'}{(r'-\ell)(r'+\ell)} =\tau+\ell\arctanh{(r/\ell)}\,. \label{int1}
\end{equation}
Taking the limit $r\rightarrow\infty$, probing the structure of asymptotic future infinity $\mathcal{I}^+$, allows one to infer that
\begin{equation}
   t_{dS}=\tau+i\frac{\pi \ell}{2} \, .
   \label{eq:dsnull}
\end{equation}
As before, the imaginary part is generated by crossing the cosmological horizon, however as opposed to black holes, it is now equal to  a quarter of its Euclidean periodicity $\beta_{dS}=1/T_{dS}=2\pi\ell$. This sign difference is related to the fact when energy $M$ is added to de Sitter space, its entropy $S$ decreases \cite{Gibbons:1976ue,Bousso:1996au,Morvan:2022ybp,morvan:2022rn,Morvan:2023tele}
\begin{equation}
    \rmd S_{b}=\beta_{b}\,\rmd M\quad,\quad\rmd S_{dS}=-\beta_{dS}\,\rmd M\,,
\end{equation}
 where $\beta$ is kept positive definite.  Equation \ref{eq:dsnull} ensures that a null radial geodesic moving outward from the symmetric center of the north or south pole, i.e. $(\tau=0, r_{\mathcal{O}}=0)$, ends up exactly at the center of asymptotic future infinity $\mathcal{I}^+$, as the real part of the complex time coordinate vanishes. Similarly, inward moving radial null geodesics arrive at the observer in the center exactly at the symmetric $\tau=0$ point when originating from the center of past infinity $\mathcal{I}^-$. This means that, with respect to the two free-falling observers at the origin of the two conjugate static patches, past infinity $\mathcal{I}^-$ and future infinity $\mathcal{I}^+$ do not bend inward or outward. Combined with the de Sitter isometries, which guarantees the equivalence of all free-falling observers, this proves that the Penrose diagram is a perfect square, with past and future infinity represented by straight horizontal orange lines in figure \ref{fig:SymmetricPenrose}. 

Let us now consider symmetric radial spacelike geodesics in pure de Sitter spacetime. With symmetric we main connecting two points at the separate poles at opposite static times ($\tau_S=-\tau_N$). As is well known, in pure de Sitter, the unique time for which a family of radial spacelike geodesics exists is at the symmetric center, i.e. $\tau_S=-\tau_N=0$, for any value of the conserved quantity $\mathcal{E}$. The turning points of these symmetric radial geodesics are found at $r_t=l\sqrt{1+\mathcal{E}^2}$, allowing to show starting from (\ref{eq:spgeot}) \cite{Aalsma:2022swk,job,Chapman:2021eyy,Galante:2022nhj}
\begin{eqnarray}
    \tau-t(r_t)=\int_{0}^{r_t} \frac{\mathcal{E}}{(1-\frac{r^2}{\ell^2})(1-\frac{r^2}{\ell^2}+\mathcal{E}^2)}dr=i\frac{\beta_{dS}}{4}
\end{eqnarray}
implying that symmetric radial geodesics, assuming the conserved quantity $\mathcal{E}$ is real, only exist when $\tau=0$. Computing the geodesic length, we then find
\begin{equation}
\mathcal{D}_{dS}(\mathcal{E})=2\int_0^{r_t}\frac{\rmd r}{\sqrt{1-\frac{r^2}{\ell^2}+\mathcal{E}^2}}=2\ell\left(\lim_{r\rightarrow r_t}\arctan\frac{r}{r_t-r}\right)=\pi\ell\,.
% =2\ell\int_0^{r_t}\frac{\rmd r}{\sqrt{r_t^2-r^2}}
\end{equation}
Notably, we see that all these symmetric geodesics have the same geodesic length, in other words the length is independent of $\mathcal{E}$. As also pointed out in \cite{Chapman:2021eyy, Galante:2022nhj}, even in the limit $\mathcal{E}\rightarrow \infty$, the geodesic length remains $\pi \ell$. This could be considered surprising as one typically expects spacelike geodesics to become null in this limit. However, one should realize that this limit, for outgoing radial spacelike geodesics in de Sitter space, is expected to be singular, since the null geodesics bounce off asymptotic (spacelike) infinity ($\mathcal{I}^{+}$ or $\mathcal{I}^{-}$) in this particular case (instead of the black hole singularity in the SAdS geometry)\footnote{Analogous to comparing outgoing null and spacelike geodesics in AdS, where the AdS boundary is at a finite distance for null geodesics but at an infinite distance for spacelike geodesics.}. We stress once more that symmetric radial spacelike geodesics  do not exist for any other points (i.e. times) on the two poles. Instead, as shown in \cite{Aalsma:2022swk, Galante:2022nhj}, one needs to introduce complex geodesics in order to connect symmetric points on the north and south pole lying that are not located at $\tau=0$.

This ends our introduction. In the rest of the paper, we will study how the pure de Sitter causal structure changes as one introduces a black hole. We will find that both the (interior) black hole singularity, as well as (exterior) asymptotic infinity, are bend inwards and outwards respectively, with respect to special, static sphere, observers. As a consequence, two conjugate static sphere observers can be connected by interior and exterior symmetric radial spacelike geodesics in a finite temporal window, whose duration is defined by the magnitude of the inward and outward bending of the interior black hole singularity and exterior asymptotic infinity respectively.

%\\\\ A non-zero real part to the integral \eqref{int1} above upon taking the limit $r\rightarrow\infty$ would indicate otherwise as we discuss below. %See for instance \cite{Fidkowski:2003nf}, where they show that the Penrose diagram of the Schwarzschild-AdS blackhole is not presented by a perfect square. Due to the presence of the blackhole the integral \eqref{eq:nullgeo} gives non zero real  value when the null ray starts from the centre of one of the boundary.
%In the context of the SdS black hole, a fascinating narrative unfolds as we see the following subsection.
%\\

\section{Geodesics in Schwarzschild-de Sitter spacetimes}	

We are now ready to explore the geodesic structure of the Schwarzschild-de Sitter (SdS) geometry. We will determine both the size of inward bending of the interior black hole singularity, as well as the exterior outward bending of asymptotic infinity, with respect to a special pair of conjugate static sphere observers. This implies corresponding changes in the size of the interior and exterior causally connected regions connected to these special observers and, related to that, a finite-size temporal window for symmetric radial spacelike geodesics going either through the interior black hole region, or the exterior de Sitter region. 

Before we delve into the detailed computations, let us first briefly sketch the SdS global geometry and how it compares to pure de Sitter. The presence of the black hole in the Sitter requires an embedding of the black hole singularity into an asymptotic de Sitter like structure, changing the topology of the global spatial slices from $S^3$ to $S^2 \otimes S^1$ for the four-dimensional SdS spacetime \cite{Bousso:1996au}. The black hole obviously breaks the dS isometries, which reduce to just time translations (connecting the north and south pole observers) and rotations. This also implies the existence of a special pair of free-falling conjugate observers, namely those that stay at rest with respect to the black hole and cosmological horizon in the two conjugate static patches of global SdS.  We will study the SdS geometry in $4$, $5$ and $6$ spacetime dimensions, for which we can clearly distinguish qualitative differences in the behavior. 

%We successfully show here that the black hole singularity in such geometry bends inward toward the center of the diagram, i.e., the Penrose diagram shrinks in the black hole region. On the other hand, the past and future infinities of the asymptotic de Sitter region bend outward (away from the cosmological horizon).\\\\

\begin{figure}[t]
\centering
\begin{subfigure}{.5\textwidth}
 \centering
\includegraphics[width=7.5cm]{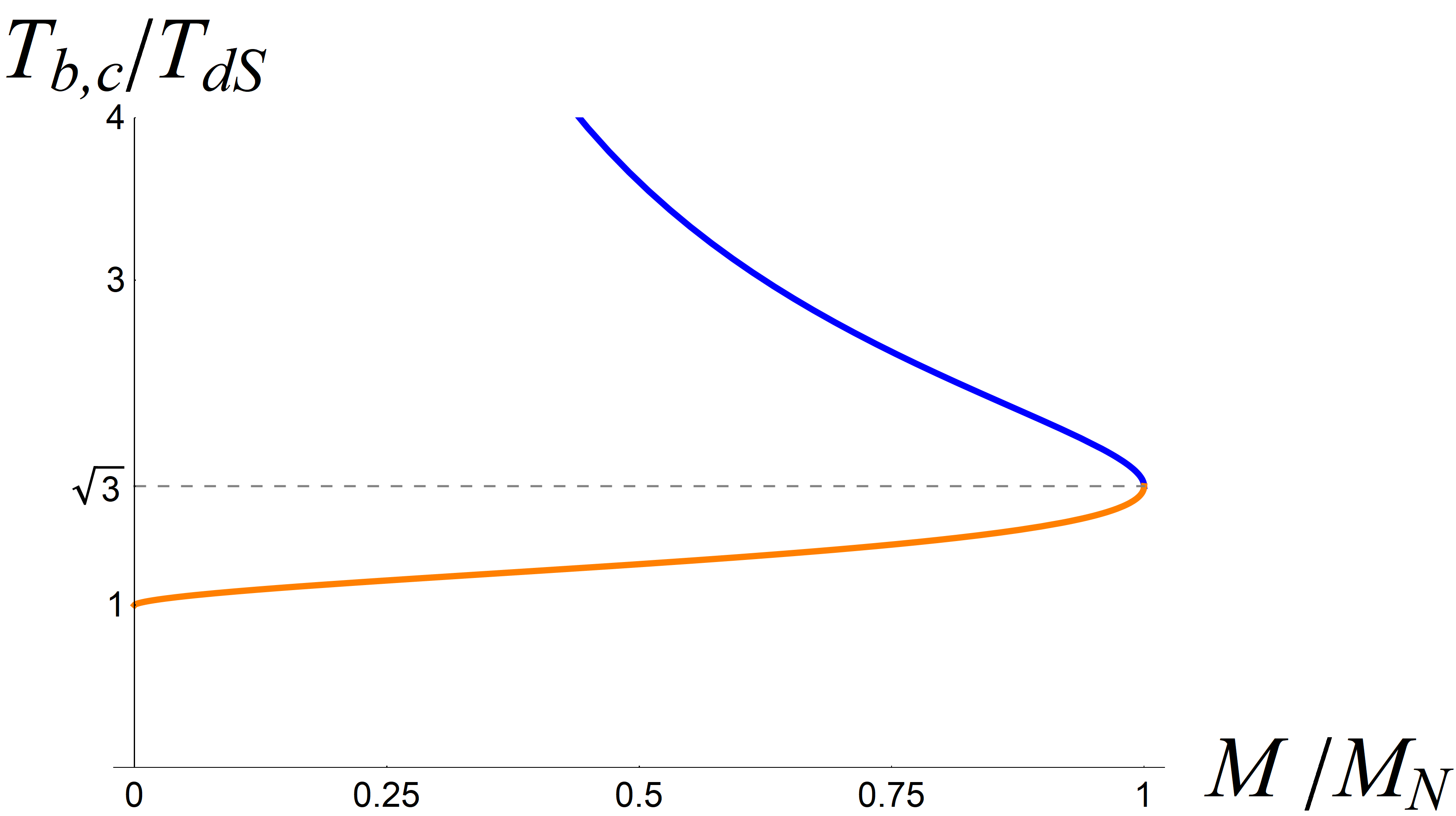}
 \caption{Normalized SdS temperatures in $D=4$}
\end{subfigure}%
\begin{subfigure}{.5\textwidth}
 \centering
\includegraphics[width=7.5cm]{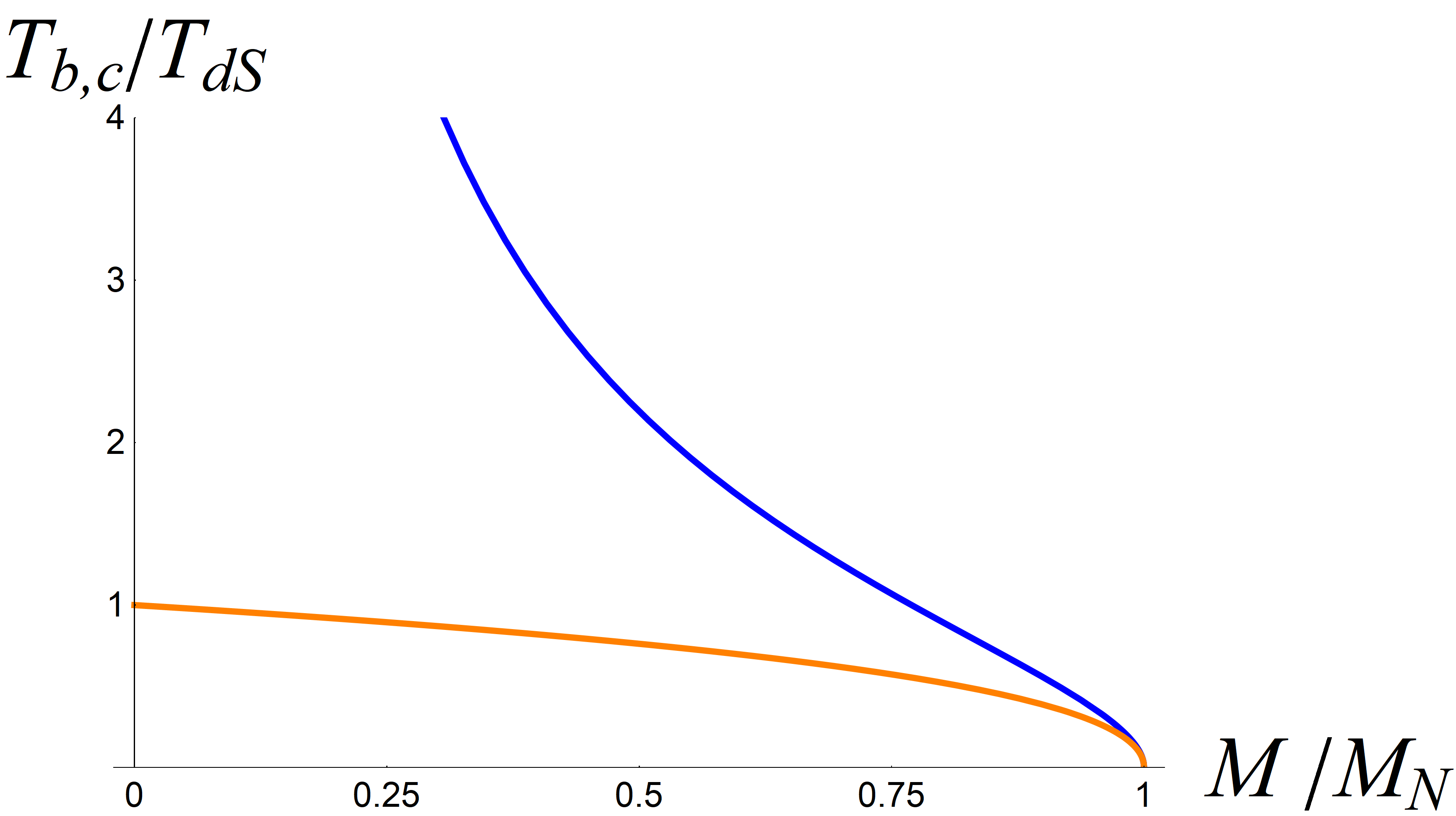}
 \caption{UN-Normalized SdS temperatures in $D=4$}
\end{subfigure}
    \caption{Left: Normalized temperatures $T_{b,c}$ as a function of the black hole mass in $D=4$. The orange and blue 
    curves correspond to the cosmological and black hole temperature respectively. Right: Un-normalized ($\gamma=1$) temperatures. %The plot is constructed from the imaginary part of the integral associated with \eqref{eq:nullgeo}. The plots exactly coincide with the horizon temperatures \cite{Morvan:2022ybp} using periodicity trick in Euclidean time. 
     }
    \label{fig:SdSradtemps}
\end{figure}

In arbitrary dimensions $D$ the blackening factor that describes a black hole in de Sitter space, a.k.a. SdS spacetime, is given \cite{Morvan:2022ybp,morvan:2022rn} by
\begin{equation}
f(r) = 1 - \frac{r^2}{\ell^2} - \alpha\frac{M}{r^{D-3}}\label{bfactorSdS} \, .
\end{equation}
Here, $\alpha$ is a constant defined as
\begin{equation}
\alpha = \frac{16\pi G}{(D-2)\Omega_{D-2}}.
\end{equation}
Where $M$ represents the mass parameter associated with the black hole geometry, and $G$ stands for Newton's constant. This metric features two (spherical) horizons: a black hole and a cosmological horizon, with radii $r_b$ and $r_c$ respectively. The existence of these horizons is due to the presence of two positive real roots of the blackening factor, i.e. where $f(r) = 0$. In contrast to Schwarzschild black holes embedded in flat space, which admit solutions for all values of the mass, in de Sitter space the range of masses is finite. There exists an upper limit for the mass of a black hole in de Sitter space, $M \leq M_N \equiv \frac{2}{\alpha(D-1)}{r_N^{D-3}}$, the Nariai mass \cite{Nariai,Morvan:2022ybp,morvan:2022rn}. The corresponding Nariai radius $r_N$ can be derived from the condition $f(r_N)=f'(r_N)=0$ and equals
\begin{equation}
r_N = \ell\sqrt{\frac{D-3}{D-1}} \, . 
\end{equation}
It follows that for all smaller values of the mass, the black hole horizon is smaller than the cosmological horizon. In the limit of vanishing mass $M=0$, one of course ends up in empty pure de Sitter spacetime. Due to the presence of a black hole, SdS not only has a cosmological temperature but also an additional one, emanating from the black hole. For masses distinct from Nariai, the two temperatures are different (as well as gravitational entropies) and as such the system is not equilibrium, which will not play an important role in our (classical) analysis.  As usual, the temperatures of the black hole and cosmological horizon are related to their respective surface gravities $\kappa_{b,c}$,
\begin{equation}
T_{b,c} = \frac{\kappa_{b,c}}{2\pi}\,.
\end{equation}
It is worth noting that only in the Nariai limit do these two temperatures coincide, while for any other value of $M$, $T_b > T_c$. As the black hole temperature is larger, ultimately this will lead to the decay to empty de Sitter space, which indeed also corresponds to the maximal entropy state. To identify the physical temperatures, corresponding to the relevant Euclidean periodicities, let us discuss the proper normalization of the Killing vectors associated with the surface gravity. The relationship between surface gravity and the Killing vector $\zeta = \gamma \,\partial_t$, which generates time translations, is expressed as follows
\begin{equation}
\zeta^\mu \nabla_\mu \zeta^\nu = \kappa \zeta^\nu\,,
\end{equation}
where the value of $\gamma$ depends on the position of the observer under consideration. In Anti-de Sitter and Minkowski spacetime, one usually considers observers infinitely far from the black hole, rendering $\gamma=1$. However, since that is not a possibility in a SdS, one needs to normalize $\gamma=1/\sqrt{f(r_{\mathcal{O}})}$, sometimes referred to as the Tolman temperature \cite{Toleman}. The natural candidate for this particular radius is the static sphere at $r_\mathcal{O}$, defined by the free-falling observer at a fixed radius in between the black hole and cosmological horizon. This unique value of the radius in a given SdS metric also reduces nicely to the special observers in the pure de Sitter ($r_\mathcal{O}=0$) and flat space limit ($r_\mathcal{O} \rightarrow \infty$). With the natural normalization fixed to observers at $r=r_{\mathcal{O}}$, we can now compute the surface gravities of the black hole and cosmological horizons
\begin{equation}
\kappa_{b,c} = \frac{\gamma}{2}f'(r)\Big|_{r=r_{b,c}}.
\end{equation}
Here, $r_{b,c}$ denote the radii of the black hole and cosmological horizons, respectively. This relation can be evaluated and expressed in arbitrary dimensions as \cite{Morvan:2022ybp},
\begin{eqnarray}
\kappa_b &= \frac{\gamma(D-1)}{2}\frac{r_N^2-r_b^2}{r_b \ell^2}\label{surfacegravity1} \\
\kappa_c &= \frac{\gamma(D-1)}{2}\frac{r_c^2-r_N^2}{r_c \ell^2}\label{surfacegravity2}
    \end{eqnarray}
Let us also stress here that using the standard value of the normalization ($\gamma=1$) actually leads to confusing results, as the temperature associated to both horizons would vanish in the Nariai limit \cite{Gibbons:1977mu} (see figure \ref{fig:SdSradtemps}), whereas the entropy does not \cite{Gibbons:1976ue}. That the more natural normalization is related to a static sphere observer was only realized relatively recently \cite{Bousso:1996au,Bousso:1997wi,Morvan:2022ybp,Svesko:2022txo}. 
%When accounting for this fact, normalization becomes suitable not only for black holes in any spacetime but also for empty de Sitter spacetime. This is due to the fact that $\zeta^2=-1$ is always ensured, both at $r\rightarrow\infty$ for black holes and at $r=0$ for the north or south pole of de Sitter spacetime, 
As already emphasized, for the remainder of this paper, we will consider the natural static sphere observers, which are free-falling at a fixed radius. This radius is determined by the stationary point of the blackening factor $f'(r_{\mathcal{O}})=0$ and in arbitrary dimensions equals \cite{Morvan:2022ybp}
\begin{eqnarray}
r_{\mathcal{O}}^{D-1}=
\frac{(D-3)}{2}r_{b,c}^{D-3}(\ell^2-r_{b,c}^2)
\label{observerpos}.
\end{eqnarray}
The blackening factor at this radius is $f(r_{\mathcal{O}})=1-(r_{\mathcal{O}}/r_N)^2$, implying that the normalization of the temperature becomes $\gamma=\sqrt{1-(r_{\mathcal{O}}/r_N)^2}$. As a consequence, the physical temperature in the Nariai limit, where $r_b=r_c=r_N$, equals \cite{Morvan:2022ybp,Svesko:2022txo} 
\begin{eqnarray}
    \kappa_N=\frac{\sqrt{D-1}}{\ell}
\end{eqnarray}
This nicely reproduces the temperature of the $dS_2 \times S^{D-2}$ near-horizon geometry that appears in the Nariai limit \cite{Bousso:2000md}, i.e. $T_N=\frac{1}{2\pi\ell}$. Interestingly, as briefly reviewed in section 2.3, we would then also expect the causal structure to be represented by a perfect square in the maixmal mass Narai limit. And indeed, that is what we will find. In Figure \ref{fig:SdSradtemps} we plotted the appropriately normalized temperatures, as measured by static sphere observers, as a function of the black hole mass in $4$ and $5$ dimensions. 

\vspace{1.5mm}

As emphasized, the temperatures obtained from \eqref{surfacegravity1} and \eqref{surfacegravity2}
are associated with the timelike Killing vector $\partial_{\tilde{t}}$ instead of $\partial_t$, related to the unique static sphere radius in SdS. Before we move on to discuss null geodesics in SdS, it is important to realize that the relevant (proper) time for a static sphere observer is of course determined by the same normalization factor 
\beq
 t\longrightarrow \tilde{t} \equiv \gamma{t}=t\,\sqrt{f(r_{\mathcal{O}})}=t\,\sqrt{1-(r_{\mathcal{O}}/r_N)^2}
\, . \label{hawking}
% =t\,\sqrt{1-(M/M_N)^{2/3}}
\eeq 
As we will be interested in computing time differences for static sphere observers, in order to understand the causal structure with respect to those observers, it will be important to use the appropriately normalized time parameter $\tilde{t}$, instead of the SdS coordinate time $t$.

\subsection{Radial null geodesics from static sphere observers in SdS}
\label{sec:example-section}

As in the basic examples, we will make use of \eqref{eq:nullgeo} to understand radial null geodesics
in SdS. Importantly, we locate an observer at the static sphere radius \eqref{observerpos}, from which an emitted radial null geodesic either goes inwards towards the black hole horizon or outwards towards the cosmological horizon. As already emphasized, the radius $r_\mathcal{O}$ corresponds to the unique radius for which free-falling observers remain at a fixed distance from the black hole and cosmological horizon respectively. In the Penrose diagrams these locations will be identified with bold (vertical) straight lines. 
 
 %It will become evident shortly that only when $M=0$ and $M=M_N$ does the causal structure of the %underlying geometry allow for representation as perfect squares in the Penrose Diagram. For any %other value of $M$, the singularity structure, future/past infinities, and the static sphere of %the geometry cannot be depicted as straight lines. If one chooses to static sphere as a straight %line in the Penrose diagram, then the singularity must be drawn as bowed in, while future/past %infinities should be represented as bowed outside of the diagram. 
 
%In the SdS spacetime, we need to systematically consider the two horizons when implementing eq. \eqref{eq:nullgeo}.

 \begin{figure}[t]
\centering
\begin{subfigure}{.5\textwidth}
 \centering
\includegraphics[width=7.5cm]{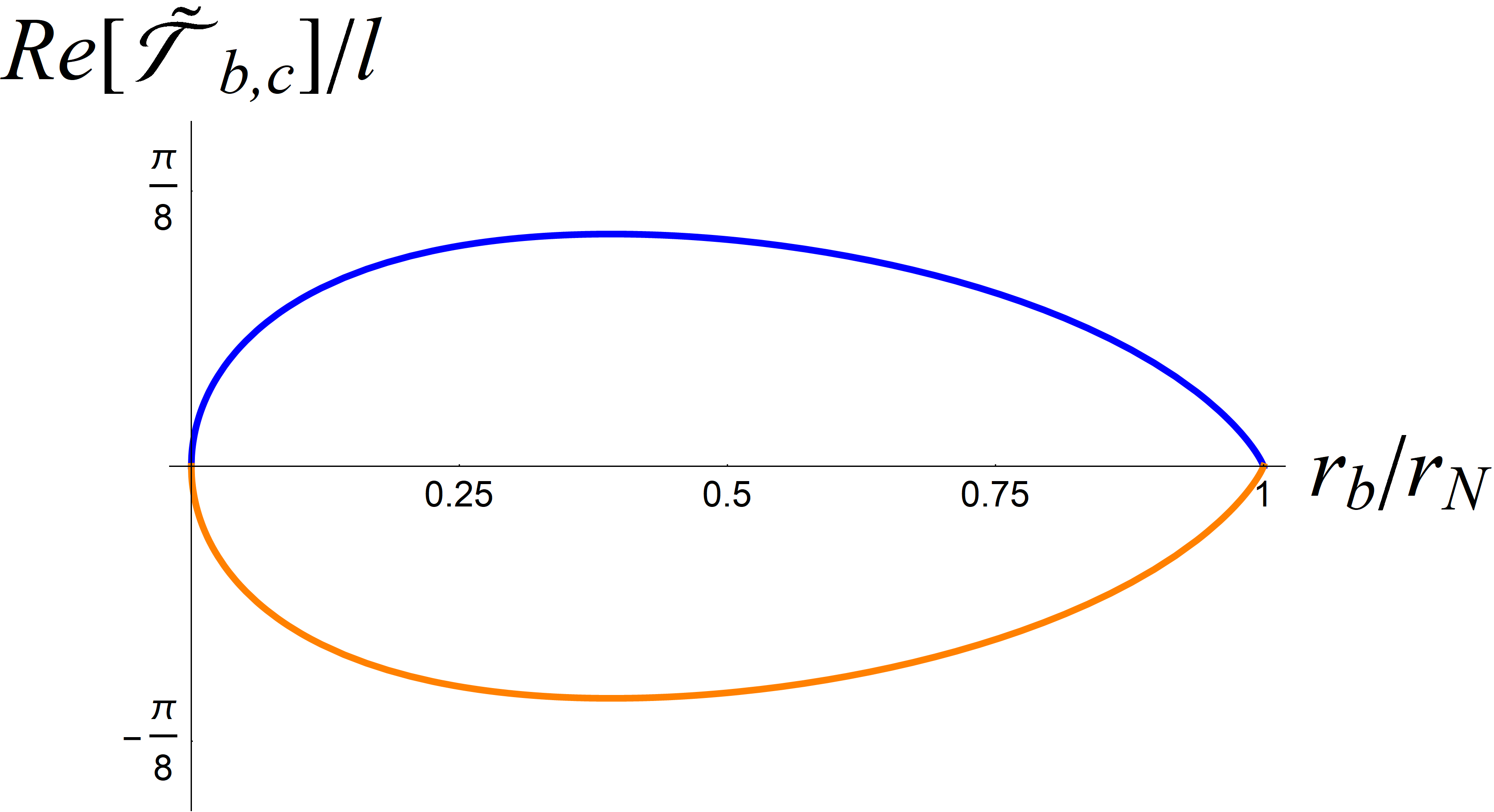}
 \caption{$\text{Re}[\tilde{\mathcal{T}}_{b,c}]=\pm\mathcal{T}_{b,c}$, temporal windows.}
\end{subfigure}%
\begin{subfigure}{.5\textwidth}
 \centering
\includegraphics[width=7.5cm]{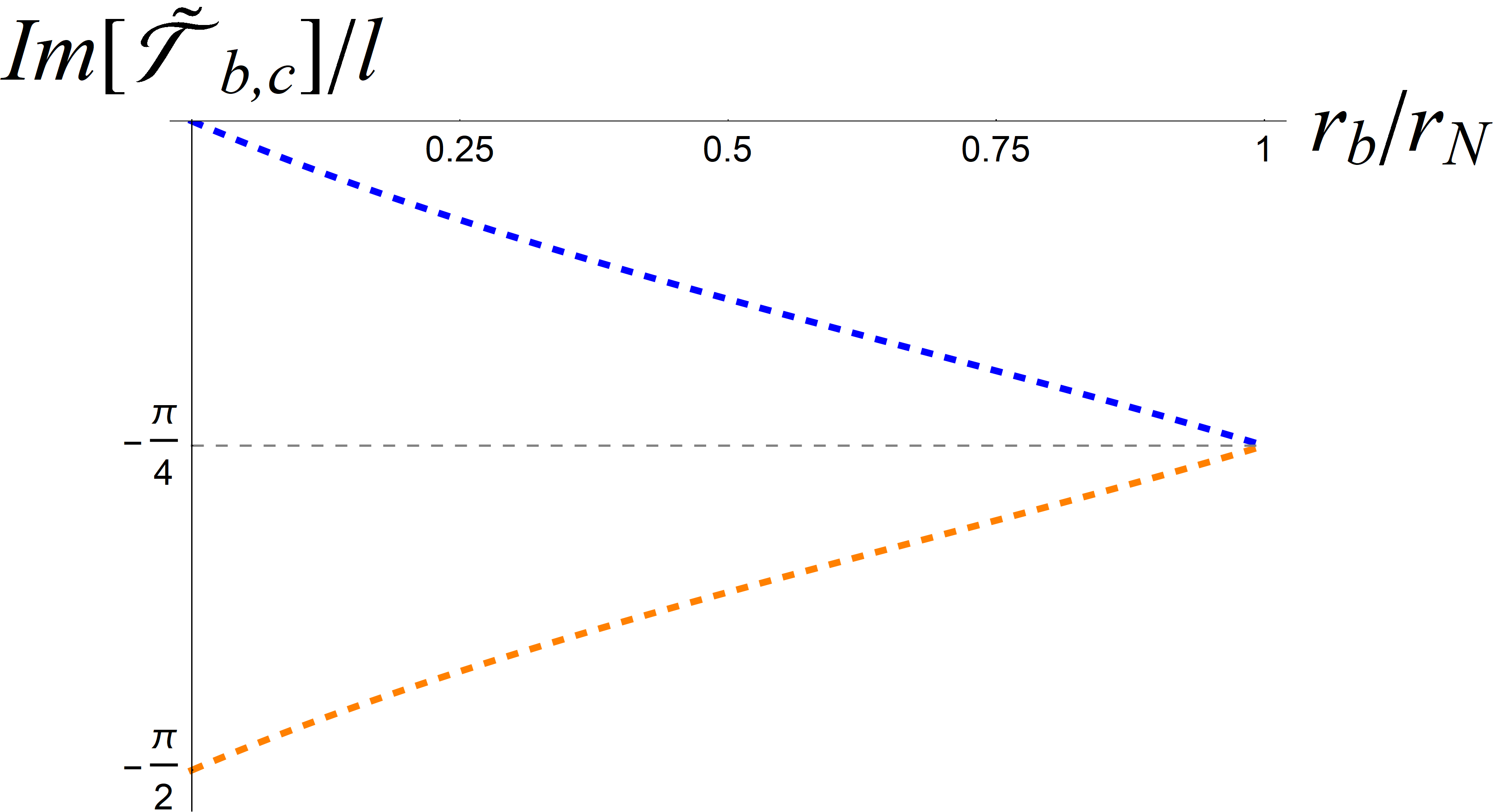}
 \caption{$\text{Im}[\tilde{\mathcal{T}}_{b,c}]=-\beta_{b,c}/4$, euclidean periodicities.}
\end{subfigure}
\caption{
The complex time $\tilde{\mathcal{T}}_{b,c}$ for $\tilde{\tau}=0$, as a function of the black hole radius $r_b$ in 5 dimensions. The blue curves stand for the black hole ($\mathcal{T}_b$) and the orange curves for the cosmological horizon ($-\mathcal{T}_c$).}
\label{fig:5dwindows}
\end{figure}

As we fix the observer at $r=r_{\mathcal{O}}$, implying we should use the appropriately rescaled (proper) time coordinate $\tilde{t}$ (\ref{hawking}), the equation for the time parameter along the null geodesic \eqref{eq:nullgeo} becomes
\begin{eqnarray}
\tilde{t}_{b,c}(r') = \tilde{\tau}_{b,c} \pm\gamma\int_{r'}^{r_{\mathcal{O}}}\frac{\rmd r}{f(r)} \,,  \label{kk}
\end{eqnarray}
where $\tilde{\tau}_{b,c}$ stands for the initial time of the null radial geodesic originating from the static sphere. The subscripts $b$ and $c$ are added to distinguish between the two different directions of the radial null geodesic: one moving inward heading towards the black hole horizon and the other moving outward towards cosmological future infinity. Imposing that the inward moving radial null geodesic reaches the black hole singularity we set $r'$ to zero, while instead setting $r'$ to infinity for an outward moving radial null geodesic. In summary, we will be interested in the following two limiting (complex) quantities  
\begin{eqnarray}
&& \tilde{\mathcal{T}}_c
=\lim_{r\rightarrow\infty}    
\tilde{t}_{c}(r)\label{xx1}\\
&& \tilde{\mathcal{T}}_b =\lim_{r\rightarrow0}    
\tilde{t}_{b}(r)\label{xx2}
\end{eqnarray}

We will evaluate these limits in $4$, $5$ and $6$ dimensions, to identify and distinguish two different types of characteristic behavior. We will start in $5$ dimensions, which is the easiest to analyse analytically and for which the results show a striking symmetry between interior black hole and exterior de Sitter behavior. In five dimensions the blackening factor takes the following form
\begin{equation}
    f(r)=\frac{\left(r^2-r_b^2\right) \left(r_c^2-r^2\right)}{\ell^2 r^2}\,,
\end{equation}
and the horizon radii can be expressed as \cite{Morvan:2022ybp},
\begin{equation}
 r_{b,c}=r_N\sqrt{1\mp\sqrt{1-M/M_N}}\, .
  \label{eq:5dhori}
\end{equation}
The static sphere observers are located at $r_{\mathcal{O}}=\sqrt{r_b\,r_c}$, such that the respective Euclidean periodicities, i.e. the temperatures, are determined by
\begin{equation}
    \gamma=\sqrt{f(r_{\mathcal{O}})}\,.
\end{equation}
We can now readily compute eq. \eqref{xx1} and eq. \eqref{xx2}. For $D=5$ SdS we find
\begin{eqnarray}
  \tilde{\mathcal{T}}_{b,c} &= \tilde{\tau}_{b,c}
  \pm \mathcal{T}_{b,c} \mp i \frac{\beta_{b,c}}{4}\,,\label{t.t}
\end{eqnarray}
where $+\mathcal{T}_{b}$ identifies the real contribution for inward moving null shells and  $-\mathcal{T}_{c}$ for outward moving shells respectively. The imaginary part in (\ref{t.t}) is connected to the Euclidean periodicity, or the inverse temperature, generated by each respective horizon.
The real contributions $\mathcal{T}_{b,c}$ are in fact the same in $D=5$, and equal to   
\begin{eqnarray}
    \mathcal{T}_{b}=\mathcal{T}_c = \ell\frac{ \left(r_b-r_c\right)}{\left(r_b+r_c\right)} \arctanh\left(\frac{r_b}{r_\mathcal{O}}\right) \, . 
\end{eqnarray}
The result for the real contribution is nonzero and in this case the same, up to a sign flip, for both the inward moving radial null geodesic reaching the black hole singularity, as well as the outward moving radial null geodesics reaching future infinity, which is distinctively different compared to empty, pure, de Sitter spacetime. 
% added a comment on the special property of D=5 %
The fact that the absolute value of this real contribution is the same for both the the inward and outward radial null geodesics is a special property of $D=5$, as we will soon discover, and can be understood as the result of the blackening factor (\ref{bfactorSdS}) in $D=5$ scaling the same in the large and small $r$ limit.  

\begin{figure}[t]
\centering
\resizebox{0.6\textwidth}{!}
{%
\begin{tikzpicture}
\draw[ dotted] (0,0) -- (2.25,2.25);
% \draw[ dashed,blue!90!black] (4.5,4.5) -- (2.25,2.25);

%\draw[] (0,2.25) -- (1.6,3.9);
% \draw[dotted] 
% (4.5,2.25) -- (2.85,0.6);
\filldraw[blue] (4.5,1.71) circle (2pt) node[anchor=west]{ \scriptsize
$\tilde{\tau}=-\mathcal{T}_b$};

\filldraw[orange] (4.5,2.84) circle (2pt) node[anchor=west]{\scriptsize $\tilde{\tau}=\mathcal{T}_c$};

\node[above] at (4.35,4.6) {\scriptsize $r=r_\mathcal{O}$};

\filldraw[] (4.5,2.25) circle (2pt) node[anchor=west]{\scriptsize $\tilde{\tau}=0$};

% \draw[dotted] 
% (0,2.25) -- (1.65,3.85);
% \draw[ dotted] 
% (0,2.25) -- (1.6,0.66);
% %\draw[ dotted] (4.5,2.25) -- (2.25,0);
% \draw[ dotted] 
% (4.5,2.25) -- (2.9,3.9);
% %\draw[ dotted] (4.5,2.25) -- (2.25,4.5);

% \draw[dotted] 
% (4.5,2.25) -- (7.3,5.085);
% \draw[dotted] 
% (9,2.25) -- (6.1,5.085);

\draw[ red ] 
(4.5,2.79) -- (6.75,5.15);
\draw[ red ] 
(9,2.80) -- (6.75,5.14);

\draw[ red ] 
(4.5,1.71) -- (6.76,-.63);

\draw[ red ] 
(9,1.71) -- (6.76,-.63);

\draw[ cyan ] 
(4.5,2.79) -- (2.25,0.66);

\draw[ cyan ] 
(4.5,1.71) -- (2.25,3.84);

\draw[ cyan] 
(0,1.71) -- (2.25,3.84);

\draw[ cyan ] 
(0,2.79) -- (2.25,0.66);

% \draw[dotted ] 
% (9,2.25) -- (6.1,-.62);

% \draw[ dotted] 
% (4.5,2.25) -- (7.47,-.620);

\draw[ dotted] (0,4.5) -- (2.25,2.25);

\draw[dotted] (4.5,0) -- (6.75,2.25);
\draw[dotted] (4.5,0) -- (2.25,2.25);
\draw[dotted] (4.5,4.5) -- (2.25,2.25);

% \draw[ dashed,blue!90!black] (4.5,0) -- (2.25,2.25);
% \draw[ dashed,orange] (4.5,0) -- (6.75,2.25);
\draw[ dotted] (4.5,4.53) -- (6.75,2.25);
% \draw[ dashed,orange] (4.5,4.53) -- (6.75,2.25);
\draw[dashed] (9,0) -- (9,4.5);
\draw[ dotted] (9,0) -- (6.75,2.25);
%\draw[thick] (6.75,4.5) -- (9,4.5); % Removed line
%\draw[densely dotted] (6.75,4.5) -- (6.75,2.25); % Modified line
\draw[dashed] (0,0) -- (0,4.5);
\draw[thick] (4.5,0) -- (4.5,4.55);
\draw[dotted] (6.75,2.25) -- (9,4.5);

% Add the label to the right-hand most center
%\node at (9.5,2.25) {\scriptsize $\tilde{\tau}=0$};

\node at (2.25,4.1) {\scriptsize $r=0$};
\node[above] at (6.8,5.2) {$\mathcal{I}^+$};
\node[below] at (6.8,-0.6) {$\mathcal{I}^-$};
\node at (2.25,.4) {\scriptsize $r=0$};

%\node at (2.45,3) {\scriptsize $r=r_b$};
\draw[densely dotted] (0,2.25) -- (9,2.25);
\draw[thick, color=orange] (4.5,4.55) to[out=29, in=151] (9,4.5);
\draw[thick, color=orange] (4.5,0) to[out=-29, in=-151] (9,0);
% Draw a cross and identify the point at (4,0)
\draw[magenta, line width=0.5pt] (4.5,1.2) -- ++(-2pt,-2pt) -- ++(4pt,4pt);
\draw[magenta, line width=0.5pt] (4.5,1.2) -- ++(-2pt,2pt) -- ++(4pt,-4pt);
\node[magenta, anchor=west] at (4.5,1.0) {\scriptsize $x$};

\draw[magenta, line width=0.5pt] (0,1.2) -- ++(-2pt,-2pt) -- ++(4pt,4pt);
\draw[magenta, line width=0.5pt] (0,1.2) -- ++(-2pt,2pt) -- ++(4pt,-4pt);
\node[magenta, anchor=west] at (0,1.0) {\scriptsize $x'$};

\draw[magenta, line width=0.5pt] (9,1.2) -- ++(-2pt,-2pt) -- ++(4pt,4pt);
\draw[magenta, line width=0.5pt] (9,1.2) -- ++(-2pt,2pt) -- ++(4pt,-4pt);
\node[magenta, anchor=west] at (9,1.0) {\scriptsize $x'$};

\draw[thick, color=blue, decoration={snake,amplitude=.15mm,segment length=2mm,
                       post length=.6mm,pre length=.2mm}, decorate] (0,4.5) to[out=-29, in=-151] (4.5,4.55);
\draw[thick, color=blue, decoration={snake,amplitude=.15mm,segment length=2mm,
                       post length=.6mm,pre length=.2mm}, decorate] (0,0) to[out=29, in=151] (4.5,0);
\end{tikzpicture}
} % resizebox
\caption{Penrose diagram for $5$-dimensional SdS spacetime, with respect to static sphere observers represented by straight vertical lines. The black hole singularity (blue wavy lines) and de Sitter future/past infinity (solid orange) bend inward and outward respectively. The dashed diagonal lines on the left stand for the black hole and the ones on the the right for the cosmological horizon respectively. Also indicated are the times for which inward or outward moving radial null shells reach the center of the future/past black hole singularity or future/past asymptotic infinity, and a pair of symmetric conjugate events $x$ and $x'$.}
%To reach the center of the black hole singularity $r=0$, the null geodesic must start from $\tilde{\tau}=-\mathcal{\tilde{T}}_b$  (solid line, light blue) from the static sphere instead of $\tilde{\tau}=0$. %Null geodesics reach the singularity "off-center" when they start from $\tilde{\tau}=0$ (black dotted line).
%In the asymptotic dS region the null geodesic 
%can reach the center of future infinity $\mathcal{I}^+$
%starting from static sphere when it starts from 
%$\tilde{\tau}=\mathcal{\tilde{T}}_c$.
%The black (bold) vertical line represents the static sphere characterized by $r=r_\mathcal{O}$. 
%The other static sphere in the opposite static patches
%(black dashed black lines) can be either identified or continued repeatedly. The point $x$ generic point on %static sphere, while $x'$ is another point on the opposite static spheres in same time slice.}
\label{fig:SdS}
\end{figure}
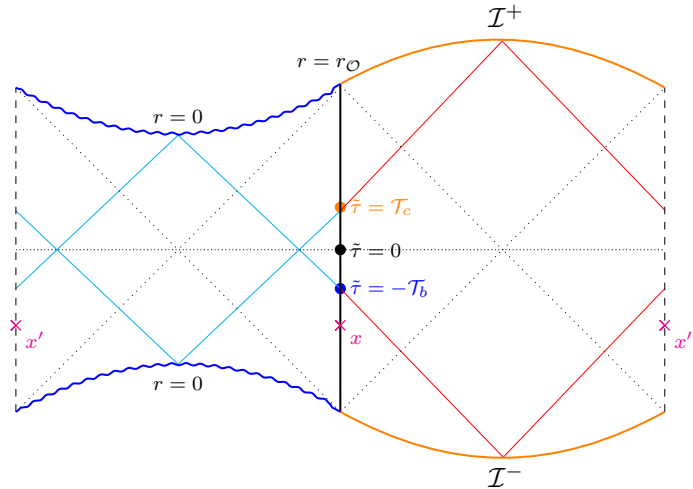

For radial null geodesics moving inward from $r=r_\mathcal{O}$ and reaching the black hole singularity the real contribution $+\mathcal{T}_b$ is positive unless $M=0$ or $M=M_N$, where it exactly vanishes as anticipated. The same happens for $D>3$ Schwarzschild black holes in AdS spacetime \cite{Fidkowski:2003nf}, where the real part of the integral in eq. \eqref{eq:nullgeo} is also positive. This means that radial null geodesics moving inwards, emitted at $\tilde{\tau}=0$ from $r=r_\mathcal{O}$, reach the singularity $r=0$ "off-center", implying that the black hole singularity is bend inwards with respect to (straight vertical line) static sphere observers. Alternatively, for a null shell to reach the singularity's center, i.e., $\text{Re}[\tilde{\mathcal{T}}_b]=0$, it should be emitted not at $\tilde{\tau}=0$, but at $\tilde{\tau}=-\mathcal{T}_b$.
%It is evident that when representing the Penrose diagram as a square, null geodesics originating from the boundary at $\tilde{\tau} = 0$ (indicated by the dashed line in the figure) would converge at the  
%centre of the black hole singularity simultaneously.
%Hence, if we depict static sphere as straight lines, then the black hole singularity is bend inwards.
Clearly, as a consequence of (global) time reversal symmetry, outward moving radial null geodesics emitted from the center of the white hole singularity $r=0$ will reach the static sphere observer at a time  $\tilde{\tau}=+\mathcal{T}_b$. Also, the same conclusions hold for the conjugate static sphere observer, implying that two inward moving radial null geodesics emitted at the symmetric events $\tilde{\tau}=-\mathcal{T}_b$ on the right, and $\tilde{\tau}=+\mathcal{T}_b$ on the left, will meet exactly in the center of the (future) black hole singularity $r=0$. 

Let us now shift our attention to radial null geodesics that move outwards towards future de Sitter infinity, i.e. towards the exterior region of the SdS spacetime, emitted from the static sphere observer. In that case the real contribution $-\mathcal{T}_c$ is negative, vanishing only for pure dS or in the maximal mass, Nariai limit. As a result, when the radial null geodesic is emitted from $r=r_\mathcal{O}$ at $\tilde{\tau}=0$ it will pass the center of future infinity, ending off-center to the right. Instead, when two radial null shells are emitted outwards from two conjugate static sphere observers at (symmetric) times $\tilde{\tau}=\mathcal{T}_c$ on the right and $\tilde{\tau}=\mathcal{T}_c$ on the left, will they meet each other, in the future infinity limit, in the center. Similarly, considering inward moving null shells, the sign of the real contribution changes, and null shells starting at the center of past infinity will reach the static sphere at $\tilde{\tau}=-\mathcal{T}_c$.

In Figure \ref{fig:5dwindows} the real and imaginary part of $\tilde{\mathcal{T}}_{b,c}$ are plotted with respect to
the black hole radius. Notably, as the real contributions $\mathcal{T}_{b,c}$ are the same, up to a sign, for $5$-dimensional SdS, we conclude that in five dimensions the magnitude of the black hole singularity bending inwards matches the magnitude of future infinity bending outwards. We summarized the results for $D=5$ in terms of the Penrose diagram in Figure \ref{fig:SdS}. 
%Added figure 5 reference, which was apparently missing
%Interestingly, the imaginary part of the integral also accurately connects the inverse temperature of each horizon. 
Note that choosing the appropriate (proper) time normalization is crucial for obtaining these results. If one would have instead picked $\gamma=1$, both the real and imaginary part of $\tilde{\mathcal{T}}_{b,c}$ would have diverged in the Nariai limit. Having determined the details of the causal structure, with respect to static sphere observers, we can already anticipate the presence of finite size temporal window for the existence of radial spacelike geodesics connecting symmetric points between two conjugate static sphere observers in SdS spacetime. These symmetric points can either be connected through the interior of the black hole, or through the exterior asymptotic de Sitter region, and in five dimensions both temporal windows have the same size.

As we move on to $4$ and $6$ dimensions we will encounter the same type of behavior, except for the fact that the magnitude of the respective interior black hole (inward) or exterior future infinity (outward) bending will no longer be the same, due to the different limiting behaviors of the blackening factor as the radius $r$ approaches the black hole singularity or asymptotic de Sitter infinity.     

% Added a sentence above, and slightly modified the text below 
Moving on to four dimensions, the horizon radii $r_{b,c}$ can be 
written as \cite{Morvan:2022ybp}
\begin{align}
r_{b,c}=r_N\left(\cos{\eta}\mp\sqrt{3}\sin{\eta}\right),\quad\text{where}\quad\eta\equiv\frac{1}{3}\arccos{M/M_N}\,.\label{eq:4dhori}
\end{align}
In terms of the event horizon radii the blackening factor equals
\begin{eqnarray}
    f(r)=\frac{(r-r_b)(r_c-r)(r+r_b+r_c)}{\ell^2 r} \, , 
\end{eqnarray}
which implies that the sphere static observers are located at the special radius 
\beq
r_{\mathcal{O}}^3= \frac{r_b r_c}{2} (r_b+ r_c)= \frac{r_{b,c}}{2} (\ell^2-r^2_{b,c})\,,\ \label{eq2}
\eeq

%It suggests that the bending of the black hole singularity toward the black hole horizon is more prominent than the bending of cosmological future or past infinity away from the cosmological horizon. This places intriguing constraints on the real or complex structure of spacelike geodesics connecting points on opposite static spheres in the same time slice. This is a distinctive feature of four dimensional SdS which will revisit   briefly later.\\\\

\begin{figure}[t]
\begin{subfigure}{.5\textwidth}
 \centering
\includegraphics[width=7.5cm]{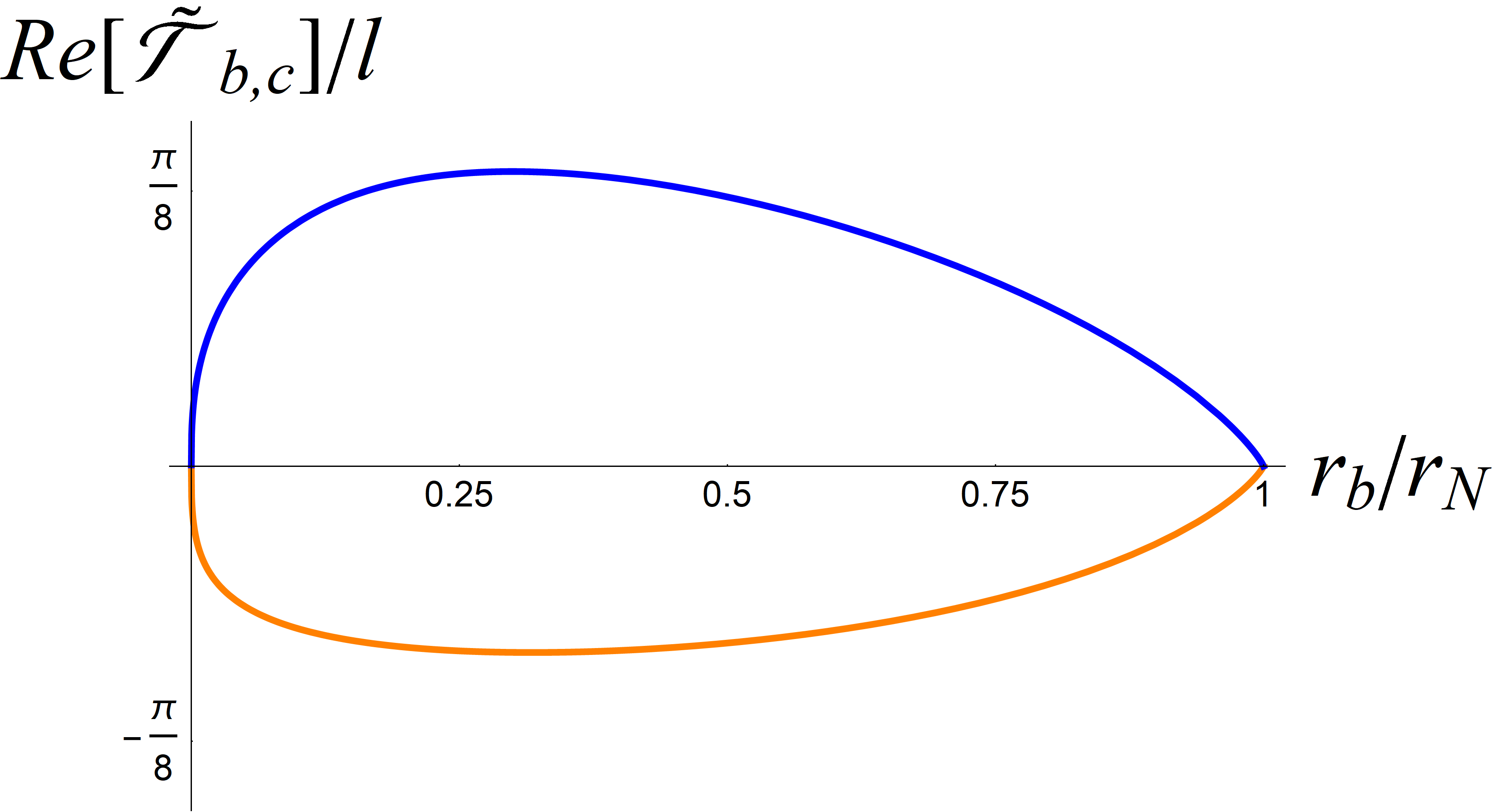}
 \caption{$\text{Re}[\tilde{\mathcal{T}}_{b,c}]=\pm\mathcal{T}_{b,c}$, 4D temporal windows.}
\end{subfigure}%
\begin{subfigure}{.5\textwidth}
 \centering
\includegraphics[width=7.5cm]{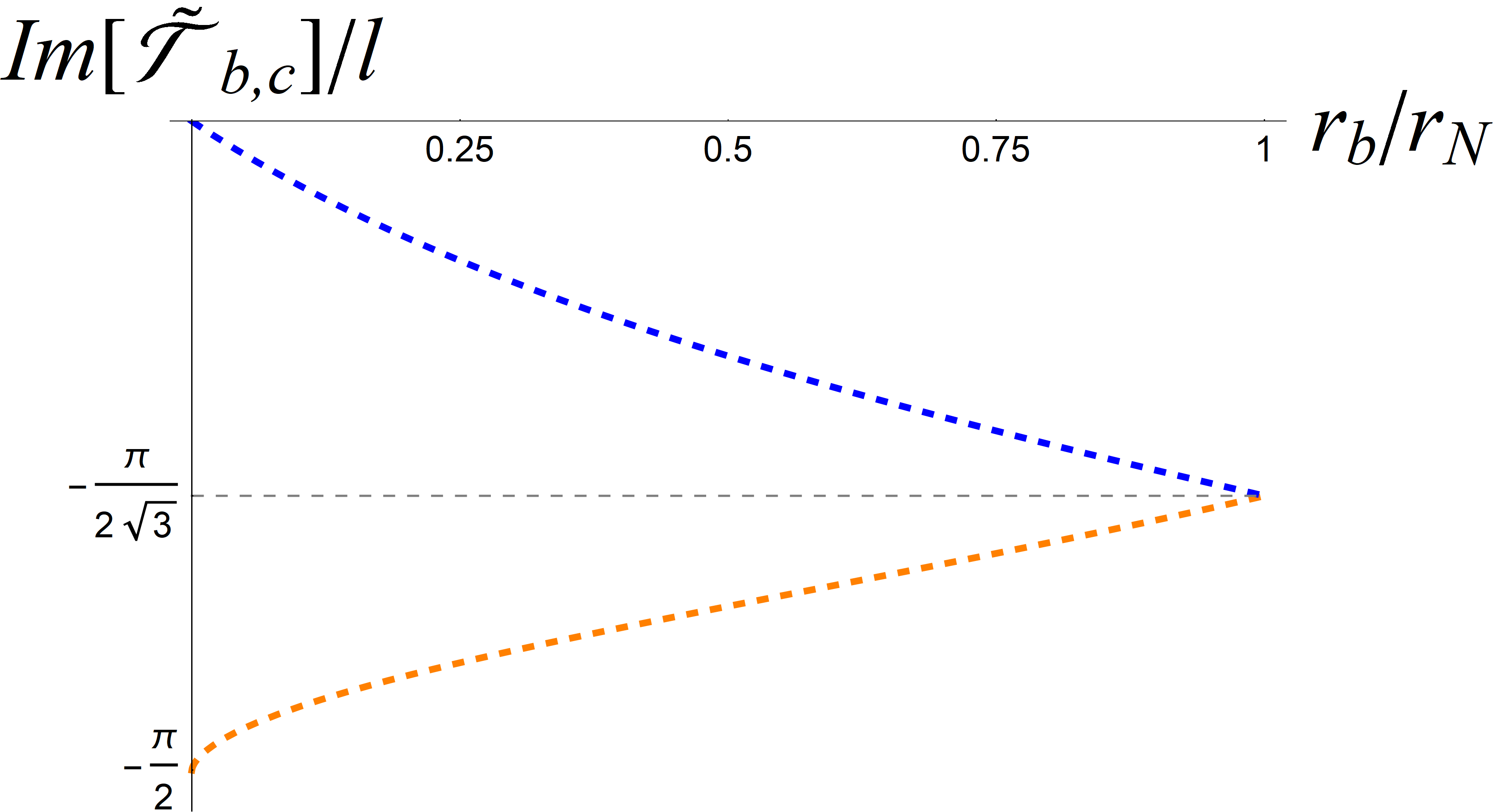}
 \caption{$\text{Im}[\tilde{\mathcal{T}}_{b,c}]=-\beta_{b,c}/4$, 4D euclidean periodicities.}
\end{subfigure}
\caption{The complex time $\tilde{\mathcal{T}}_{b,c}$ for $\tilde{\tau}=0$, as a function of the black hole radius $r_b$ in $4$ dimensions. The blue curve represents $\tilde{\mathcal{T}}_{b}$, and the orange curve represents $\tilde{\mathcal{T}}_{c}$.}
\label{4dw}
\end{figure}

Let us now again use \eqref{kk} to evaluate the (complex) time parameter along a radial null geodesic emitted from the static sphere observer to any point $r'$. As compared to $5$ dimensions, applying \eqref{eq:4dhori} and \eqref{eq2} and evaluating the result is much more involved. We refer to the appendix for the details, but in the end we can find the following expression
\begin{eqnarray}
\hspace{-10mm}\tilde{t}_{b,c}(r') 
&=\tilde{\tau}\pm \gamma\ell^2\left(\frac{\left(r_b-r_c\right)\left(r_b+r_c\right) \ln \left(r_b+r_c+r\right)+r_c \left(2 r_b+r_c\right) \ln \left(r-r_c\right)-r_b \left(r_b+2 r_c\right) \ln\left(r-r_b\right)}{\left(r_b-r_c\right) \left(2 r_b+r_c\right) \left(r_b+2 r_c\right)}\bigg|_{r'}^{r_{\mathcal{O}}}\right)\,.
   \label{eq:longeqSdS4}
\end{eqnarray}
Here, as before, the plus sign is selected for the inward moving null shells and the minus sign corresponds to outward moving null shells. Taking the relevant limits we can evaluate the (real) times $\mathcal{T}_{b,c}$ for null geodesics to reach the black hole singularity or dS future infinity. 
Not surprisingly, we find expressions similar to \eqref{t.t} for $\mathcal{T}_{b,c}$, which we plotted in figure \ref{4dw}. Again, the real contribution $+\mathcal{T}_b$ 
is positive, while it is negative for $-\mathcal{T}_c$, except in the pure de Sitter or Nariai limit, where it vanishes. So as should be expected, with respect to static sphere observers, the black hole singularity is bending inwards and asymptotic de Sitter infinity is bending outwards. The important difference with respect to five dimensions is however that in four dimensions $|\mathcal{T}_b|\geq|\mathcal{T}_c|$, implying that the magnitude of the inward bending of the black hole singularity exceeds the outward bending of future infinity, for any value of the mass of the black hole. 

Except for this important difference, for $D=4$ we can draw similar conclusions as in five dimensions: an inward null shell emitted from the static sphere reaches the center of the black hole singularity only when emitted at $\tilde{\tau}= -\mathcal{T}_b$. While to reach the center of de Sitter future infinity from the static sphere the null shell should be emitted at a positive (later) time, i.e. $\tilde{\tau}= \mathcal{T}_c$. Again, we can use the time reversal symmetry of the SdS spacetime to identify the other null shells of interest (arriving from the past white hole singularity or dS past infinity). 
%This sets up an interesting constraint on the points that are causally connected with each other via null geodesics. There seems to be an increment of causal connectivity through the asymptotic dS region. \\\\
%Null geodesics emitted inward or outward from two conjugate static sphere observers at symmetric times within the temporal windows, can either intersect or they share a common origin in the interior black hole region or the exterior asymptotic dS region. In other words 
%For the asymptotic dS region this is possible for $-\infty<\tilde{\tau}\leq\mathcal{\tilde{T}}_c$.  
The inward bending of the interior black and white hole singularity imply a decrease of the size of the interior causal diamond, whereas the outward bending of future and past infinity define an increase of the exterior causal diamond, with a magnitude measured by the two temporal windows $\mathcal{T}_b$ and $\mathcal{T}_c$. Clearly, if exterior future infinity was 'straight', null shells emitted outward from symmetric (positive time for the right, negative time for the left) static sphere observers  would never intersect. In other words, for times within the temporal window $-\mathcal{T}_c < \tilde{\tau} < \mathcal{T}_c$ the two conjugate static sphere observers share the same causal region. 

%In fact in that scenario these two points would be in causal contact when $x$ lies within $-\infty<\tilde{\tau}\leq0$.
%A non-zero $\mathcal{\tilde{T}}_c$ for arbitrary value of blackhole mass indicates increment in casually connected regions in the asymptotic dS region. 

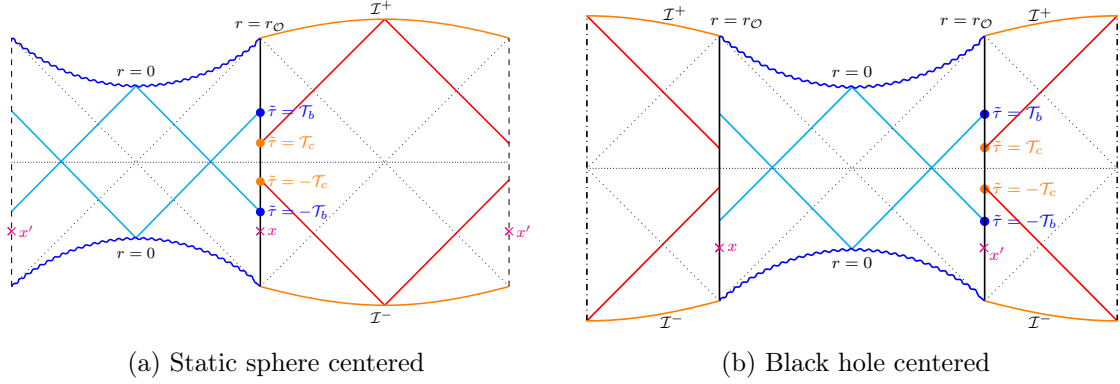
\begin{figure}[t]
\centering
\begin{subfigure}{.46\textwidth}
    \centering
    \resizebox{\linewidth}{!}{%
        \begin{tikzpicture}[scale=1]
              \node at (2.25,3.9) {\scriptsize $r=0$};
            \node[above] at (6.7,4.8) {\scriptsize $\mathcal{I}^+$}; % Adjusted position
            \node[above] at (6.7,-0.8) {\scriptsize $\mathcal{I}^-$}; % Adjusted position
              \node at (2.25,0.6) {\scriptsize $r=0$};

            \draw[cyan,thick]  (0,1.36) -- (2.25,3.63);
            \draw[cyan,thick]  (0,3.18) -- (2.25,.88);
            \draw[cyan,thick]  (4.5,3.18) -- (2.25,.88);
            \draw[cyan,thick]  (4.5,1.35) -- (2.25,3.63);
            \draw[red,thick]  (4.5,2.59) -- (6.75,4.84);
            \draw[red,thick]  (9,2.59) -- (6.75,4.84);
            \draw[red,thick]  (9,1.93) -- (6.75,-.34);
            \draw[red,thick]  (4.5,1.93) -- (6.75,-.34);
            \draw[dashed] (9,0) -- (9,4.5);
            \draw[dashed] (0,0) -- (0,4.5);
            \draw[thick] (4.5,0) -- (4.5,4.5);
            \draw[dotted] (9,0) -- (4.5,4.5);
            \draw[dotted] (4.5,0) -- (9,4.5);
            \draw[dotted] (0,0) -- (4.5,4.5);
            \draw[dotted] (4.5,0) -- (0,4.5);
            \node[above] at (4.5,4.5) {\scriptsize $r=r_{\mathcal{O}}$}; 
            \draw[thick, color=blue, decoration={snake,amplitude=.17mm,segment length=1.5mm, post length=.5mm}, decorate] (0,4.5) to[out=-40, in=-140] (4.5,4.5);
            \draw[thick, color=blue, decoration={snake,amplitude=.17mm,segment length=1.5mm, post length=.5mm}, decorate] (0,0) to[out=40, in=140] (4.5,0);
            \draw[densely dotted] (0,2.25) -- (9,2.25);
            \draw[thick, color=orange] (4.5,4.5) to[out=15, in=165] (9,4.5);
            \draw[thick, color=orange] (4.5,0) to[out=-15, in=-165] (9,0);
              \filldraw[orange] (4.5,2.6) circle (2pt) node[anchor=west]{\scriptsize $\tilde{\tau}=\mathcal{T}_c$};
              \filldraw[orange] (4.5,1.9) circle (2pt) node[anchor=west]{\scriptsize $\tilde{\tau}=-\mathcal{T}_c$};
               \filldraw[blue] (4.5,3.15) circle (2pt) node[anchor=west]{\scriptsize $\tilde{\tau}=\mathcal{T}_b$};
            \filldraw[blue] (4.5,1.35) circle (2pt) node[anchor=west]{\scriptsize $\tilde{\tau}=-\mathcal{T}_b$};
            \draw[magenta, line width=0.5pt] (4.5,01) -- ++(-2pt,-2pt) -- ++(4pt,4pt);
\draw[magenta, line width=0.5pt] (4.5,01) -- ++(-2pt,2pt) -- ++(4pt,-4pt);
\node[magenta, anchor=west] at (4.5,01) {\scriptsize $x$};

\draw[magenta, line width=0.5pt] (0,01) -- ++(-2pt,-2pt) -- ++(4pt,4pt);
\draw[magenta, line width=0.5pt] (0,01) -- ++(-2pt,2pt) -- ++(4pt,-4pt);
\node[magenta, anchor=west] at (0,01) {\scriptsize $x'$};

\draw[magenta, line width=0.5pt] (9,01) -- ++(-2pt,-2pt) -- ++(4pt,4pt);
\draw[magenta, line width=0.5pt] (9,01) -- ++(-2pt,2pt) -- ++(4pt,-4pt);
\node[magenta, anchor=west] at (9,01) {\scriptsize $x'$};

        \end{tikzpicture}
    }
    \caption{Static sphere centered}
\end{subfigure}
\hspace{0.3cm} % Adjust the space here
\begin{subfigure}{.46\textwidth}
    \centering
    \resizebox{\linewidth}{!}{%
        \begin{tikzpicture}[scale=1]
            \clip (-2.25,-.5) rectangle + (9,5.5);
            \draw[cyan,thick]  (0,1.36) -- (2.25,3.63);
            \draw[cyan,thick]  (0,3.18) -- (2.25,.88);
            \draw[cyan,thick]  (4.5,3.18) -- (2.25,.88);
            \draw[cyan,thick]  (4.5,1.35) -- (2.25,3.63);
            \filldraw[orange] (4.5,1.9) circle (2pt) node[anchor=west]{\scriptsize $\tilde{\tau}=-\mathcal{T}_c$};
            \filldraw[orange] (4.5,2.6) circle (2pt) node[anchor=west]{\scriptsize $\tilde{\tau}=\mathcal{T}_c$};
            \node at (2.25,3.9) {\scriptsize $r=0$};
            \node at (2.25,0.65) {\scriptsize $r=0$};
            \node[above] at (5.45,4.61) {\scriptsize $\mathcal{I}^+$}; % Adjusted position
            \node[above] at (5.45,-0.6) {\scriptsize $\mathcal{I}^-$}; % Adjusted position
            \node[above] at (-0.8,-0.6) {\scriptsize $\mathcal{I}^-$}; % Adjusted position
  \node[above] at (-0.75,4.63) {\scriptsize $\mathcal{I}^+$}; 
  \node[above] at (0.3,4.5) {\scriptsize $r=r_{\mathcal{O}}$}; 
  \node[above] at (4.2,4.5) {\scriptsize $r=r_{\mathcal{O}}$}; 
            \filldraw[blue] (4.5,3.17) circle (2pt) node[anchor=west]{\scriptsize $\tilde{\tau}=\mathcal{T}_b$};
            \filldraw[blue] (4.5,1.35) circle (2pt) node[anchor=west]{\scriptsize $\tilde{\tau}=-\mathcal{T}_b$};

            \draw[red,thick]  (4.5,2.59) -- (6.75,4.84);
            \draw[red,thick]  (0,2.59) -- (-2.25,4.84);
            \draw[red,thick]  (0,1.93) -- (-2.25,-.34);
            \draw[red,thick]  (4.5,1.93) -- (6.75,-.34);
            \draw[dash dot,thick] (6.75,-0.34) -- (6.75,4.84);
            \draw[dash dot,thick] (-2.25,-0.34) -- (-2.25,4.84);
            \draw[thick] (0,0) -- (0,4.5);
            \draw[thick] (4.5,0) -- (4.5,4.5);
            \draw[dotted] (0,4.5) -- (-2.25,2.25);
            \draw[dotted] (0,0) -- (-2.25,2.25);
            \draw[dotted] (4.5,4.5) -- (6.75,2.25);
            \draw[dotted] (4.5,0) -- (6.75,2.25);
            \draw[dotted] (0,0) -- (4.5,4.5);
            \draw[dotted] (4.5,0) -- (0,4.5);
            \draw[densely dotted] (-2.25,2.25) -- (9,2.25);
            \draw[thick, color=blue, decoration={snake,amplitude=.17mm,segment length=1.5mm, post length=.5mm}, decorate] (0,4.5) to[out=-40, in=-140] (4.5,4.5);
            \draw[thick, color=blue, decoration={snake,amplitude=.17mm,segment length=1.5mm, post length=.5mm}, decorate] (0,0) to[out=40, in=140] (4.5,0);
            \draw[thick, color=orange] (4.5,4.5) to[out=15, in=165] (9,4.5);
            \draw[thick, color=orange] (4.5,0) to[out=-15, in=-165] (9,0);
            \draw[thick, color=orange] (-4.5,4.5) to[out=15, in=165] (0,4.5);
            \draw[thick, color=orange] (-4.5,0) to[out=-15, in=-165] (0,0);

\draw[magenta, line width=0.5pt] (0,0.9) -- ++(-2pt,-2pt) -- ++(4pt,4pt);
\draw[magenta, line width=0.5pt] (0,0.9) -- ++(-2pt,2pt) -- ++(4pt,-4pt);
\node[magenta, anchor=west] at (0,0.9) {\scriptsize $x$};

\draw[magenta, line width=0.5pt] (4.49,0.9) -- ++(-2pt,-2pt) -- ++(4pt,4pt);
\draw[magenta, line width=0.5pt] (4.49,0.9) -- ++(-2pt,2pt) -- ++(4pt,-4pt);
\node[magenta, anchor=west] at (4.45,0.8) {\scriptsize $x'$};

        \end{tikzpicture}
    }
    \caption{Black hole centered}
\end{subfigure}%
\caption{Causal Penrose diagram for $4$-dimensional SdS spacetime, with respect to static sphere observers represented by straight vertical lines, centered around a static sphere observer or the black hole. The black hole singularity (blue wavy lines) and de Sitter future/past infinity (solid orange) bend inward and outward respectively. The dashed blue and orange lines are the black hole and cosmological horizon respectively. We also indicated some inward and outward radial null shells, as well as a pair of symmetric conjugate events.} 
%Also indicated are the times for which inward or outward moving radial null shells reach the center of the future/past black hole singularity or future/past asymptotic infinity, and a pair of symmetric conjugate events $x$ and $x'$.}
\label{fig:SdSa}
\end{figure}

The opposite behavior is observed in the four-dimensional black hole region. The causally connected region between the black and white hole singularity shrinks. This means that null shells emitted from conjugate static sphere observers at a symmetric (global) time never intersect in the interior if emitted after $\tilde{\tau} > -\mathcal{T}_b$. Equivalently, two null shells cannot have a common origin in the interior of the black hole if arriving before $\tilde{\tau} < \mathcal{T}_b$ by a conjugate pair of static sphere observers.     
%Null geodesics starting from two points, $x$ and $x'$, located in conjugate static spheres can come into causal contact with each other through the black hole region if $x$ lies within $-\infty<\tilde{\tau}\leq-\mathcal{\tilde{T}}_b$. If the singularity of a black hole were represented by a straight line, all points between $-\infty < \tilde{\tau} \leq 0$ on the static sphere would be causally connected to the points on the same time slice of the opposite static sphere through the black hole region. However, $\mathcal{\tilde{T}}_b$ is actually a nonzero positive quantity; therefore, causally connected regions shrink in the black hole section of SdS geometry.  \\\\
In summary, as expected the general behavior in four dimensions is the same, but the size of the temporal window for inward radial null geodesics, related to the magnitude of the inward bending of the black and white hole singularity, exceeds the size of the temporal window for outward radial null geodesics, related to the magnitude of the outward bending of asymptotic de Sitter future and past infinity, for any value of the mass. %Added a couple of sentences below
We refer to Figure \ref{fig:SdSa} for the graphical illustrations of these results for $D=4$ in terms of the causal Penrose diagram. One clearly observes that the larger inward bending of the black and white hole singularity, as compared to the outward bending of future and past de Sitter infinity, implies that the largest interior black hole causal region and largest exterior de Sitter causal region are disconnected, whereas in five dimensions they touch. We have sketched the four-dimensional causal Penrose diagrams using two different perspectives, either centered around the static sphere observer or centered around the black hole.

%we note that null geodesics originating from $x$ can be in causal contact with the other geodesic starting from $x'$ only through the asymptotic de Sitter (dS) region when $x$ falls in the range $-\mathcal{\tilde{T}}_b < \tilde{\tau} < \mathcal{\tilde{T}}_c$. They will be in causal contact with each other through both the black hole and dS region when $x$ lies between $-\infty < \tilde{\tau} < -\mathcal{\tilde{T}}_b$.
%They will not be in causal with either of the two regions when   $x$
%is between $\mathcal{\tilde{T}}_c<\tilde{\tau}<\infty$. The main essence of the above discussion also holds for five dimensional SdS except $\mathcal{\tilde{T}}_b=\mathcal{\tilde{T}}_c$ in that case. 

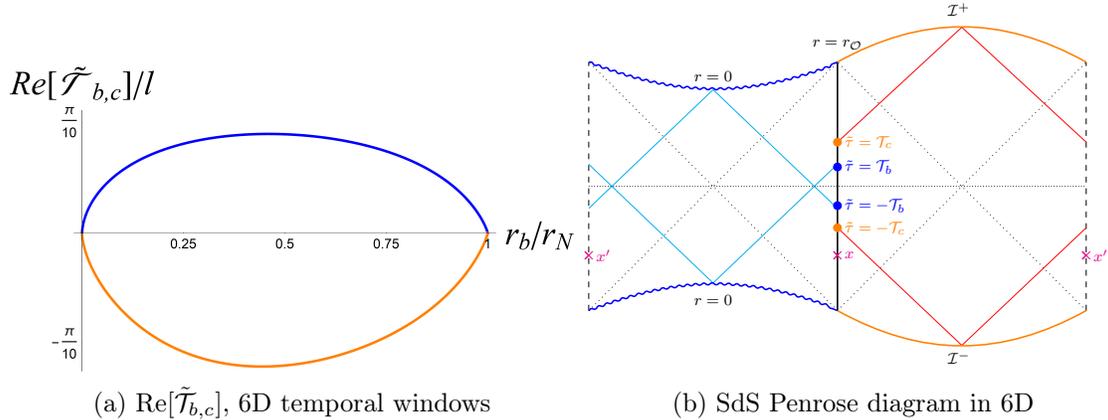
\begin{figure}[t]
\centering
\begin{subfigure}{.5\textwidth}
 \centering
\includegraphics[width=7.5cm]{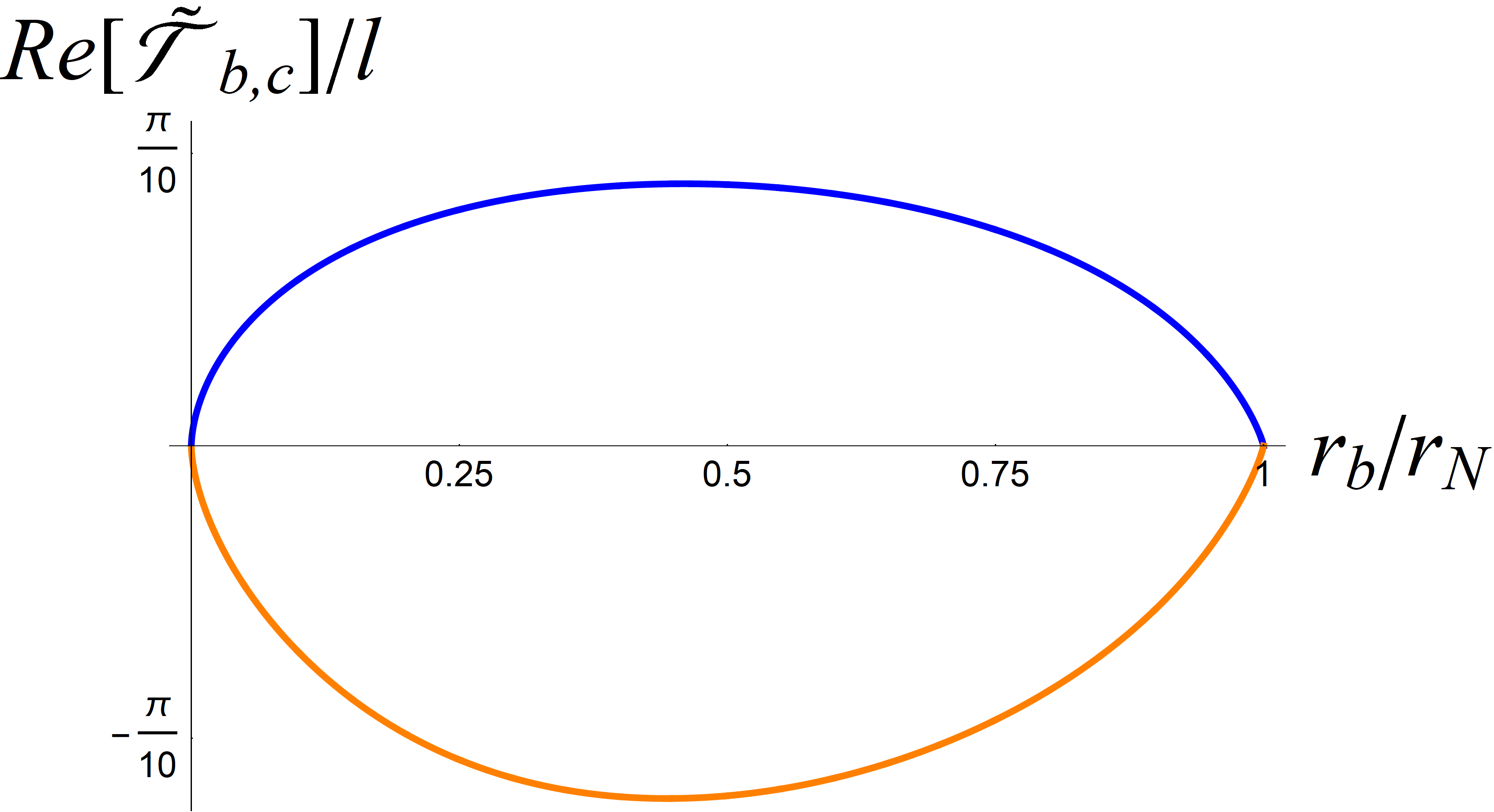}
 \caption{$\text{Re}[\tilde{\mathcal{T}}_{b,c}]$, 6D temporal windows}
\end{subfigure}%
\begin{subfigure}{.46\textwidth}
    \centering
    \resizebox{\linewidth}{!}{%
        \begin{tikzpicture}[scale=1]
              \node at (2.25,4.25) {\scriptsize $r=0$};
            \node[above] at (6.7,5.2) {\scriptsize $\mathcal{I}^+$}; % Adjusted position
            \node[above] at (6.7,-1.1) {\scriptsize $\mathcal{I}^-$}; % Adjusted position
              \node at (2.25,0.2) {\scriptsize $r=0$};

            \draw[dashed] (9,0) -- (9,4.5);
            \draw[dashed] (0,0) -- (0,4.5);
            \draw[thick] (4.5,0) -- (4.5,4.5);
            \draw[dotted] (9,0) -- (4.5,4.5);
            \draw[dotted] (4.5,0) -- (9,4.5);
            \draw[dotted] (0,0) -- (4.5,4.5);
            \draw[dotted] (4.5,0) -- (0,4.5);
            \node[above] at (4.5,4.6) {\scriptsize $r=r_{\mathcal{O}}$};

  \draw[cyan]  (4.5,1.85) -- (2.25,4);

              \draw[cyan]  (0,1.85) -- (2.25,4);

              \draw[cyan]  (0,2.65) -- (2.25,0.5);

              \draw[cyan]  (4.5,2.65) -- (2.25,0.5);

  \draw[red] (6.75,5.13) -- (9,3.05);
  \draw[red] (6.75,5.13) -- (4.5,3.05);

  \draw[red] (6.75,-0.63) -- (9,1.5);
  \draw[red] (6.75,-0.63) -- (4.5,1.5);
            
            \draw[thick, color=blue, decoration={snake,amplitude=.17mm,segment length=1.5mm, post length=.5mm}, decorate] (0,4.5) to[out=-21, in=-159] (4.5,4.5);
            \draw[thick, color=blue, decoration={snake,amplitude=.17mm,segment length=1.5mm, post length=.5mm}, decorate] (0,0) to[out=21, in=159] (4.5,0);
              
            \draw[thick, color=orange] (4.5,4.5) to[out=29, in=151] (9,4.5);
            \draw[thick, color=orange] (4.5,0) to[out=-29, in=-151] (9,0);

                \draw[densely dotted] (0,2.25) -- (9,2.25);

                \draw[magenta, line width=0.5pt] (4.5,01) -- ++(-2pt,-2pt) -- ++(4pt,4pt);

\draw[magenta, line width=0.5pt] (4.5,01) -- ++(-2pt,2pt) -- ++(4pt,-4pt);
\node[magenta, anchor=west] at (4.5,01) {\scriptsize $x$};

\draw[magenta, line width=0.5pt] (0,01) -- ++(-2pt,-2pt) -- ++(4pt,4pt);
\draw[magenta, line width=0.5pt] (0,01) -- ++(-2pt,2pt) -- ++(4pt,-4pt);
\node[magenta, anchor=west] at (0,01) {\scriptsize $x'$};

\draw[magenta, line width=0.5pt] (9,01) -- ++(-2pt,-2pt) -- ++(4pt,4pt);
\draw[magenta, line width=0.5pt] (9,01) -- ++(-2pt,2pt) -- ++(4pt,-4pt);
\node[magenta, anchor=west] at (9,01) {\scriptsize $x'$};
 \filldraw[blue] (4.5,2.6) circle (2pt) node[anchor=west]{\scriptsize $\tilde{\tau}=\mathcal{T}_b$};
              \filldraw[blue] (4.5,1.9) circle (2pt) node[anchor=west]{\scriptsize $\tilde{\tau}=-\mathcal{T}_b$};
               \filldraw[orange] (4.5,3.05) circle (2pt) node[anchor=west]{
               \scriptsize $\tilde{\tau}=
               \mathcal{T}_c$};
            \filldraw[orange] (4.5,1.5) circle (2pt) node[anchor=west]{\scriptsize $\tilde{\tau}=-\mathcal{T}_c$};
        \end{tikzpicture}
    }
    \caption{SdS Penrose diagram in 6D}
\end{subfigure}
\caption{(a) $\mathcal{T}_{b}$ (blue curve) and $-\mathcal{T}_{c}$ (orange curve) as a function of the black hole radius $r_b$ in $6$ dimensions. 
(b) Causal Penrose diagram of $6$-dimensional SdS, with respect to static sphere observers, showing larger outward bending of future (past) infinity as compared to the inward bending of the black (white) hole singularity.}
\label{fig:6dwindows}
\end{figure}

To finish this section on radial null geodesics in SdS spacetime, let us discuss the results for the static sphere time parameter along inward and outward radial null geodesics in $6$-dimensional SdS, which we have plotted in fig \ref{fig:6dwindows}. For the six-dimensional case we only plotted the real part of the (complex) static sphere time parameter, and in addition sketched the corresponding $6$-dimensional Penrose diagram. Although technically more involved (for the detailed expression we refer to the appendix), the general results are of course similar, except that in $6$ dimensions we conclude that $|\mathcal{T}_c| \geq |\mathcal{T}_b|$ for any value of the black hole mass. 
%This indicates a complete different trend in the causal structure of six dimensional SdS geometry than the above mentioned cases.
In other words, in $6$ dimensions the size of the outward bending of future (or past) infinity, as measured in terms of the temporal window at the static sphere, exceeds the size of the inward bending of the black (or white) hole singularity, for any black hole mass. Qualitatively this is therefore opposite to the situation in $4$ dimensions, with $5$ dimensions sitting exactly in between the two types of behavior, where the interior (inward) and exterior (outward) temporal window are exactly the same. This means that in $D=6$ the largest interior black hole causal diamond overlaps with the largest exterior de Sitter causal diamond, as can be readily checked in \ref{fig:6dwindows}. 
% modified and extended the text below
In dimensions more than $6$ we expect the larger exterior (outward) bending of future (past) infinity to become more pronounced as the number of dimensions grows, i.e. that the largest interior and exterior causal diamonds will overlap more and more. One might anticipate a simplification of the results in the large $D$ limit, but we were unable to verify this. 
 
\begin{figure}[h]
\centering
\resizebox{0.6\textwidth}{!}
{%
\begin{tikzpicture}
\draw[ dotted] (0,0) -- (2.25,2.25);
% \draw[ dashed,blue!90!black] (4.5,4.5) -- (2.25,2.25);
\draw[dotted] (4.5,4.5) -- (2.25,2.25);

%\draw[] (0,2.25) -- (1.6,3.9);

\node[above] at (4.37,4.55) {\scriptsize $r=r_\mathcal{O}$};

\filldraw[] (4.5,2.25) circle (2pt) node[anchor=west]{\scriptsize $\tilde{\tau}=0$};

 \draw[cyan] (4.5,2.25) -- (2.25,0.0);
 \draw[cyan] (0,2.25) -- (2.25,0);
 \draw[cyan] (0,2.25) -- (2.25,4.5);
\draw[cyan] (4.5,2.25) -- (2.25,4.5);
\draw[red] (4.5,2.25) -- (6.75,4.495);
\draw[red] (9,2.25) -- (6.75,4.495);
\draw[red ] (9,2.25) -- (6.75,0.0);
\draw[red] (4.5,2.25) -- (6.75,-.0);

\draw[ dotted] (0,4.5) -- (2.25,2.25);

% \draw[ dashed,blue!90!black] (4.5,0) -- (2.25,2.25);
\draw[dotted] (4.5,0) -- (2.25,2.25);
% \draw[ dashed,orange] (4.5,0) -- (6.75,2.25);
\draw[dotted] (4.5,0) -- (6.75,2.25);
\draw[dotted] (4.5,4.53) -- (6.75,2.25);
% \draw[ dashed,orange] (4.5,4.53) -- (6.75,2.25);
\draw[dashed] (9,0) -- (9,4.5);
\draw[dotted] (9,0) -- (6.75,2.25);
%\draw[thick] (6.75,4.5) -- (9,4.5); % Removed line
%\draw[densely dotted] (6.75,4.5) -- (6.75,2.25); % Modified line
\draw[dashed] (0,0) -- (0,4.5);
\draw[thick] (4.5,0) -- (4.5,4.55);
\draw[dotted] (6.75,2.25) -- (9,4.5);

% Add the label to the right-hand most center
%\node at (9.5,2.25) {\scriptsize $\tilde{\tau}=0$};

\node at (2.25,4.8) {\scriptsize $r=0$};
\node[above] at (6.83,4.65) {$\mathcal{I}^+$};
\node[below] at (6.83,-0.1) {$\mathcal{I}^-$};
\node at (2.25,-0.3) {\scriptsize $r=0$};

%\node at (2.45,3) {\scriptsize $r=r_b$};
\draw[densely dotted] (0,2.25) -- (9,2.25);
\draw[thick, color=orange] (4.5,4.55) to (9,4.5);
\draw[thick, color=orange] (4.5,0) to (9,0);
% Draw a cross and identify the point at (4,0)
\draw[magenta, line width=0.5pt] (4.5,1.2) -- ++(-2pt,-2pt) -- ++(4pt,4pt);
\draw[magenta, line width=0.5pt] (4.5,1.2) -- ++(-2pt,2pt) -- ++(4pt,-4pt);
\node[magenta, anchor=west] at (4.5,1.0) {\scriptsize $x$};

\draw[magenta, line width=0.5pt] (0,1.2) -- ++(-2pt,-2pt) -- ++(4pt,4pt);
\draw[magenta, line width=0.5pt] (0,1.2) -- ++(-2pt,2pt) -- ++(4pt,-4pt);
\node[magenta, anchor=west] at (0,1.0) {\scriptsize $x'$};

\draw[magenta, line width=0.5pt] (9,1.2) -- ++(-2pt,-2pt) -- ++(4pt,4pt);
\draw[magenta, line width=0.5pt] (9,1.2) -- ++(-2pt,2pt) -- ++(4pt,-4pt);
\node[magenta, anchor=west] at (9,1.0) {\scriptsize $x'$};

\draw[thick, color=blue, decoration={snake,amplitude=.15mm,segment length=2mm,
                       post length=.6mm,pre length=.2mm}, decorate] (0,4.5) to (4.5,4.55);
\draw[thick, color=blue, decoration={snake,amplitude=.15mm,segment length=2mm,
                       post length=.6mm,pre length=.2mm}, decorate] (0,0) to (4.5,0);
\end{tikzpicture}
} % resizebox
\caption{$D$-dimensional Penrose diagram of SdS in the Nariai limit.}
%The dashed blue and orange lines are the blackhole horizon $r=r_b$ and cosmological horizons $r=r_c$, respectively.
%Null geodesic starting from $\tilde{\tau}=0$
%reaches the centre blackhole
%singularity or the cosmological future infinity in the Nariai case.}
\label{penrose.nariai}
\end{figure}
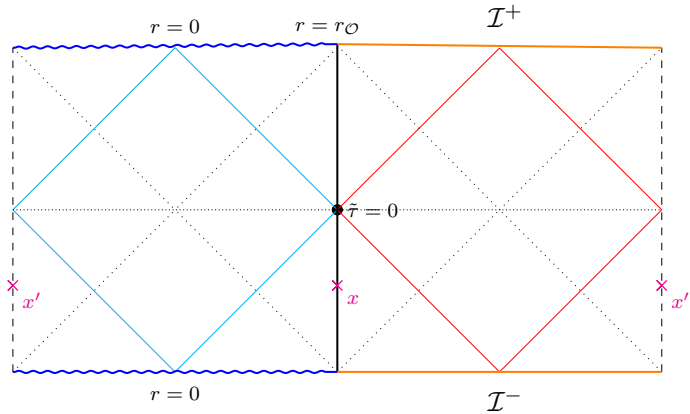

As before we conclude that symmetric conjugate static sphere observers, on the same global time slice, are part of the same exterior causal region as long as $-\mathcal{T}_c < \tilde{\tau} < \mathcal{T}_c$. Correspondingly, the temporal window $-\mathcal{T}_b < \tilde{\tau} < \mathcal{T}_b$ informs us that the interior black hole region has decreased, i.e. for all times within that window symmetric emitted or received null shells from the pair of conjugate static sphere observers cannot intersect in the interior black hole region. 
%can intersect causal contact before reaching the singularity or future infinity through both the blackhole region or asymptotic dS region if $x$ falls within $-\infty<\tilde{\tau}<-\mathcal{\tilde{T}}_c$. If on the other hand 
%$x$ falls between $-\mathcal{\tilde{T}}_c<\tilde{\tau}<\mathcal{\tilde{T}}_c$, these points can only communicate through the  asymptotic dS %region.
%When $x$ is located $\tilde{\tau}>\mathcal{\tilde{T}}_c$
%on the static sphere,
%the null geodesics never meet. However, in six dimensions, $\mathcal{{\tilde{T}}}_c> \mathcal{\tilde{T}}_b$ for any mass of the black hole beside the Nariai limit. \\\\
%Because of the outward bending of $\mathcal{I}^\pm$ in the asymptotic dS sector and the inward bending of the black hole singularity, we observe intriguing aspects concerning the causal connectivity of points lying on opposite static spheres from which null geodesics originate.
Finally, let us emphasize that our conclusions for the relative size of the interior (inward) and exterior (outward) bending of the black hole singularity and asymptotic de Sitter infinity hold for any value of the mass. Moreover, for any number of dimensions, in the Nariai limit, the causal structure for radial null geodesics reduces again to that of empty, pure, de Sitter, as anticipated from the near-horizon de Sitter structure of the geometry. In the Nariai limit the causal Penrose diagram therefore effectively reduces to that of $2$-dimensional de Sitter spacetime, as portrayed in Figure \ref{penrose.nariai}. 
%This is noticeable  for any dimensions, with  non-zero mas of an SdS black hole except in the Nariai limit,  as visible
%from figure \ref{fig:5dwindows}, \ref{4dw} and \ref{fig:6dwindows}.
%In the Nariai limit there is no bending as  
%\(\text{Re}[\mathcal{T}_c]=\text{Re}[\mathcal{T}_b]=0\) generating  Penrose diagram with perfect square in all dimensions. 
%This makes the causal structure trivial, i.e., the null rays originating from the opposite static spheres can be in causal contact with each other through either of the regions only when $x$ is between $-\infty<\tilde{\tau}\leq 0$. 
%So in any number of dimensions, the increase of the exterior (outward) causal diamond, with respect to conjugate static sphere observers, disappears again in the Nariai limit.

This ends our discussion of null geodesics and the corresponding bending inwards of the black (white) hole singularity, as well as the bending outwards of future/past infinity. We can rephrase these results as follows: as compared to pure de Sitter spacetime, where the causally connected regions (defined by the center of future and past infinity) of two conjugate observers touch at their boundaries, in the SdS geometry the interior black hole (defined by the center of the black/white hole singularity) and the exterior de Sitter (defined by the center of future/past infinity)  causal regions are disconnected in $D=4$, they touch (as in pure de Sitter) in $D=5$, and they overlap in $D=6$. 

This structure should have consequences for the symmetric correlations between conjugate (large mass) operators. In \cite{Fidkowski:2003nf} the modified causal structure of the interior black hole region signalled a `phase transition' for symmetric correlations between conjugate (large mass) operators exactly as one crosses the boundary of the temporal window, when the correlator starts to probe the singularity. Using a geodesic approximation for the symmetric correlators of (large mass) conjugate operators this transition is signalled by the presence (and disappearance) of spacelike geodesics between the conjugate operators. The spatial compactness of the de Sitter and SdS geometry implies that two conjugate operators can be connected by spacelike geodesics either through the black hole interior or the de Sitter exterior. As is well known, in pure de Sitter, except for the symmetric point at $\tau=0$, no spacelike geodesics exist that connect symmetric pairs of conjugate operators, but the appearance of (two) temporal windows suggests otherwise for SdS. In the next section we will generalize our analysis of radial null geodesics in SdS to study the presence (and length) of spacelike geodesics between symmetric pairs of conjugate static sphere observers. 
%%%%%%%%%%%

\subsection{Spacelike geodesics between static sphere observers in SdS}
\label{sec2}

%In the previous section we studied the causal structure in SdS geometry with the help of null geodesics.
In the previous section, we observed that the causal connectivity in the SdS background exhibits some interesting features. In general, the interior and exterior causal regions shrink and expand respectively, opening up two temporal windows for conjugate static sphere observers. 
% Modified text below %
The details, conveniently summarized in the figures \ref{fig:SdS}, \ref{fig:SdSa} and \ref{fig:6dwindows}, depend on the dimension of the SdS spacetime under consideration, qualitatively changing from disconnected interior and exterior causal regions in $D=4$ to overlapping causal regions in $D=6$. In this section, we will study the related appearance of radial spacelike geodesics connecting symmetric conjugate points $x$ and $x'$, located on opposite static spheres at the same global time slice. 

\begin{figure}[t]
\begin{subfigure}{.5\textwidth}
\centering
\includegraphics[width=7.5cm]{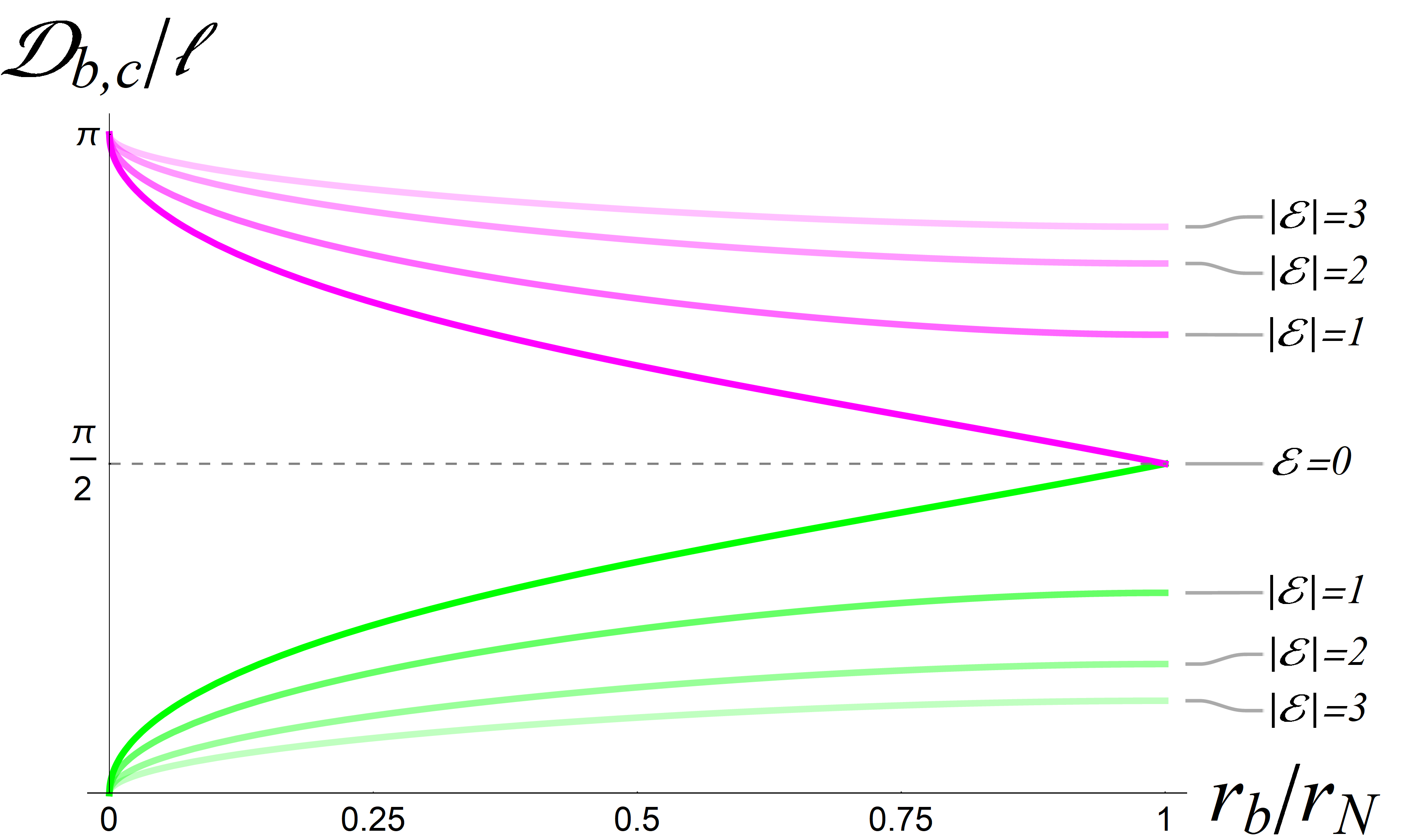}
\caption{5D spatial lengths}
\label{subfigure1.ax}
\end{subfigure}%
\begin{subfigure}{.5\textwidth}
\centering
\includegraphics[width=7.5cm]{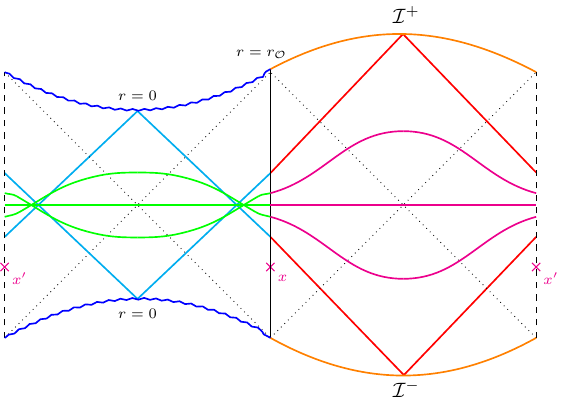}
\caption{Penrose diagram in $D=5$}
\label{subfigure2.ax}
\end{subfigure}
\caption{Left: different lengths of interior and exterior spacelike geodesics in $D=5$ as a function of the black hole radius $r_b$, for different $\mathcal{E}=0,1,2,3$, indicated by their respective opacity. Right: Penrose diagram with (approximated) spacelike geodesics connecting symmetric conjugate static sphere points (interior in green, exterior in magenta) for different $\mathcal{E}$.}
\label{fig:LeoComp5}
\end{figure}

\begin{figure}[h!]
\centering
\begin{subfigure}{.48\textwidth}
\centering
\includegraphics[width=\linewidth]{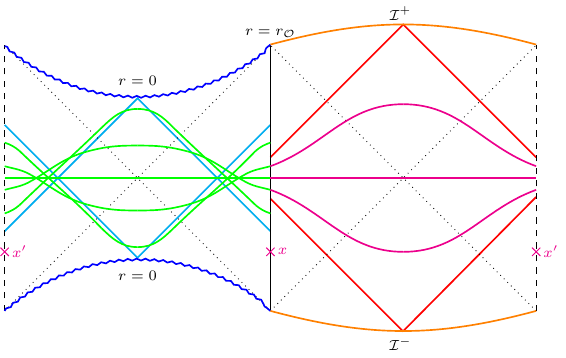}
\caption{4D Penrose diagram}
\label{subfigure2.a}
\end{subfigure}%
\hfill
\begin{subfigure}{.48\textwidth}
\centering
\includegraphics[width=\linewidth]{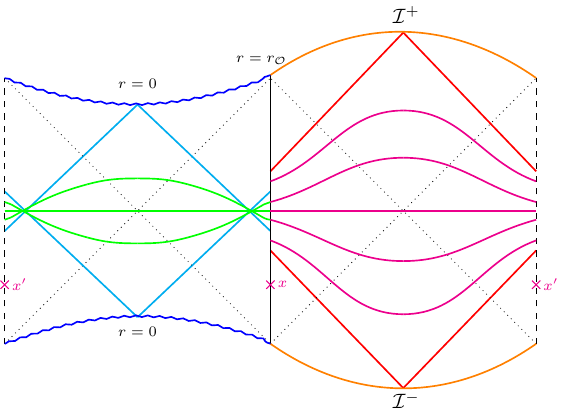}
\caption{6D Penrose diagram}
\label{subfigure3.a}
\end{subfigure}
\caption{Penrose diagram with (approximated) spacelike geodesics connecting symmetric conjugate static sphere points (interior in green, exterior in magenta) for different $\mathcal{E}$, in $D=4$ (left) and $D=6$ (right). More geodesics were added to better highlight the different temporal window sizes.}
\label{fig:TimeComp466}
\end{figure}

We start our discussion again in $D=5$, but will also briefly discuss the $4$- and $6$-dimensional case. All the results are graphically summarized in Figures \ref{fig:LeoComp5} (for $D=5$) and \ref{fig:TimeComp466} (for $D=4$ and $D=6$). For radial spacelike geodesics the conserved quantity $\mathcal{E}$ plays an important role, as is evident from \eqref{eq:spgeot} and \eqref{eq:spgeoL}. For example, in empty dS space, two (symmetric) points lying at the north and south poles at $\tau_N = \tau_S = 0$ are connected by a family of real spacelike geodesics characterized by $\mathcal{E}$ (see \cite{Chapman:2021eyy} for a nice review with some specific examples). When the conserved quantity $\mathcal{E}$ approaches infinity, the spacelike geodesics from the north pole at $\tau_N = 0$ get ever closer to future infinity before they turn around and reach the south pole at $\tau_S = 0$. For any other (symmetric) time,  no (real) spacelike geodesics exist that connect two conjugate points \cite{Lars:2022geo, Galante:2022nhj}. As our analysis of the SdS causal structure has shown, introducing a black hole in de Sitter significantly affects the above situation for a pair of conjugate static sphere observers.  
%Surprisingly, in SdS geometry any two points on the opposite static spheres, (let's say $x$ and $x'$ in figure \ref{fig:SdS}), can be connected to each other through asymptotic dS region with real spacelike geodesics if $x$ lie between $-\mathcal{\tilde{T}}_c < \tilde{\tau} < \mathcal{\tilde{T}}_c$.\\\\
For SdS one should instead be able to construct spacelike geodesics connecting two symmetric conjugate points for some (finite) temporal window, both through the interior black hole region, as well as the exterior de Sitter region. Below we will work out some of the details, keeping in mind that we have two horizons, giving rise to an interior and exterior connection between conjugate points and that we should use the appropriately normalized static sphere time \eqref{hawking} in our analysis. 

\begin{figure}[t]
\centering
  \includegraphics[width=0.65\linewidth]{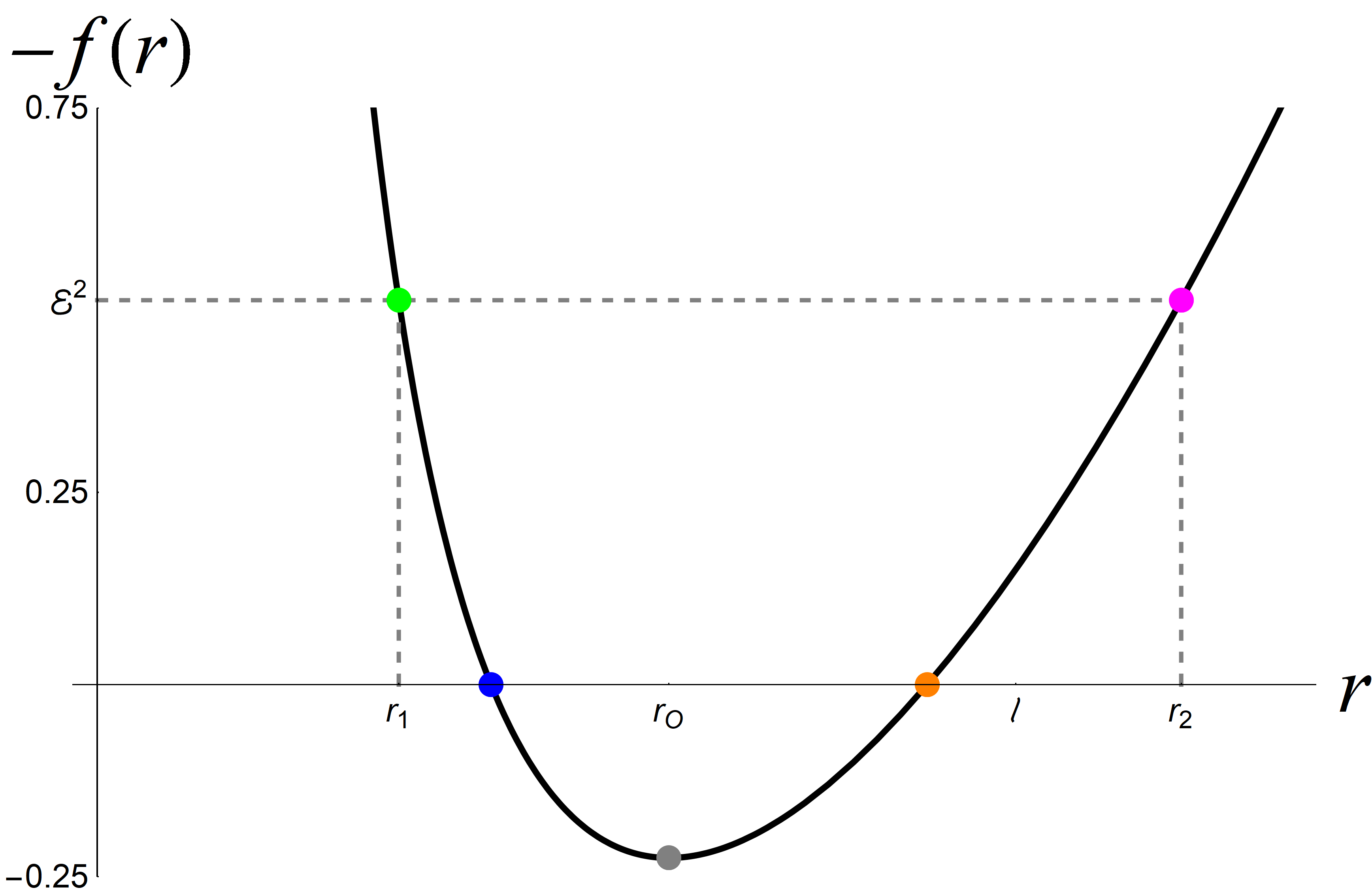}
\caption{Turning points for spacelike geodesics in the SdS geometry, for some value of $\mathcal{E}$. The green and magenta points correspond to the turning points, while the blue and orange points are the black hole and cosmological horizon respectively. The grey point denotes the position of the static sphere.}
\label{fig:5dpotential}
\end{figure}

We will start with the general expressions for the proper lengths of spacelike radial geodesics connecting two conjugate symmetric points
\begin{eqnarray}
&&\mathcal{D}_{b}=2\int_{r_{1}}^{r_{\mathcal{O}}}\frac{\rmd r}{\sqrt{f(r)+\mathcal{E}^2}}\label{L_b}\\
&&\mathcal{D}_{c}=2\int_{r_{\mathcal{O}}}^{r_{2}}\frac{\rmd r}{\sqrt{f(r)+\mathcal{E}^2}}\label{L_c}.
\end{eqnarray}
Here, $\mathcal{D}_b$ and $\mathcal{D}_c$ are the lengths of radial spacelike geodesics connecting opposite symmetric points on the static sphere through the interior black hole and exterior de Sitter regions. Note that $\mathcal{E}$ will be a function of the initial (proper) time on the static sphere. In other words, for the case at hand: a spacelike geodesic connecting symmetric points labeled by some time $\tilde{t}$ only exists for a specific value of $\mathcal{E}$, or does not exist at all. This relation is governed by \eqref{eq:spgeot}. The radii $r_1$ and $r_2$ are the turning points for the radial spacelike geodesics inside the interior black hole and the exterior asymptotic dS region, respectively. They correspond to the positive roots of $f(r)+\mathcal{E}^2$, assuming $\mathcal{E}$ is real, where the two conjugate parts of the radial spacelike geodesic are glued together in a consistent, continuous, way\footnote{This means that the turning points always lie in the 'center' $\text{Re}[\tilde{t}]=0$ and the derivative $d\tilde{t}/dr$ vanishes there as well.}. The presence of turning points can be nicely understood graphically by plotting the blackening factor, with an opposite sign, behaving as a potential barrier \cite{Fidkowski:2003nf}, as shown in Figure \ref{fig:5dpotential}, and where the conserved quantity $\mathcal{E}$ can then be interpreted as the energy determining the location of the turning points. In $4$- and $5$-dimensional SdS spacetime they are given by
 \begin{equation}
 r_{1,2}=
    \begin{cases}
        r_N\sqrt{(1+\mathcal{E}^2)\mp\sqrt{(1+\mathcal{E}^2)^2-M/M_N}} &,\, D=5 \\
r_N\sqrt{1+\mathcal{E}^2}\left(\cos{\eta'}\mp\sqrt{3}\sin{\eta'}\right) &,\, D=4 \, , 
    \end{cases}
\label{eq:tprE}
\end{equation}
where
\begin{equation}
\eta'=\frac{1}{3}\arccos{\frac{M}{M_N(1+\mathcal{E}^2)^{3/2}}}\,,
\end{equation}
Note that the initial derivative of the radial spacelike geodesic $(d\tilde{t}/dr)|_{r_\mathcal{O}}$ also depends on $\mathcal{E}$, and that it should approach the (inward or outward) radial null geodesic in the $\mathcal{E} \rightarrow \pm \infty$ limit. 
%Let us now clarify the story before solving the integrals of equations \eqref{L_b} and \eqref{L_c}. 
%First, considering the spacelike geodesic starting from the static sphere, passing through the interior of the black hole to reach the symmetric point of the opposite static sphere. 

Let us rewrite \eqref{eq:spgeot} for the specific case of a spacelike radial geodesic passing into the black hole interior from a particular time $\tilde{\tau}_b$ at the static sphere radius
\beq
\tilde{t}_{\mkern2mu b}(r)-\tilde{\tau}_b
% =\int_{\tau}^{t(r)}dt
=\int^{r_{\mathcal{O}}}_r\frac{\gamma \,\mathcal{E}\,\rmd r'}{f(r')\sqrt{f(r')+\mathcal{E}^2}}\,.
\label{eq:spgeott}
\eeq
Introducing the interior, black hole, turning point $r=r_1$ (at which $\text{Re}[\tilde{t}_b]=0$), to allow for a (symmetric) connection to the other conjugate static sphere observer, one obtains  
\begin{equation}
-\tilde{\tau}_b=i\frac{\beta_{b}}{4}+\int_{r_{1}}^{r_{\mathcal{O}}}\frac{\gamma\,\mathcal{E}\,\rmd r}{f(r)\sqrt{f(r)+\mathcal{E}^2}}\,.
\label{eq:windows1}
\end{equation}
The above equation relates the conserved quantity $\mathcal{E}$ to the initial time $\tilde{\tau}_b$ on the static sphere for interior, black hole, spacelike geodesics reaching an (interior) turning point. Similarly, we find a relation between $\mathcal{E}$ and the initial time $\tilde{\tau}_c$ on the static sphere for the other, exterior de Sitter, spacelike geodesic reaching an (exterior) turning point
\begin{equation}
-\tilde{\tau}_c=-i\frac{\beta_{c}}{4}+\int_{r_{\mathcal{O}}}^{r_{{2}}}\frac{\gamma\,\mathcal{E}\,\rmd r}{f(r)\sqrt{f(r)+\mathcal{E}^2}}\,,
\label{eq:windows2}
\end{equation}
 As we swapped the boundaries of the integral when switching from interior to exterior spacelike radial geodesics, we will pick up a relative minus sign in comparing interior and exterior geodesics (related to the distinction between inward and outward radial geodesics) that can also be incorporated in the sign of the conserved quantity $\mathcal{E}$. However, to more easily distinguish the results for the interior and exterior geodesics in our plots, we will keep this relative minus sign for the static sphere time comparing interior and exterior spacelike geodesics, for positive values of $\mathcal{E}$. Fixing one overall consistent sign for the two different integrals will remove this relative sign difference. In other words, the negative values we will find for the exterior static sphere time, for positive $\mathcal{E}$, are useful for plotting them in the same graph, but actually correspond to negative $\mathcal{E}$. As long as we are aware of this small subtlety, we can now go ahead and evaluate the integrals \eqref{eq:windows1} and \eqref{eq:windows2} in $D=5$, for which we obtain the following (analytic) results for the relation between (real) static sphere times and the conserved quantity $\mathcal{E}$

\begin{align}
 \tilde{\tau}_b&=\text{Re} \left[ \frac{\ell^2\,\mathcal{E}}{r_b+r_c} \left(\frac{r_b^2 \arctanh\left(\frac{\sqrt{r_{\mathcal{O}}^2-r_1^2}}{\sqrt{r_2^2-r_{\mathcal{O}}^2}}\frac{\sqrt{r_2^2-r_b^2}
}{\sqrt{r_b^2-r_1^2}
}\right)}{\sqrt{r_2^2-r_b^2}
\sqrt{r_b^2-r_1^2}}-\frac{r_c^2 \arctan\left(\frac{\sqrt{r_{\mathcal{O}}^2-r_1^2}}{\sqrt{r_2^2-r_{\mathcal{O}}^2}}\frac{\sqrt{r_c^2-r_2^2}}{\sqrt{r_c^2-r_1^2}}\right)}{\sqrt{r_c^2-r_1^2}
\sqrt{r_c^2-r_2^2}}\right) \right]
%-i\frac{\beta_b}{4}
\,,\\
 \tilde{\tau}_c&= - \tilde{\tau}_b \,.
 %\frac{i \pi \ell ^2 \, \mathcal{E}}{2
%\left(r_b+r_c\right)}\left(\frac{r_c^2}{\sqrt{r_2^2-r_c^2}
%\sqrt{r_c^2-r_1^2}}-\frac{r_b^2}{\sqrt{r_2^2-r_b^2} \sqrt{r_b^2-r_1^2}}\right) 
%-i\frac{\beta_c}{4}
%\nonumber\\
%&-\frac{\ell^2\,\mathcal{E}}{r_b+r_c} \left(\frac{r_b^2 \arctanh\left(\frac{\sqrt{r_{\mathcal{O}}^2-%r_1^2}}{\sqrt{r_2^2-r_{\mathcal{O}}^2}}\frac{\sqrt{r_2^2-r_b^2}
%}{\sqrt{r_b^2-r_1^2}
%}\right)}{\sqrt{r_2^2-r_b^2}
%\sqrt{r_b^2-r_1^2}}-\frac{r_c^2 \arctan\left(\frac{\sqrt{r_{\mathcal{O}}^2-r_1^2}}{\sqrt{r_2^2-%r_{\mathcal{O}}^2}}\frac{\sqrt{r_c^2-r_2^2}}{\sqrt{r_c^2-r_1^2}}\right)}{\sqrt{r_c^2-r_1^2}
%\sqrt{r_c^2-r_2^2}}\right)\,,
\end{align}%i.e.   $\mathcal{\tilde{T}}_b=\mathcal{\tilde{T}}_c$.
\begin{figure}[t]
\centering
\includegraphics[width=0.65\linewidth]{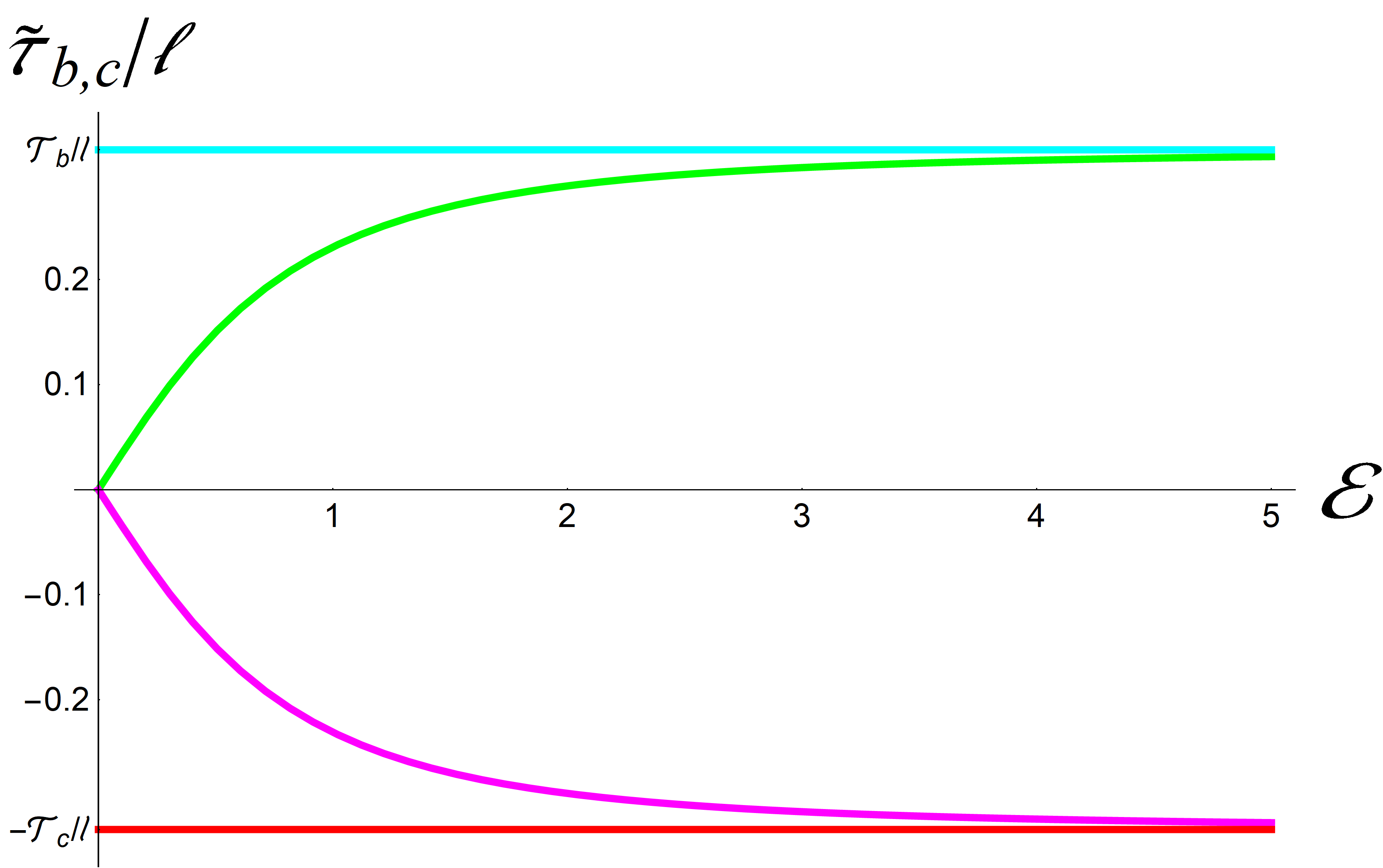} % Adjust the width (0.8 here) to make it smaller
\caption{Static sphere times as a function of $\mathcal{E}$ for interior (green) and exterior (magenta) radial spacelike geodesics in $5$-dimensional SdS at fixed $r_b/r_N=1/2$.}
%$\tilde{\tau}_b$ (green) and $\tilde{\tau}_c$ (magenta) in five-dimensional SdS black hole as a function of the energy $\mathcal{E}$. We have fixed $r_b/r_N=1/2$.}
\label{fig:TimeComp46a}
\end{figure}

%We have just quoted the final result here, for some of the technical subtleties related to the cancellation of the imaginary parts
We refer to the appendix for the full expression. 
The final $5$-dimensional result is conveniently expressed in terms of the turning point radii, whose dependence on $\mathcal{E}$ is determined by (\ref{eq:tprE}). As for the radial null geodesics, the absolute values are the same, meaning that the interior and exterior static sphere time are the same for a fixed $\mathcal{E}$. We plotted the results for the real part of the static sphere times as a function of the conserved quantity $\mathcal{E}$ in figure \ref{fig:TimeComp46a}. 

For the spacelike radial geodesic passing through the interior black hole region, as the conserved quantity $\mathcal{E}$ goes to infinity, the initial time for the spacelike geodesic approaches $\tilde{\tau}_{b} \rightarrow \mathcal{T}_b$. Similarly, in the opposite limit $\mathcal{E}\rightarrow-\infty$, we find that $\tilde{\tau}_{b} \rightarrow -\mathcal{T}_b$. For $\mathcal{E}=0$ the spacelike radial geodesic connects two conjugate static sphere points at the center $\tilde{t}=0$, never entering the black hole interior, connecting through the bifurcation point (where the past and future event horizon intersect). 
%Thus, for every specific value of $\mathcal{E}$ a symmetric time in between $-\mathcal{T}_b \leq \tilde{\tau}^{\mkern2mu} < \mathcal{T}_b$ is identified for which a symmetric spacelike radial geodesic can be constructed. 
%the spacelike radial geodesic will start from a point . For a unique value of real $\mathcal{E}$, there exists only one geodesic connecting two symmetric points of opposite static spheres.

\begin{figure}[t]
\begin{subfigure}{.48\textwidth}
\centering
\includegraphics[width=\linewidth]{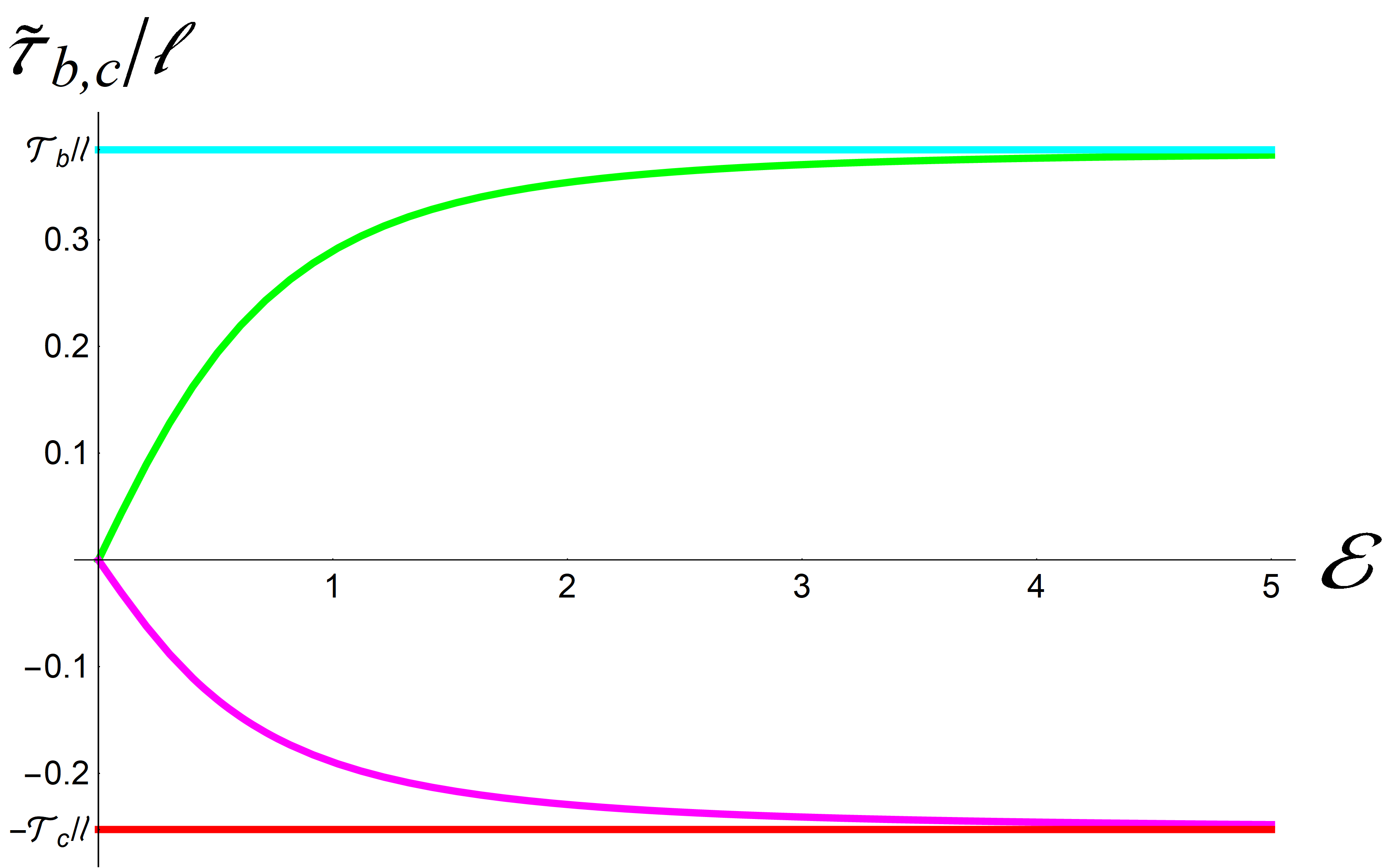}
\caption{$D=4$}
\label{subfigure2}
\end{subfigure}%
\hfill
\begin{subfigure}{.48\textwidth}
\centering
\includegraphics[width=\linewidth]{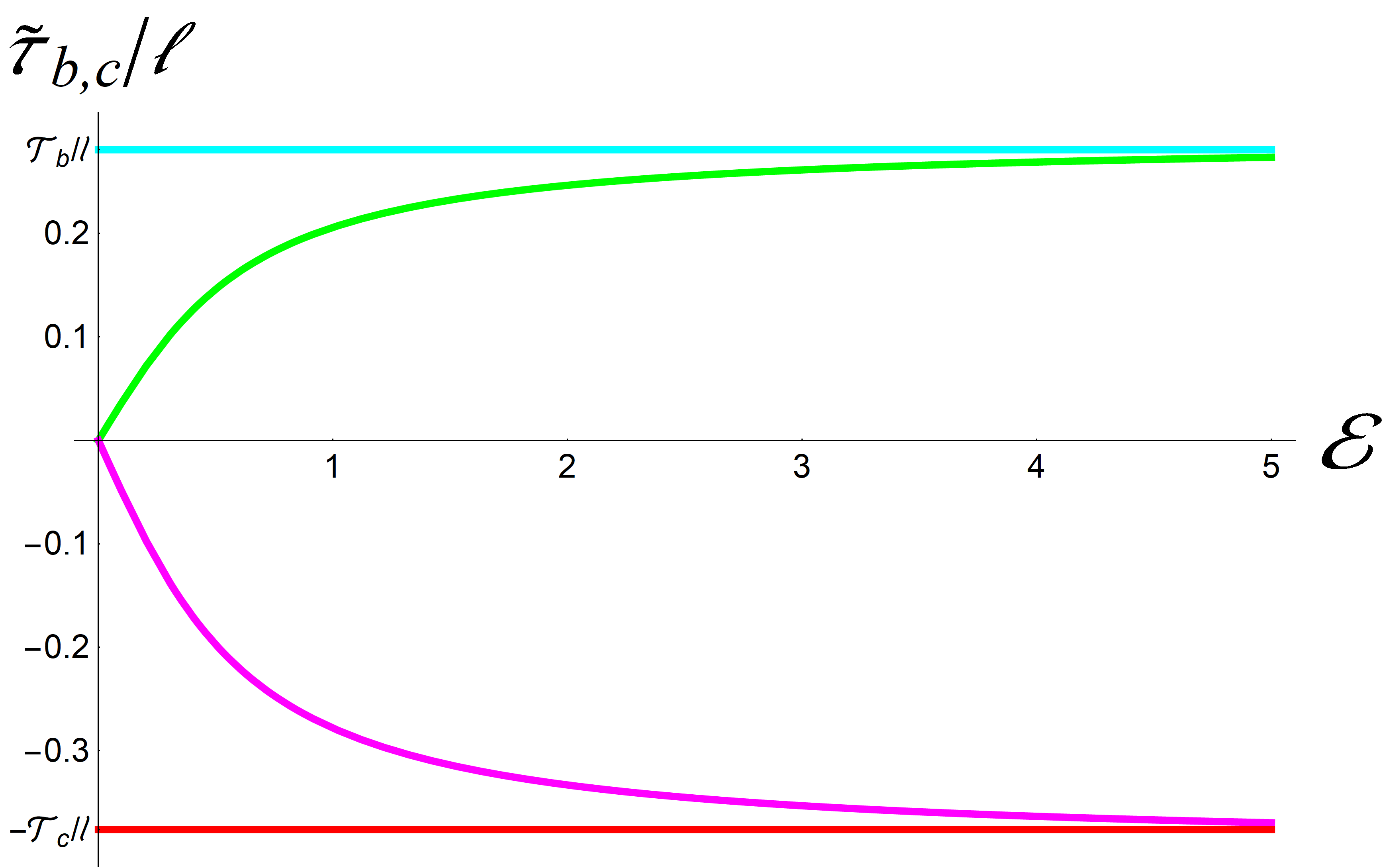}
\caption{$D=6$}
\label{subfigure3}
\end{subfigure}
\caption{Static sphere times as a function of $\mathcal{E}$ for interior (green) and exterior (magenta) radial spacelike geodesics in a) $4$-dimensional and b) $6$-dimensional SdS at fixed $r_b/r_N=1/2$.}
%$\tilde{\tau}_b$ (green) and $\tilde{\tau}_c$ (magenta) in  (b) four and (c) six-dimensional SdS black hole as a function of the energy $\mathcal{E}$. We have fixed $r_b/r_N=1/2$.}
\label{fig:TimeComp46}
\end{figure}
A similar story holds for the exterior radial spacelike geodesic passing through the asymptotic dS region, as is evident from figure \ref{fig:TimeComp46a}. In the limit $\mathcal{E}\rightarrow\pm\infty$, we find, as expected, that the radial spacelike geodesics approach $\tilde{\tau}_c\rightarrow\mp\mathcal{T}_c$. As we emphasized before, the $5$-dimensional SdS causal structure is special in the sense that the inward interior (black/white hole) bending equals the exterior (future/past infinity) bending, implying that the interior and exterior temporal windows are the same. This also implies that the map between static sphere times and $\mathcal{E}$ is the same for inward (interior) and outward (exterior) spacelike geodesics, i.e. the inward and outward radial spacelike geodesics can be smoothly connected (it is differentiable). In Figure \ref{fig:TimeComp46a} this property is made evident through the symmetry around the $\mathcal{E}$ axis. %the both geodesics approach their asymptotic windows at the same rate when we increase the value of $\mathcal{E}$. 

% Modified and extended below paragraph %
In $4$ and $6$ dimensions this symmetry between interior and exterior time windows is absent, as we saw in the previous section, see Figure \ref{fig:TimeComp46}. This also causes the integrals in eq. (\ref{eq:windows1}) and (\ref{eq:windows2}) to be more difficult to solve in four and six dimensions. We did find analytic results in $4$ dimensions, for which we mainly refer to the Appendix \eqref{4dstartingtime}, and we will briefly discuss those below. In $6$ dimensions we were only able to derive approximate results that we have plotted in Figure \ref{fig:TimeComp46}. We have not studied the large $D$ limit, where perhaps the expressions might simplify and allow for an analytical treatment. We will comment on the qualitative differences between the different dimensions and what we expect to happen when we increase the number of dimensions. For a quick and convenient graphical summary of the results we refer to the Figures \ref{fig:TimeComp46} and \ref{fig:TimeComp46}.

In four dimensions we concluded that the interior (black/white hole) temporal window exceeds the exterior (asymptotically de Sitter) window $|\mathcal{T}_b|>|\mathcal{T}_c|$, implying that the relation between static sphere time and $\mathcal{E}$ is not the same for interior and exterior radial spacelike geodesics. As a corollary, for a given static sphere time, the interior and exterior (conserved) quantity $\mathcal{E}$ is different, meaning that the interior and exterior radial spacelike geodesic do not connect in a differentiable way (as the inward and outward initial slope are not the same, as they are set by different $\mathcal{E}$). In four dimensions, for any finite mass below the Nariai limit, and for a fixed $\mathcal{E}$, the interior static sphere time is always bigger than the exterior static sphere time, i.e. $|\tilde{\tau}_b|>|\tilde{\tau}_c|$. And no symmetric radial spacelike geodesics exist between two conjugate static sphere observers beyond the maximum of the two temporal windows. Presumably, as for the AdS black hole \cite{Fidkowski:2003nf} and in pure de Sitter \cite{Aalsma:2022swk, Galante:2022nhj}, one should instead consider complex geodesics for static sphere times outside the temporal windows (related to complex values of the conserved quantity $\mathcal{E}$). In five dimensions this point coincides for interior (black hole) and exterior (de Sitter) geodesics, but in four dimensions one first crosses the exterior temporal window before reaching the interior temporal boundary.   

%The limiting behavior of energy $\mathcal{E}\rightarrow \pm \infty$ is similar in all dimensions. But, as we have seen in Section Three, in four %dimensions, the black hole singularity bending is more prominent compared to cosmological future infinity, as 
%$|\mathcal{\tilde{T}}_b|>|\mathcal{\tilde{T}}_c|$. With any fixed energy $\mathcal{E}$, the initial time on the static sphere for both geodesics is different, and $|\tilde{\tau}_b|>|\tilde{\tau}_c|$ for any non-zero black hole mass other than the Nariai limit. \\\\

Since we now understand when and how to connect two symmetric static sphere points by a radial spacelike geodesics either going through the interior black hole region or the exterior asymptotic de Sitter region, in $4$ and $5$ dimensions for any real value of $\mathcal{E}$, we can proceed and compute the proper lengths \eqref{L_b} and \eqref{L_c}. In $5$ dimensions one then obtains relatively nice analytic expressions for the interior and exterior geodesic lengths
\begin{align}
\mathcal{D}_b&=2\ell \arcsin{\sqrt{\frac{r_1^2-r_{\mathcal{O}}^2}{r_1^2-r_2^2}}}\,,\\
% \ell \left(\pi -2 \sin^{-1} \left(\frac{\sqrt{\sqrt{\frac{\left(\ell ^2-2r_b^2\right){}^2}{\ell ^4}+\mathcal{E}^2(\mathcal{E}^2+2)}+\mathcal{E}^2+1-2 \frac{r_b}{\ell}\sqrt{1-\frac{r_b^2}{\ell ^2}}}}{\sqrt{2} \sqrt[4]{\frac{\left(\ell ^2-2r_b^2\right){}^2}{\ell ^4}+\mathcal{E}^2(\mathcal{E}^2+2)}}\right)\right)\,,\\
\mathcal{D}_c&=\pi\ell-\mathcal{D}_b\,.
 % \ell \left(\pi -2 \sin ^{-1}\left(\frac{\sqrt{\sqrt{\frac{\left(\ell ^2-2 r_b^2\right){}^2}{\ell ^4}+\mathcal{E}^2(\mathcal{E}^2+2)}-\mathcal{E}^2-1+2 \frac{r_b}{\ell} \sqrt{1-\frac{r_b^2}{\ell ^2}}}}{\sqrt{2} \sqrt[4]{\frac{\left(\ell ^2-2r_b^2\right){}^2}{\ell ^4}+\mathcal{E}^2(\mathcal{E}^2+2)}}\right)\right)\,.
\end{align}Remarkably, we conclude that the sum of the (interior plus exterior) lengths for $5$-dimensional SdS equals
\begin{eqnarray}
    \mathcal{D}_b+
    \mathcal{D}_c
=\pi \ell \label{5dL}
\end{eqnarray}
for any energy $\mathcal{E}$, which is the same as the length of any spacelike geodesic connecting conjugate observers at $\tau_N=\tau_S=0$ in empty, pure, de Sitter spacetime. Apparently the introduction of a black hole in $5$-dimensional de Sitter does not change the total length of the spacelike geodesics, it just splits it up in interior black hole and exterior de Sitter parts, the division of the split depending on the mass of the black hole, where the interior black hole length is the shortest for generic (non-vanishing and not maximal) black hole mass. 

In four dimensions the final expressions for the integrals eq. \eqref{L_b} and \eqref{L_c} are much more involved, but can still be written elegantly in terms of special functions of the horizon and turning point radii
  \begin{align}
\mathcal{D}_c&=\frac{4 i \ell}{\sqrt{r_2} \sqrt{2 r_1+r_2}} \left\{r_1 F(A|B)+\left(r_1-r_2\right) \Pi \left(\frac{
   r_2}{r_1}B;A|B\right)\right\}\,\label{yy1},\\
\mathcal{D}_b&=\mathcal{D}_c(r_1\longleftrightarrow r_2)\,,\label{yy2}
\end{align}  

where $F$ is the elliptic function of the first kind and $\Pi$ is the elliptic function of the second kind. Furthermore, we have identified,
\begin{align}A&\equiv\arcsin{\left(\sqrt{\frac{\left(2 r_1+r_2\right) \left(r_{\mathcal{O}}-r_2\right)}{\left(r_1+2 r_2\right) \left(r_{\mathcal{O}}-r_1\right)}}\right)}\,,\quad B\equiv \frac{r_1 \left(r_1+2 r_2\right)}{r_2 \left(2 r_1+r_2\right)}\,.
\end{align}

\begin{figure}[h!]
\begin{subfigure}{.5\textwidth}
\centering
\includegraphics[width=7.5cm]{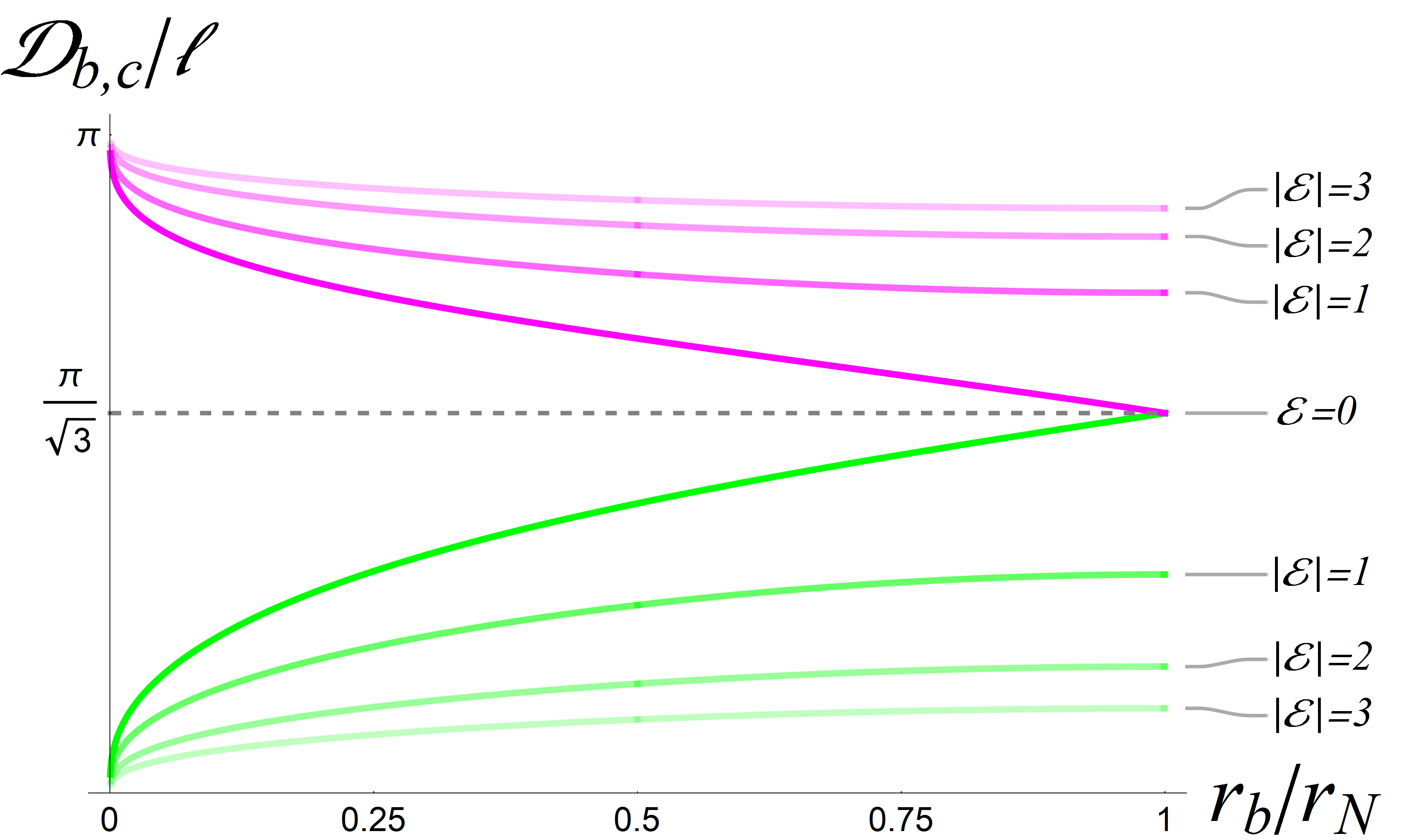}
\caption{4D spatial lengths.}
\end{subfigure}%
\begin{subfigure}{.5\textwidth}
\centering
\includegraphics[width=7.5cm]{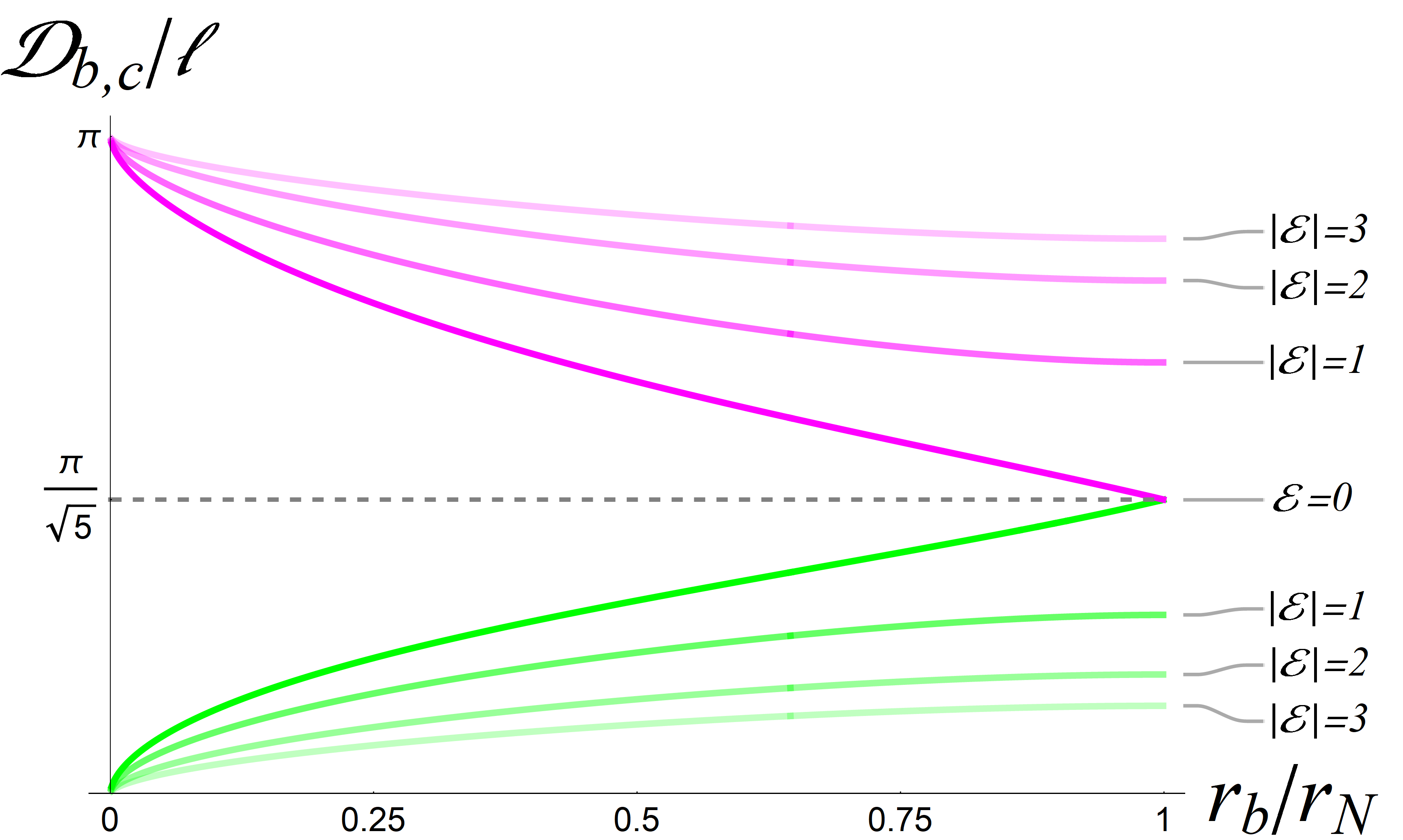}
\caption{6D spatial lengths.}
\end{subfigure}
\caption{Different lengths of interior and exterior spacelike geodesics in $D=4$ (left) and $D=6$ (right) as a function of the black hole radius $r_b$. Plotted for different $\mathcal{E}=0,1,2,3$, indicated by their respective opacity.}
\label{fig:LeoComp}
\end{figure}
In this case the sum of the two lengths, for arbitrary $\mathcal{E}$, no longer satisfies \eqref{5dL}. Instead, in $D=4$ their sum is larger, whereas in $D=6$ it is smaller, for generic (non-vanishing) black hole mass. We also note that the total geodesic length connecting the symmetric points on the static sphere becomes less and less sensitive to the presence of the black hole in the limit $|\mathcal{E}|\rightarrow\infty$ and approaches the empty de Sitter result 
\begin{equation}
\mathcal{D}_b\xrightarrow{\mathcal{E}\rightarrow\infty}0\,,\quad\mathcal{D}_c\xrightarrow{\mathcal{E}\rightarrow\infty}\pi\ell=\mathcal{D}_{dS}\,.\label{L_cinf}
\end{equation}

The vanishing of $\mathcal{D}_b$ in that limit is not surprising, as the radial spacelike geodesic crossing the black hole interior becomes null, but the other finite limit \eqref{L_cinf} might at first be considered surprising. However, as for empty de Sitter spacetime, this can be understood by realizing that the limit to a radial null geodesic through the exterior cannot be regular, as null geodesics reach (or `bounce off') future infinity in de Sitter, which is an infinite distance away for any spacelike geodesic.

%Interestingly in SdS geometry the spacelike geodesic the length of spacelike geodesic going through the interior of the cosmology is given by $\mathcal{D}_{dS}$, when $\mathcal{E}$ approaches infinity. 

Having understood the details of the symmetric radial spacelike geodesics between a pair of conjugate static sphere observers, as well as the lengths of these geodesics, we can sketch them in the Penrose diagram of SdS spacetime. The results can be found in Figure \ref{fig:LeoComp5}, Figure \ref{fig:LeoComp}, and Figure \ref{fig:TimeComp466} for the five, four, and six-dimensional examples, respectively.

This ends our (partially qualitative, especially in $D=6$) discussion on radial spacelike geodesics. The most important conclusions are that in the presence of a black hole, symmetric radial spacelike geodesics between conjugate static sphere observers, probing the interior black hole, and the exterior de Sitter region, exist within two generically different temporal windows. Only in $D=5$ are the interior and exterior windows the same, implying that the inward and outward radial spacelike geodesics can be smoothly connected (have the same value of $\mathcal{E}$), and their combined length is independent of $\mathcal{E}$ and equal to the length of spacelike geodesics in empty de Sitter. 

In $D=4$ the interior/exterior symmetry is absent, with the exterior spacelike geodesics absent before the interior spacelike geodesics, meaning that the inward and outward radial spacelike geodesics cannot be connected smoothly, and that the interior black hole (centered) causal diamond and the exterior de Sitter (centered) causal diamond are disconnected. In $D=6$ this situation is reversed, implying that the interior black hole (centered) causal diamond is overlapping with the exterior de Sitter (centered) causal diamond. 
% Added below sentence %
Extending to higher dimensions we expect the overlap between the interior and exterior causal diamonds to increase, and one could speculate that in the $D\rightarrow \infty$ limit the exterior causal diamond extends maximally into the black hole interior, possibly implying that a spatial section (at $t=0$) becomes a full Cauchy slice. It would be interesting to check this. 
These modifications to the causal structure are bound to have interesting consequences for (symmetric) correlators between a pair of conjugate static sphere observers.

\section{Conclusions and discussion}

In this paper we analyzed the (changes in the) causal structure of SdS spacetime, with respect to special static sphere observers, in $D=4,5$ and $6$. We traced radial null geodesics as they cross the interior black hole horizon or the exterior cosmological horizon, introducing imaginary contributions to the (static) time parameter along the radial null geodesic. The generated imaginary parts are related to the Euclidean periodicity of the corresponding horizon and as long as one separates the calculations for geodesics going through the interior black hole from geodesics going through the exterior de Sitter region, this procedure is completely well-defined and regular.  
Using these standard techniques, we explicitly derived the inward bending of the black hole singularity, as expected \cite{Fidkowski:2003nf}, and the outward bending of spacelike (asymptotic) de Sitter infinity, with respect to free-falling, static sphere, observers. Related to this we then computed the temporal windows at the static sphere observer marking the relevant interior (black hole) and exterior (de Sitter) causal regions, in $D=4,5$ and $6$ dimensions. 
% modified below sentence (modified again)
The $5$-dimensional SdS spacetime is special, due to the particular form for the blackening factor that relates the small and large $r$ limit. Specifically the $1/r^2$ and $r^2$ factors have the same limits for $r\rightarrow0$ and $r\rightarrow\infty$ respectively, implying that the relevant interior and exterior causal regions just connect to cover a complete Cauchy slice of the SdS spacetime. Whereas in $D=4$ the interior and exterior causal regions are disconnected, and in $D=6$ they share a finite size region. We expect that this behavior can be extrapolated into higher dimensions, increasing the finite size region as the number of dimensions grow. In all cases the inward and outward bending increases to a maximum size, as a function of the black hole mass, and then decreases again to vanish in the maximal black hole Nariai limit. 

This modified structure plays an important role when considering protocols for information recovery and exchange in de Sitter space times \cite{Aalsma:2021kle, Morvan:2023tele}. In particular, the outward bending of de Sitter future infinity implies that signals crossing the de Sitter exterior have more time to be intersect a (positive energy) shockwave emitted \cite{Gao:2000ga,Galante:2023uyf} from the conjugate static sphere observer, and be scattered into the causal region of that same conjugate static sphere observer. Black holes play a natural role of providing an energy reservoir for the positive energy shockwaves in these scenarios and it would be interesting to study the details of those information exchange protocols \cite{Freivogel:2019lej,Freivogel:2019whb}, especially regarding the question of how much information can be transmitted this way \cite{Morvan:2023tele}.  

The existence of temporal windows for static sphere observers also has interesting consequences for the presence of radial spacelike geodesics connecting conjugate static sphere observers at the same (global) time, which should be of interest when applying a geodesic approximation of correlators between two conjugate static sphere observers. By introducing a black hole into de Sitter spacetime conjugate static sphere observers are connected by spacelike geodesics for some finite amount of time, depending on the mass of the black hole, and whether one probes the interior (crossing the black hole) or the exterior (crossing the de Sitter region). In $D=5$ the interior and exterior times and conserved quantities $\mathcal{E}$ match, implying the connected geodesic is everywhere smooth, but in $D=4$ and $D=6$ the slope at the static sphere time is discontinuous. We also computed their lenghts, finding the particularly interesting result that the combined interior and exterior total length in $D=5$ exactly equals the length of all radial spacelike geodesics in pure (empty) de Sitter spacetime. In $D=4$ and $D=6$ the sum of interior black hole and exterior de Sitter length is bigger, respectively smaller, as compared to empty de Sitter spacelike geodesics. 
% Added sentence below %
In $D=6$ (and higher) we have not been able to solve the relevant equations analytically. In the large $D$ limit one might expect the relevant equations to simplify and this could be interesting to study further.   
In addition, beyond some finite time, that we quantified analytically in $D=4$ and $D=5$, either the interior or the exterior spacelike geodesic disappears and one expects a phase transition that introduces complex geodesics (labeled by complex $\mathcal{E}$) connecting static sphere observers. We expect this to play an important role in a geodesic approximation to the correlator \cite{Fidkowski:2003nf, Aalsma:2022swk, Galante:2022nhj}. 

For example, consider the geodesic approximation to the (equal time) correlation function of a scalar field operators $\phi$ of (large) mass $m \gg 1/l$
\begin{equation}
    \braket{\phi \phi}(t)=\sum_i^N e^{-m\,\mathcal{D}_i(t)}\,,
\end{equation}
 which is a sum is over all lengths of the spacelike geodesics connecting the two points in the correlator. Of course, except for the Nariai limit, for the SdS geometry the spacelike geodesic crossing the black interior is always shorter and therefore dominates the correlator, as illustrated in figures (\ref{fig:LeoComp5}) and (\ref{fig:LeoComp}).   
 Recently, the geodesic approximation was used to establish a connection between $2$-dimensional de Sitter space-time and a particular implementation of (an entangled pair of) Sachdev–Ye–Kitaev (SYK) models \cite{VerlindeH:2023SYKdS, Verlinde:2024znh}, claiming to be a potential unitary holographic description of de Sitter that is correctly able to reproduce the known bulk correlators. Generalising to SdS spacetimes, which arguably should be considered as a finite energy excitation in a holographic description of de Sitter \cite{Dinsmore:2019elr, Chakrabhavi:2023avi}, it would be of interest to construct the relevant generalised SdS correlators in a geodesic approximation. Generalising \cite{Fidkowski:2003nf}, after some particular time that we identified in this work, one requires complex values for the parameter $\mathcal{E}$, rendering the lengths $\mathcal{D}_i$ complex as well, suggesting the following structure of the SdS (conjugate static sphere) correlator
\begin{equation}
    \sum_i^N \mathcal{D}_i=\begin{cases}
        \mathcal{D}_b+\mathcal{D}_c
    &,\,0\leq |t|\leq \mathcal{T}_{min}\\
(\mathcal{D}_{c,b}+\mathcal{D}_{c,b}^{*})+\mathcal{D}_{b,c} &,\,\mathcal{T}_{min}< |t|\leq \mathcal{T}_{max}\\
(\mathcal{D}_{b,c}+\mathcal{D}_{b,c}^{*})+(\mathcal{D}_{c,b}+\mathcal{D}_{c,b}^{*}) &, \,\mathcal{T}_{max}< |t|\leq \infty
    \end{cases} \,,
\end{equation}
where we should use the $c$ index in the second line for $D=4$ and the $b$ index for $D>5$. For $D=5$, the black hole interior and exterior de Sitter temporal windows are the same  $\mathcal{T}_{min}=\mathcal{T}_{max}$, avoiding a distinction between interior and exterior (real or complex) geodesics. Although we have not studied the complex geodesic regime in this paper, with the help of the detailed results for radial null and spacelike geodesics in SdS spacetimes we hope to soon report on progress for complex geodesics in SdS as well. 

Finally, let us make a final comment on the appropriate SdS vacuum state that should be selected by these static sphere correlators. As a consequence of the semi-classical non-equilibrium nature of SdS the identification of an appropriate vacuum state is problematic. However, our analysis clearly suggests, in line with earlier observations \cite{Draper:2022ofa, Morvan:2022ybp}, that it should be consistent to split SdS spacetime, for instance at the static sphere, into an interior black hole part and and exterior de Sitter part that each correspond to smooth (complex) geometries for which a natural Euclidean (Hartle-Hawking type) vacuum can be selected. Physically one can presumably understand this in terms of a adding a reflective sphere at the static sphere radius \cite{Morvan:2023tele}, keeping both the black hole and cosmological horizon in equilibrium. It would be interesting to work out the details of such a SdS vacuum state construction \cite{Brum:2014nea,Witten:2023xze} and corresponding correlation functions.

\acknowledgments
We would like to thank
Ben Freivogel, Lars Aalsma and Damian Galante for useful discussion. 
This work is part of the Delta ITP consortium, a program of the Netherlands Organisation for Scientific Research (NWO) that is funded by the Dutch Ministry of Education, Culture and Science (OCW).
MMF is partly supported by funds from NSERC.

\appendix
\section{Appendix}
The $D$-dimensional SdS metric is given by
\begin{equation}
    ds^2=-f(r)dt^2+\frac{1}{f(r)}dr^2+ r^2 \, d\Omega^2_{D-2}\,,\label{general blackapp}
\end{equation}
where the blackening factor $f(r)$ for dimensions $D>3$ equals
\begin{equation}
f(r) = 1 - \frac{r^2}{\ell^2} - \alpha\frac{M}{r^{D-3}}\nonumber\label{bfactorSdSapp} \, .
\end{equation}
The roots of blackening factors can be found to follow the relations \cite{Morvan:2022ybp}
\begin{align}\label{desitterradiirelation}
    \sum_{i=1}^{D-1}r_i^n&=\left\{
    \begin{array}{ll}
        0\quad\,\,\,\,\,\text{for $0<n\le D-2$\, odd,}\\
        2\ell^n\,\,\,\,\text{for $0<n\le D-2$\, even,}
    \end{array}\right.\, ,\quad\ell^2=\frac{r_c^{D-1}-r_b^{D-1}}{r_c^{D-3}-r_b^{D-3}}\,,
\end{align}
 providing $D-2$ relations between the $D-1$ roots and enabling to rewrite every root $r_i$ as a function of only a single one. We will consider $r_b$ and $r_c$ to be real positive roots in this discussion.\\\\
For more readability, in section \ref{sec:example-section}, we omitted analytic formulae for \eqref{xx1} and \eqref{xx2} determining the finiteness of the temporal windows. Nevertheless, the sizes of the windows can be explicitly obtained in function of the black hole and cosmological horizon radius thanks to \eqref{desitterradiirelation}. In this appendix, we give the explicit relations in four and six dimensions and also provide some of rather long analytic expressions from \ref{sec2}.
\subsection{4D windows}
In four dimensions, we find that the temporal windows are given by
\begin{align}
    \tilde{{\mathcal{T}}}_{c}=\frac{\gamma  \ell ^2}{\left(r_b-r_c\right) \left(2 r_b+r_c\right) \left(r_b+2 r_c\right)}& \Big\{i r_c \left(2 r_b+r_c\right) \Big(i \ln\left(r_c-r_{\mathcal{O}}\right)-\pi\Big)\nonumber\\
   +r_b \left(r_b+2 r_c\right) &\ln\left(r_{\mathcal{O}}-r_b \right)+\left(r_c^2-r_b^2\right) \ln\left(r_b+r_c+r_{\mathcal{O}}\right)\Big\}\,,\\
   %  \textbf{t}_{b}=\frac{\gamma  \ell ^2}{\left(r_b-r_c\right) \left(2 r_b+r_c\right) \left(r_b+2 r_c\right)}&\Big(r_b^2 \left(-\ln \left(\left(r_b+r_c\right) \left(r_{\mathcal{O}}-r_b\right)\right)\right)\nonumber\\
   %  -2 r_b r_c \ln \left(r_c
   % \left(r_b-r_{\mathcal{O}}\right)\right)+&2 r_b r_c \ln \left(r_c-r_{\mathcal{O}}\right)+\left(r_b-r_c\right)
   % \left(r_b+r_c\right) \ln \left(r_b+r_c+r_{\mathcal{O}}\right)\nonumber\\
   % +i \pi  r_c \left(2 r_b+r_c\right)+&r_b \left(r_b+2
   % r_c\right) \ln \left(-r_b\right)\nonumber\\
   % +2 r_c^2 \arctanh&\left(\frac{2 r_c}{r_b}+1\right)+r_c^2 \ln
   % \left(r_c-r_{\mathcal{O}}\right)\Big)
   \tilde{\mathcal{T}}_{b}=\,\tilde{\mathcal{T}}_{c}(r_{\mathcal{O}}\longrightarrow 0)-\tilde{\mathcal{T}}_{c}\,.\quad\quad\,\quad\,\,\quad&\,
\end{align}
Evaluating the real part of those expressions leads to the left panel of figure \eqref{4dw}, whereas taking the imagery part yields the right panel.
\subsection{6D windows}
In six dimensions, the temporal windows can be expressed as
\begin{align}
    \tilde{\mathcal{T}}_{c}=\gamma  \ell ^2 &\bigg\{-\frac{r_a^3 \ln\left(r_a-r_{\mathcal{O}}\right)}{\left(r_a-r_b\right) \left(r_a-r_c\right)\left(r_a-r_d\right) \left(r_a-r_e\right)}+\frac{r_b^3 \ln\left(r_b-r_{\mathcal{O}}\right)}{\left(r_a-r_b\right) \left(r_b-r_c\right)\left(r_b-r_d\right) \left(r_b-r_e\right)}\nonumber\\
    +&\frac{r_c^3 \ln\left(r_c-r_{\mathcal{O}}\right)}{\left(r_a-r_c\right) \left(r_c-r_b\right)\left(r_c-r_d\right) \left(r_c-r_e\right)}+\frac{r_d^3 \ln\left(r_d-r_{\mathcal{O}}\right)}{\left(r_a-r_d\right) \left(r_d-r_b\right)\left(r_d-r_c\right) \left(r_d-r_e\right)}\nonumber\\
    +&\frac{r_e^3 \ln
   \left(r_e-r_{\mathcal{O}}\right)}{\left(r_a-r_e\right) \left(r_e-r_b\right)
   \left(r_e-r_c\right) \left(r_e-r_d\right)}\bigg\}\label{eq:6dwinc}\\
    % \frac{1}{4} \gamma  \ell ^2 \bigg\{&\frac{2 \left(\tilde{r}_d-i \tilde{r}_e\right){}^3 \arctan
   % \left(\frac{\tilde{r}_e}{r_{\mathcal{O}}-\tilde{r}_d}\right)}{\tilde{r}_e \left(-\tilde{r}_d+i\tilde{r}_e+r_b\right) \left(\tilde{r}_d-i \tilde{r}_e-r_c\right) \left(3 \tilde{r}_d-i\tilde{r}_e+r_b+r_c\right)}\nonumber\\
   % &-\frac{i \left(\tilde{r}_d-i \tilde{r}_e\right){}^3 \ln\left(\left(\tilde{r}_d-r_{\mathcal{O}}\right){}^2+\tilde{r}_e^2\right)}{\tilde{r}_e \left(-\tilde{r}_d+i\tilde{r}_e+r_b\right) \left(\tilde{r}_d-i \tilde{r}_e-r_c\right) \left(3 \tilde{r}_d-i\tilde{r}_e+r_b+r_c\right)}\nonumber\\
   % &+\frac{i \left(\tilde{r}_d+i \tilde{r}_e\right){}^3 \ln\left(\left(\tilde{r}_d-r_{\mathcal{O}}\right){}^2+\tilde{r}_e^2\right)}{\tilde{r}_e \left(\tilde{r}_d+i\tilde{r}_e-r_b\right) \left(-\tilde{r}_d-i \tilde{r}_e+r_c\right) \left(3 \tilde{r}_d+i\tilde{r}_e+r_b+r_c\right)}\nonumber\\
   % &+\frac{2 \left(\tilde{r}_d+i \tilde{r}_e\right){}^3 \arctan\left(\frac{\tilde{r}_e}{r_{\mathcal{O}}-\tilde{r}_d}\right)}{\tilde{r}_e \left(\tilde{r}_d+i\tilde{r}_e-r_b\right) \left(-\tilde{r}_d-i \tilde{r}_e+r_c\right) \left(3 \tilde{r}_d+i\tilde{r}_e+r_b+r_c\right)}\bigg\}\,,\\
    \tilde{\mathcal{T}}_{b}=\,\tilde{\mathcal{T}}_{c}(&r_{\mathcal{O}}\rightarrow 0)-\tilde{\mathcal{T}}_{c}
    \label{eq:6dwinb}\,,
\end{align}
where $r_a,r_d$ and $r_e$ are the additional roots of the blackening factor $f(r)$. $r_a$ is the negative root while $r_d$ and $r_e$ are the complex roots.  Taking the real part of equations \eqref{eq:6dwinc}  and \eqref{eq:6dwinb} yields the left panel of figure \ref{fig:6dwindows}. 
\section{5D spatial geodesic lengths $\mathcal{D}_{b,c}$ and starting times $\tilde{\tau}_{b,c}$}
The five dimensional spatial lengths can be compactly written in terms of $\mathcal{E}$, we find
\begin{align}
     \mathcal{D}_{b,c}&=
     % 2\ell \arcsin{\sqrt{\frac{r_1^2-r_{\mathcal{O}}^2}{r_1^2-r_2^2}}}\\
     \pi\ell-2\ell \arcsin \sqrt{\frac{\sqrt{\left(1-2\frac{r_b^2}{\ell^2}\right)^2+\mathcal{E}^2(\mathcal{E}^2+2)}\pm \mathcal{E}^2\pm1\mp2 \frac{r_b}{\ell}\sqrt{1-\frac{r_b^2}{\ell ^2}}}{2 \sqrt{\left(1-2\frac{r_b^2}{\ell^2}\right)^2+\mathcal{E}^2(\mathcal{E}^2+2)}}}\,.
    %  ,\\
    % \mathcal{D}_c&=
    % % \pi\ell-\mathcal{D}_b\\
    %  \ell  \left(\pi -2 \sin ^{-1}\left(\frac{\sqrt{\sqrt{\frac{\left(\ell ^2-2 r_b^2\right){}^2}{\ell ^4}+\mathcal{E}^2(\mathcal{E}^2+2)}-\mathcal{E}^2-1+2 \frac{r_b}{\ell} \sqrt{1-\frac{r_b^2}{\ell ^2}}}}{\sqrt{2}  \sqrt[4]{\frac{\left(\ell ^2-2r_b^2\right){}^2}{\ell ^4}+\mathcal{E}^2(\mathcal{E}^2+2)}}\right)\right)\,.
\end{align}
The dependence of 
starting time over energy $E$,
\begin{align}
\tilde{\tau}_{b}&=\frac{\ell^2\,E}{r_b+r_c} \left(\frac{r_b^2 \arctanh\left(\frac{\sqrt{r_{\mathcal{O}}^2-r_1^2}}{\sqrt{r_2^2-r_{\mathcal{O}}^2}}\frac{\sqrt{r_2^2-r_b^2}
}{\sqrt{r_b^2-r_1^2}
}\right)}{\sqrt{r_2^2-r_b^2}
\sqrt{r_b^2-r_1^2}}-\frac{r_c^2 \arctan\left(\frac{\sqrt{r_{\mathcal{O}}^2-r_1^2}}{\sqrt{r_2^2-r_{\mathcal{O}}^2}}\frac{\sqrt{r_c^2-r_2^2}}{\sqrt{r_c^2-r_1^2}}\right)}{\sqrt{r_c^2-r_1^2}
\sqrt{r_c^2-r_2^2}}\right)-i\frac{\beta_b}{4}\,,\\
 \tilde{\tau}_{c}&=\frac{i \pi \ell ^2 \, E}{2
\left(r_b+r_c\right)}\left(\frac{r_c^2}{\sqrt{r_2^2-r_c^2}
\sqrt{r_c^2-r_1^2}}-\frac{r_b^2}{\sqrt{r_2^2-r_b^2} \sqrt{r_b^2-r_1^2}}\right)-i\frac{\beta_c}{4}\nonumber\\
&-\frac{\ell^2\,E}{r_b+r_c} \left(\frac{r_b^2 \arctanh\left(\frac{\sqrt{r_{\mathcal{O}}^2-r_1^2}}{\sqrt{r_2^2-r_{\mathcal{O}}^2}}\frac{\sqrt{r_2^2-r_b^2}
}{\sqrt{r_b^2-r_1^2}
}\right)}{\sqrt{r_2^2-r_b^2}
\sqrt{r_b^2-r_1^2}}-\frac{r_c^2 \arctan\left(\frac{\sqrt{r_{\mathcal{O}}^2-r_1^2}}{\sqrt{r_2^2-r_{\mathcal{O}}^2}}\frac{\sqrt{r_c^2-r_2^2}}{\sqrt{r_c^2-r_1^2}}\right)}{\sqrt{r_c^2-r_1^2}
\sqrt{r_c^2-r_2^2}}\right)\,,
\end{align}
\section{4D starting times $\tilde{\tau}_{b,c}$}
\subsection{Starting times $\tilde{\tau}_{b,c}$}
\begin{align}
    \tilde{\tau}_{c}&=\frac{2 i \mathcal{E} \ell ^3 \gamma\left(r_1-r_2\right) }{\sqrt{r_2} \sqrt{2 r_1+r_2}}
    \left\{\frac{r_1^2 F(A|B)}{\left(r_1-r_2\right) \left(r_1-r_b\right) \left(r_1-r_c\right) \left(r_b+r_c+r_1\right)}\right.\nonumber\\
   +&\frac{r_b^2 \Pi \left(\frac{\left(r_b-r_1\right)}{\left(r_b-r_2\right)}\frac{
   r_2}{r_1}B;A|B\right)}{\left(r_b-r_1\right)
   \left(r_b-r_2\right) \left(r_b-r_c\right) \left(2 r_b+r_c\right)}-\frac{r_c^2 \Pi \left(\frac{ \left(r_c-r_1\right)}{\left(r_c-r_2\right)}\frac{
   r_2}{r_1}B;A|B\right)}{\left(r_c-r_1\right) \left(r_c-r_2\right) \left(r_b-r_c\right) \left(r_b+2 r_c\right)}\nonumber\\
   +&\left.\frac{\left(r_b+r_c\right){}^2 \Pi \left(\frac{ \left(r_1+r_b+r_c\right)}{
   \left(r_2+r_b+r_c\right)}\frac{
   r_2}{r_1}B;A|B\right)}{\left(r_b+r_c+r_1\right) \left(r_b+r_c+r_2\right) \left(5 r_b r_c+2 r_b^2+2 r_c^2\right)}\right\}\,,\\
    \tilde{\tau}_{b}&=\tilde{\tau}_{c}(r_1\longleftrightarrow r_2)\,,\label{4dstartingtime}
\end{align}
where $F$ is the elliptic function of the first kind and $\Pi$ is the elliptic function of the second kind. Furthermore, we have defined
\begin{align}
    A&\equiv\arcsin{\left(\sqrt{\frac{\left(2 r_1+r_2\right) \left(r_{\mathcal{O}}-r_2\right)}{\left(r_1+2 r_2\right) \left(r_{\mathcal{O}}-r_1\right)}}\right)}\,,\quad B\equiv \frac{r_1 \left(r_1+2 r_2\right)}{r_2 \left(2 r_1+r_2\right)}\,.
\end{align}
This was used to produce the left panel of figure \ref{fig:TimeComp46}.
\section{Approximating the starting times $\tilde{\tau}_{b,c}$}
We find that the dependence of the connected times  $\tilde{\tau}_{b,c}$ on \textit{real} values of the energy parameter $\mathcal{E}$ can be well approximated by
\begin{equation}
    \tilde{\tau}_{b,c}(\mathcal{E})\approx\pm \frac{\mathcal{T}_{b,c}}{\pi/2}\arctan{2\mathcal{E}}\,,
\end{equation}
especially for $|\mathcal{E}|\ll1$ and $|\mathcal{E}|\gg1$. This approximation was used to produce the right panel of figure \ref{fig:TimeComp46}.

\bibliographystyle{JHEP}
\bibliography{refs}

\end{document}